\documentclass[usenatbib]{mn2e}
\usepackage{textcomp,graphicx,amsmath,pdflscape,amssymb}
\usepackage{mathptmx}
\usepackage{color}

\newcommand{\todo}{\ifmmode {\Huge \bullet} \else {\Huge$\bullet$}\fi}

\def \NMc08 {19 }
\def \NB04 {59 }
\def \NSh07 {22 }
\def \NM09 {30 }
\def \NSul07 {90 }
\def \NN07 {44 }
\def \NT11 {40 }
\def \ND09 {10 }

\newcommand{\NSDSS}{66653}	
\newcommand{\NSDSSraw}{74517}
\newcommand{\NSDSShb}{20894} 	
\newcommand{\NSDSSmg}{43995} 	
\newcommand{\NSDSShbmg}{6731}	
\newcommand{\NSDSSmgc}{10910}	
\newcommand{\NSDSSbal}{1764}	

\newcommand{\NQZ}{8873}
\newcommand{\NQZraw}{20315} 

\newcommand{\NQZmg}{9111} 
\newcommand{\NQZhbmg}{238}
\newcommand{\NQZmgc}{489}

\newcommand{\NSLAQ}{2287}
\newcommand{\NSLAQraw}{5873} 

\newcommand{\NSLAQhbmg}{39}
\newcommand{\NSLAQmgc}{217}
\newcommand{\NSLAQXSDSSraw}{639}
\newcommand{\NSLAQXSDSS}{262}

\newcommand{\ltsim}{\raisebox{-.5ex}{$\;\stackrel{<}{\sim}\;$}}
\newcommand{\gtsim}{\raisebox{-.5ex}{$\;\stackrel{>}{\sim}\;$}}
\newcommand{\vFWHM}{\ifmmode V_{\mbox{\tiny FWHM}} \else $V_{\mbox{\tiny FWHM}}$ \fi}

\newcommand{\mic}{\ifmmode \mu{\rm m} \else $\mu$m\fi}
\newcommand{\ld}{\ifmmode {\rm lt-days} \else lt-days \fi}
\newcommand{\kms}{\ifmmode {\rm km\,s}^{-1} \else km\,s$^{-1}$ \fi}

\newcommand{\ergs}{\ifmmode {\rm erg\,s}^{-1} \else erg\,s$^{-1}$ \fi}
\newcommand{\ergcms}{\ifmmode {\rm erg\,cm}^{-2}\,{\rm s}^{-1} \else erg\,cm$^{-2}$\,s$^{-1}$\fi}
\newcommand{\ergcmsA}{\ifmmode{\rm erg}\, {\rm cm}^{-2}\,{\rm s}^{-1}\,{\rm\AA}^{-1} \else erg\, cm$^{-2}$\, s$^{-1}$\, \AA$^{-1}$\fi}
\newcommand{\ergcmsHz}{\ifmmode{\rm erg\,cm}^{-2}\,{\rm s}^{-1}\,{\rm Hz}^{-1} \else erg\,cm$^{-2}$\,s$^{-1}$\,Hz$^{-1}$\fi}
\newcommand{\phcms}{\ifmmode {\rm ph\,cm}^{-2}\,{\rm s}^{-1} \else ,ph\,cm$^{-2}$\,s$^{-1}$\fi}
\newcommand{\phcmsA}{\ifmmode {\rm ph\,cm}^{-2}\,{\rm s}^{-1}\,{\rm\AA}^{-1} \else ph\,cm$^{-2}$\,s$^{-1}$\,\AA$^{-1}$\fi}

\newcommand{\Msun}{\ifmmode M_{\odot} \else $M_{\odot}$\fi}
\newcommand{\msun}{\ifmmode M_{\odot} \else $M_{\odot}$\fi}
\newcommand{\Lsun}{\ifmmode L_{\odot} \else $L_{\odot}$\fi}

\newcommand{\mpyr}{\ifmmode \Msun\,{\rm yr}^{-1} \else $\Msun\,{\rm yr}^{-1}$ \fi}

\newcommand{\auvo}{\ifmmode \alpha_{\nu,{\rm UVO}} \else $\alpha_{\nu,{\rm UVO}}$\fi}
\newcommand{\Ledd}{\ifmmode L_{\rm Edd} \else $L_{\rm Edd}$\fi}
\newcommand{\Luv}{\ifmmode L_{1450} \else $L_{1450}$\fi}
\newcommand{\Lop}{\ifmmode L_{5100} \else $L_{5100}$\fi}
\newcommand{\Lthree}{\ifmmode L_{3000} \else $L_{3000}$\fi}
\newcommand{\lLthree}{\ifmmode \log\left(\Lthree/\ergs\right) \else $\log\left(\Lthree/\ergs\right)$\fi}
\newcommand{\lledd}{\ifmmode L/L_{\rm Edd} \else $L/L_{\rm Edd}$\fi}
\newcommand{\lamLlam}{\ifmmode \lambda L_{\lambda} \else $\lambda L_{\lambda}$\fi}
\newcommand{\Lbol}{\ifmmode L_{\rm bol} \else $L_{\rm bol}$\fi}
\newcommand{\lLbol}{\ifmmode \log\left(\Lbol/\ergs\right) \else $\log\left(\Lbol/\ergs\right)$\fi} 

\newcommand{\Fthree}{\ifmmode F_{3000} \else $F_{3000}$\fi}

\newcommand{\fuv}{\ifmmode f_{\lambda}\left(1450{\rm \AA}\right) \else $f_{\lambda}\left(1450 {\rm \AA}\right)$\fi}
\newcommand{\fthree}{\ifmmode f_{\lambda}\left(3000{\rm \AA}\right) \else $f_{\lambda}\left(3000{\rm \AA}\right)$\fi}
\newcommand{\fH}{\ifmmode f_{\lambda}\left(1.65\micron\right) \else
$f_{\lambda}\left(1.65\micron\right)$\fi}

\newcommand{\mbh}{\ifmmode M_{\rm BH} \else $M_{\rm BH}$\fi}
\newcommand{\lmbh}{\ifmmode \log\left(\mbh/\Msun\right) \else $\log\left(\mbh/\Msun\right)$\fi} 
 
\newcommand{\mseed}{\ifmmode M_{\rm seed} \else $M_{\rm seed}$\fi}
\newcommand{\mbul}{\ifmmode M_{\rm Bulge} \else $M_{\rm Bulge}$\fi} 
\newcommand{\mstar}{\ifmmode M_{*} \else $M_{*}$\fi} 
\newcommand{\mhost}{\ifmmode M_{\rm Host} \else $M_{\rm Host}$\fi}
\newcommand{\mm}{\ifmmode M_{*}/M_{\rm BH} \else $M_{*}/M_{\rm BH}$\fi}
\newcommand{\mmwp}{\ifmmode \left(M_{*}/M_{\rm BH}\right) \else $\left(M_{*}/M_{\rm BH}\right)$\fi}
\newcommand{\ml}{\ifmmode M_{*}/L_{*} \else $M_{*}/L_{*}$\fi}
\newcommand{\mlwp}{\ifmmode \left(M_{*}/L\right) \else $\left(M_{*}/L\right)$\fi}
\newcommand{\mlk}{\ifmmode \left(M_{*}/L_{K}\right) \else $\left(M_{*}/L_{K}\right)$\fi}
\newcommand{\sigs}{\ifmmode \sigma_{*} \else $\sigma_{*}$\fi}
\newcommand{\fbol}{\ifmmode f_{\rm bol} \else $f_{\rm bol}$\fi}
\newcommand{\fbolwv}{\ifmmode f_{\rm bol}\left(\lambda\right) \else $f_{\rm bol}\left(\lambda\right)$\fi}
\newcommand{\fbolopt}{\ifmmode f_{\rm bol}\left(5100{\rm \AA}\right) \else $f_{\rm bol}\left(5100{\rm \AA}\right)$\fi}
\newcommand{\fbolthree}{\ifmmode f_{\rm bol}\left(3000{\rm \AA}\right) \else $f_{\rm bol}\left(3000{\rm \AA}\right)$\fi}
\newcommand{\fboluv}{\ifmmode f_{\rm bol}\left(1450{\rm \AA}\right) \else $f_{\rm bol}\left(1450{\rm \AA}\right)$\fi}

\newcommand{\zfpe}{\ifmmode z\simeq4.8 \else $z\simeq4.8$\fi}
\newcommand{\znetprev}{$z\simeq2.4$ and $\simeq3.3$}

\newcommand{\zsix}{$z \simeq 6.2$}

\newcommand{\hkband}{\textit{H}- and \textit{K}-band}
\newcommand{\bj}{\ifmmode b_{\rm J} \else $b_{\rm J}$\fi}

\newcommand{\Lya}{\ifmmode {\rm Ly}\alpha \else Ly$\alpha$\fi}
\newcommand{\Halpha}{\ifmmode {\rm H}\alpha \else H$\alpha$\fi}
\newcommand{\Hbeta}{\ifmmode {\rm H}\beta \else H$\beta$\fi}
\newcommand{\ha}{\ifmmode {\rm H}\alpha \else H$\alpha$\fi}
\newcommand{\hb}{\ifmmode {\rm H}\beta \else H$\beta$\fi}
\newcommand{\MgII}{\ifmmode {\rm Mg}\,\textsc{ii}\,\lambda2798 \else Mg\,{\sc ii}\,$\lambda2798$\fi}
\newcommand{\mgii}{\ifmmode {\rm Mg}{\textsc{ii}} \else Mg\,{\sc ii}\fi} 
\newcommand{  \ciii     }{\ifmmode {\rm C}\,\textsc{iii}\right] \else C\,\textsc{iii}]\fi}
\newcommand{  \CIII     }{\ifmmode {\rm C}\,\textsc{iii}\right]\,\lambda1909 \else C\,\textsc{iii}]\,$\lambda1909$\fi}
\newcommand{\CIV}{\ifmmode {\rm C}\,\textsc{iv}\,\lambda1549 \else C\,{\sc iv}\,$\lambda1549$\fi}
\newcommand{\civ}{\ifmmode {\rm C}\,\textsc{iv} \else C\,{\sc iv}\fi}
\newcommand{\feii}{Fe\,{\sc ii}}
\newcommand{\feiii}{Fe\,{\sc iii}}
\newcommand{\oi}{\ifmmode \left[{\rm O}\,\textsc{i}\right] \else [O\,{\sc i}]\fi}
\newcommand{\OI}{\ifmmode \left[{\rm O}\,\textsc{i}\right]\,\lambda6300 \else [O\,{\sc i}]$\,\lambda6300$ \fi}

\newcommand{\oii}{\ifmmode \left[{\rm O}\,\textsc{ii}\right] \else [O\,{\sc ii}]\fi}
\newcommand{\OII}{\ifmmode \left[{\rm O}\,\textsc{ii}\right]\,\lambda3727 \else [O\,{\sc ii}]\,$\lambda3727$ \fi}
\newcommand{\oiii}{\ifmmode \left[{\rm O}\,\textsc{iii}\right] \else [O\,{\sc iii}]\fi}
\newcommand{\OIII}{\ifmmode \left[{\rm O}\,\textsc{iii}\right]\,\lambda5007 \else [O\,{\sc iii}]\,$\lambda5007$\fi}

\newcommand{\HeIIuv}{He\,{\sc ii}\,$\lambda1640$}
\newcommand{\OIIIuv}{O\,{\sc iii}]\,$\lambda1663$}
\newcommand{\NIVuv}{N\,{\sc iv}]\,$\lambda1486$}

\newcommand  {\RBLR}        {\hbox{$ {R_{\rm BLR}} $}}

\newcommand{\Lhb}{\ifmmode L\left(\hb\right) \else $L\left(\hb\right)$\fi}
\newcommand{\fwhb}{\ifmmode {\rm FWHM}\left(\hb\right) \else FWHM(\hb)\fi}
\newcommand{\Lmg}{\ifmmode L\left(\mgii\right) \else $L\left(\mgii\right)$\fi}
\newcommand{\fwmg}{\ifmmode {\rm FWHM}\left(\mgii\right) \else FWHM(\mgii)\fi}
\newcommand{\Lciv}{\ifmmode L\left(\civ\right) \else $L\left(\civ\right)$\fi}
\newcommand{\fwciv}{\ifmmode {\rm FWHM}\left(\civ\right) \else FWHM(\civ)\fi}
\newcommand{\fwhm}{\ifmmode {\rm FWHM} \else FWHM\fi} 
\newcommand{\voff}{\ifmmode v_{\rm off} \else $v_{\rm off}$\fi} 

\newcommand{\mumg}{\ifmmode \mu\left(\mgii\right) \else $\mu\left(\mgii\right)$\fi}
\newcommand{\fmg}{\ifmmode f\left(\mgii\right) \else $f\left(\mgii\right)$\fi}
\newcommand{\muciv}{\ifmmode \mu\left(\civ\right) \else $\mu\left(\civ\right)$\fi}
\newcommand{\fciv}{\ifmmode f\left(\civ\right) \else $f\left(\civ\right)$\fi}

\newcommand{\hbXmg}{``\hb$\times$\mgii''}
\newcommand{\mgXciv}{``\mgii$\times$\civ''}
\newcommand{\hbXciv}{``\hb$\times$\civ''}

\title[Measuring \mbh\ \& \lledd]{Black Hole Growth to $z=2$ - I: Improved Virial Methods for Measuring \mbh\ \& \lledd}

\author[Trakhtenbrot \&\, Netzer]
{Benny Trakhtenbrot\thanks{E-mail: trakht@wise.tau.ac.il} and Hagai~Netzer\\
School of Physics and Astronomy, The Raymond and Beverly Sackler
 Faculty of Exact Sciences, Tel-Aviv University, Tel-Aviv 69978, Israel\\}

\date{Accepted 2012 September 4.  Received 2012 September 3; in original form 2012 April 19}

\pagerange{\pageref{firstpage}--\pageref{lastpage}} \pubyear{2012}

\begin{document}

\label{firstpage}

\maketitle

\begin{abstract}
\noindent
We analyze several large samples of Active Galactic Nuclei (AGN) in order to establish the best tools required to study the evolution of black hole mass (\mbh) and normalized accretion rate (\lledd). 
The data include spectra from the SDSS, 2QZ and 2SLAQ public surveys at $z<2$, and a compilation of smaller samples with $0<z<5$. 
We critically evaluate the usage of the \MgII\ and \CIV\ lines, and adjacent continuum bands, as estimators of \mbh\ and \lledd, by focusing on sources where one of these lines is observed together with \hb.
We present a new, luminosity-dependent bolometric correction for the monochromatic luminosity at 3000\AA, \Lthree, which is lower by a factor of $\sim1.75$ than those used in previous studies.
We also re-calibrate the use of \Lthree\ as an indicator for the size of the broad emission line region (\RBLR) and find that $\RBLR\propto\Lthree^{0.62}$, in agreement with previous results.
We find that $\fwmg\simeq\fwhb$ for all sources with $\fwmg\ltsim6000\,\kms$.  
Beyond this \fwhm, the \mgii\ line width seems to saturate.
The spectral region of the \mgii\ line can thus be used to reproduce \hb-based estimates of \mbh\ and \lledd, with negligible systematic differences and a scatter of $\sim$0.3 dex.
The width of the \civ\ line, on the other hand, shows no correlation with either that of the \hb\ or the \mgii\ lines and we could not identify the reason for this discrepancy. 
The scatter of \mbh(\civ), relative to \mbh(\hb) is of almost 0.5 dex.
Moreover, 46\% of the sources have $\fwciv\ltsim\fwhb$, in contrast with the basic premise of the virial method, which predicts $\fwciv/\fwhb\simeq\sqrt{3.7}$, based on reverberation mapping experiments.
This fundamental discrepancy cannot be corrected based on the continuum slope or any \civ-related observable.
Thus, the \civ\ line cannot be used to obtain precise estimates of \mbh. 
We conclude by presenting the observed evolution of \mbh\ and \lledd\ with cosmic epoch.
The steep rise of \lledd\ with redshift up to $z\simeq1$ flattens towards the expected maximal value of $\lledd\simeq1$, with lower-\mbh\ sources showing higher values of \lledd\ at all redshifts. 
These trends will be further analyzed in a forthcoming paper.
\end{abstract}


\section{Introduction}
\label{sec_intro}

The growth of Super-Massive Black Holes (SMBHs), which reside in the centers of most large galaxies, progresses through episodes of radiatively-efficient accretion, during which such systems appear as Active Galactic Nuclei (AGN). 
Many details of this growth process are still unknown.
In the local Universe most AGN are powered by lower-mass BHs, with $\mbh\sim10^{6}-10^{8}\,\Msun$, growing at very slow rates \cite[e.g., ][hereafter NT07]{Marconi2004,Hasinger2005,Netzer_Trakht2007}. It is thus clear that the more massive BHs accreted at higher rates in the past. 
Indeed, several studies suggest that the typical normalized accretion rates (\lledd) increase with redshift \cite[e.g., ][NT07]{Fine2006}.
In contrast, the few luminous $z\sim6$ QSOs for which \mbh\ was measured show extremely high masses, of about $\sim10^9\,\Msun$, and high accretion rates, near their Eddington limit. 
These systems could have grown to be the most massive BHs known ($\mbh\sim10^{10}\,\Msun$) as early as $z\sim4$ \cite[][hereafter T11]{Willott2010,Kurk2007,DeRosa2011,Trakhtenbrot2011}.

For un-obscured, type-I AGN, \mbh\ can be estimated by so-called ``single-epoch'' or ``virial'' estimators. These methods rely on the results of reverberation-mapping (RM) experiments, which provide empirical relations between the emissivity-weighted radius of the Broad Line Region (BLR) and the source luminosity $L$.
These $\RBLR-L$ relations are parametrized as $\RBLR \propto \left(\lamLlam\right)^{\alpha}$.
Assuming the motion of the BLR gas is virialized, the single-epoch \mbh\ estimators take the general form
\begin{equation}
 \mbh=f\,G^{-1}\,\left(\lamLlam\right)^{\alpha} V_{\rm BLR}^{2}\,\,\, ,
 \label{eq:MBH_virial}
\end{equation}
where $V_{\rm BLR}$ is a probe of the BLR velocity field and $f$ is a factor that depends on the geometrical distribution of the BLR gas.
A common estimator of this type is the ``\hb\ method'' (hereafter \mbh[\Lop]). 
Here \RBLR\ is estimated from \lamLlam\ at 5100\AA\ (hereafter \Lop), \fwhb\ is the velocity proxy, and $\alpha=0.6\pm0.1$ \cite[][]{Kaspi2000,Kaspi2005,Bentz2009_RL_host}. 
Another method is based on the \MgII\ line and the adjacent continuum luminosity (\Lthree). 
Although the few \mgii-dedicated reverberation campaigns have not yet revealed a robust $\RBLR-\Lthree$ relation \cite[e.g.,][]{Clavel1991,Metzroth2006}, 
several studies calibrated the \mgii\ relation by using UV spectra of sources that have \hb\ reverberation data. 
One particular example is the relation presented in \cite{McLure2002}, and later refined by \citet[hereafter MD04]{McLure_Dunlop2004}.
The ``\hb'' and ``\mgii'' estimators were used in numerous papers to derive \mbh\ for many thousands of sources.
This means focusing on either $z\lesssim0.75$ or $0.75\lesssim z \lesssim 2$ AGN \cite[e.g.,][]{Croom2004,Fine2006,Shen_dr5_cat_2008,Rafiee2011}. 
Other studies used these estimators for small samples of sources at higher redshifts, where the lines are observed in one of the NIR bands \cite[][]{Shemmer2004,Netzer2007_MBH,Kurk2007,Marziani2009,Dietrich2009_Hb_z2,Willott2010}.

In principal, \mbh\ can also be estimated from the broad \CIV\ line\footnotemark,
\footnotetext{Hereafter we refer to the UV lines under study simply as \mgii\ and \civ.}
since \RBLR(\civ) is known from RM experiments \cite[e.g.,][]{Kaspi2007}.
A specific calibration of this type is given in \cite{Vester_Peterson2006} and several other papers.
Such methods would potentially enable the study of large samples of AGN at $1.5\ltsim z\ltsim 5$ for which \civ\ is observed in large optical surveys \cite[e.g.,][]{Vestergaard2008,Vestergaard2009}.
However, there is evidence that such \civ-based estimates are highly uncertain.
In particular, \citet[][hereafter BL05]{Baskin2005} found that the \civ\ line is often blue-shifted with respect to the AGN rest-frame, and that \fwciv\ is often smaller than \fwhb, both indicating that the dynamics of the \civ-emitting gas may be dominated by non-virial motion 
A later study by \cite{Vester_Peterson2006} claimed that at least some of these findings may be due to the inclusion of narrow-line objects and low-quality spectra in the BL05 sample.
Despite this reservation, several subsequent studies of large, flux-limited samples clearly demonstrated that the relation between the widths of the \civ\ and \mgii\ lines shows considerable scatter and, as a result, the \civ-based estimates of \mbh\ can differ from those deduced from \mgii\ by up to an order of magnitude \cite[][]{Shen_dr5_cat_2008,Fine2010_CIV}. 
Moreover, studies of small samples of $z\sim2-3.5$ AGN by, e.g., \cite[][N07 hereafter]{Netzer2007_MBH}, \cite[][S04 hereafter]{Shemmer2004} and \cite[][D09 hereafter]{Dietrich2009_Hb_z2}, show that the large discrepancies between \civ\ and \hb\ mass estimates persist even in high-quality spectra of broad-line AGN (i.e. where $\fwhb \gtsim 4000\,\kms$).
These issues were investigated in a recent study by \cite{Assef2011}, which suggested an empirical correction for the \civ-based estimators that is based on the shape of the observed UV-optical spectral energy distribution (SED). This correction, however, may turn impractical for large optical surveys of high redshift AGN, where only  the rest-frame UV regime is observed.
In this paper, we re-examine the methods used to derive \mbh\ and \lledd\ of high-redshift type-I AGN. 
We discuss both large and small samples, including some that were never presented in this context.
The various samples are described in \S\ref{sec_samples} and the measurement procedures in \S\ref{sec_fitting}.
We discuss the monochromatic luminosities and bolometric corrections in \S\ref{sec_L_bol_corr}.
We briefly discuss the premise of the virial assumption in for estimating \mbh\ in \S\ref{sec_mbh_general}.
In \S\ref{sec_mbh_mgii} we examine how the \mgii\ emission line complex can be used to measure \mbh, 
and in \S\ref{sec_civ_prob} we provide new evidence regarding the fundamental limitations of the \civ\ method.
Finally, in \S\ref{sec_diss_con} we briefly describe the observed evolution of \mbh\ and \lledd, as measured with the tools developed in this paper, and summarize our conclusions.
A more detailed analysis of the evolutionary trends is deferred to a forthcoming paper.
Throughout this work we assume a cosmological model with $\Omega_{\Lambda}=0.7$, $\Omega_{\rm M}=0.3$, and
$H_{0}=70\,\kms\,{\rm Mpc}^{-1}$.

\section{Samples Selection}
\label{sec_samples}

The main goal of the present work is to test how the \MgII\ and \CIV\ emission line complexes can be used to estimate \mbh, \Lbol\ and \lledd, and to apply these methods to probe the evolution of \lledd\ to $z\simeq2$.
While \mgii\ can be compared to \hb\ or \civ\ within the same spectroscopic window (at $z\sim0.6$ or $z\sim1.7$, respectively), the comparison between \civ\ and \hb\ has to be based on a combination of separate observations.
This dictates two distinct types of samples: 
(1) large samples drawn from the Sloan Digital Sky Survey \cite[SDSS;][]{York2000}, 
2dF QSO Redshift survey \cite[2QZ;][]{Croom2004} and the 
2dF SDSS LRG And QSO survey \cite[2SLAQ;][]{Richards2005_2SLAQ, Croom2008} catalogs; 
and (2) several smaller AGN samples, with publicly available data. 
The following details all the samples used in the paper, which are summarized in Table~\ref{tab_samples}.
We note that the 2QZ and 2SLAQ surveys provided relatively few high-quality spectra, which are usable for the comparisons between the different lines (see below). 
All the spectra analyzed in the present work were corrected for galactic extinction using the maps of
\cite{Schlegel1998} and the model of \cite{Cardelli1989}.

\subsection{SDSS DR7}
\label{subsec_sample_sdss}

We queried the public catalogs of the seventh data release of the SDSS \cite[DR7;][]{Abazajian2009} for all sources that are classified as ``QSO'' and have a high confidence redshift determination (i.e. \texttt{zconf}$>0.7$) in the range of $z<2$.
This resulted in \NSDSSraw\ sources.
First, we omitted from this sample 7845 objects for which the line fitting procedures resulted in poor-quality fits, similarly to our criteria in NT07 (see more details in \S\ref{sec_fitting}).
Next, we omitted from our analysis all radio-loud sources.
The radio loudness of each source, $R_{\rm L}\equiv f_{\nu}\left(5\,\rm{GHz}\right)/f_{\nu}\left(4400\rm{\AA}\right)$ \cite[following ][]{Kellermann1989}, was derived from the \texttt{FIRST} data \cite[][]{Becker1995, White1997} incorporated in the SDSS public archive. 
For sources at $z>1$, we estimated $f_{\nu}\left(4400\rm{\AA}\right)$ using the continuum flux density near 2200\AA\ and by assuming $f_{\nu}\propto\nu^{-0.44}$ \cite[][VdB01 hereafter]{VandenBerk2001}.
We removed all sources with $R_{\rm L}>10$ from our sample. 

Finally, we omitted \NSDSSbal\ sources that are classified as broad absorption line QSOs (BALQSOs), based on their \mgii\ and/or \civ\ spectral regions. 
For this, we used the relevant flags in the SDSS/DR7 catalog of \citet[][see below]{Shen_dr7_cat_2011} that, in turn, is largely based on the SDSS/DR5 catalog of BALQSOs of \cite{Gibson2009_BAL_cat}, with further (manual and hence partial) classification of the post-DR5 spectra. 
Although the \cite{Shen_dr7_cat_2011} BALQSO classification is probably incomplete, its combination with our fit quality criteria provide a sample which is almost completely BALQSO-free.

The final SDSS sample comprises \NSDSShb\ for which \hb\ is observed ($z<0.75$) and \NSDSSmg\ sources for which \textit{only} the \mgii\ line is observed (i.e. $0.75<z<2$).
There are \NSDSShbmg\ sources for which both the \hb\ and \mgii\ lines are observed ($0.5<z<0.75$; the SDSS \hbXmg\ sub-sample hereafter) and \NSDSSmgc\ sources where both the \mgii\ and \civ\ lines are observed ($1.5<z<1.95$; the SDSS \mgXciv\ sub-sample hereafter).

The general SDSS sample significantly overlaps with the sample studied by \cite{Shen_dr7_cat_2011}. 
Virtually all our $z>0.75$ sources are found within the \cite{Shen_dr7_cat_2011} catalog, in particular about 98\% of our SDSS \hbXmg\ and \mgXciv\ sub-samples, which have \mbh\ derived from more than one emission line in that catalog. 
At lower redshifts the fractions are lower, and only $\sim68\%$ of our \hb-only SDSS sources can be found within the \cite{Shen_dr7_cat_2011} catalog.
This is due to our choice to include relatively faint sources that were excluded from the \cite{Shen_dr7_cat_2011} catalog, namely sources with $M_{i}\ge-22$ \cite[see also][]{Schneider2010_QSOCAT_DR7}. We verified that these faint sources have high-quality \hb\ fits, according to our criteria (see \S\ref{sec_fitting}).
In addition, we note that our SDSS \hbXmg\ sub-sample includes 354 of the 495 sources ($\sim75\%$) analyzed in the study by \cite{Wang_MgII_2009}. 
The discrepancy is due to the fact that \cite{Wang_MgII_2009} chose to include in their sample sources with slightly lower redshifts ($0.45<z<0.75$). Our general (\hb) SDSS sample includes 473 of their sources ($\sim95\%$).


\begin{table*}
  \caption{Summary of Samples}
  \label{tab_samples}
  \begin{center}
    \begin{tabular}{lcccccccc} \hline \hline
     & & &
    \multicolumn{3}{c}{Lines used$^{\rm a}$} & 
    \multicolumn{3}{c}{Sizes of cross-calibration samples} \\
    Name (Acc.) &
    ${\rm N}_{\rm Total}$ &
    $z$ &
    H$\beta$ &
    Mg\,{\sc ii} &
    C\,{\sc iv} &
    \hb$\times$\mgii &
    \mgii$\times$\civ &
    \hb$\times$\civ \\

    \hline
    
    SDSS  			& \NSDSS  & $0-2$        & $s$ & $s$ & $s$ & \NSDSShbmg\ & \NSDSSmgc\ & $-$ \\
    2QZ   			& \NQZmg    & $0.35-1.7$  & $s$ & $s$ & $s$ & \NQZhbmg\   & \NQZmgc\   & $-$ \\
    2SLAQ 			& \NSLAQ  & $0.45-1.6$  & $s$ & $s$ & $s$ & \NSLAQhbmg\ & \NSLAQmgc\ & $-$ \\

\hline
\\
\citet[][BL05]{Baskin2005}	& \NB04  & $0.25-0.5$	& $s$ & $-$ & $t$ & $-$	& $-$	& 81\\
\citet[][N07]{Netzer2007_MBH}$^{\rm b}$
				& \NN07  & $2.2-3.5$	& $s$ & $-$ & $s$ & $-$	& $-$	& 44\\
\citet[][Sh07]{Shang2007}	& \NSh07 & $0.1-0.4$    & $t$ & $t$ & $t$ & 22	& 22	& 22\\
\citet[][Sul07]{Sulentic2007} 	&\NSul07 & $<1$		&$t/s$& $s$ & $t$ &  9	& 11	& 57\\
\citet[][Mc08]{McGill2008}	& \NMc08 & $\sim0.36$	& $t$ & $t$ & $-$ & 19	& $-$	&$-$\\ 
\citet[][D09]{Dietrich2009_Hb_z2}& \ND09 & $1-2.2$	& $t$ & $t$ & $t$ &  7 	& 7	& 9 \\
\citet[][M09]{Marziani2009}	& \NM09  & $1-2.4$	& $t$ & $s$ & $s$ & 16	& 12	& 6 \\
\citet[][T11]{Trakhtenbrot2011}	& \NT11  & $\sim4.8$	& $-$ & $s$ & $s$ & $-$	& 40	&$-$\\

\hline
  
  \multicolumn{9}{l}{$^{\rm a}$ Type of data used: ``t'' - tabulated data; ``s'' - spectra fitted by our team 
\cite[i.e., this work,][N07 and T11]{Shemmer2004}.}  \\
  \multicolumn{5}{l}{$^{\rm b}$ The majority of sources in this sample were presented in \cite{Shemmer2004}.}\\
    \end{tabular}
  \end{center}
\end{table*}


\subsection{2QZ}
\label{subsec_sample_2qz}

We used the public 2QZ catalog of \cite{Croom2004}, initially choosing all $0.5<z<2$ sources which are classified as ``QSO'' and have high confidence redshift determination (i.e. \texttt{ZQUAL} flag better than 22).
Our query naturally omits BALQSOs, which are classified as such within the 2QZ catalog \cite[based on visual inspection of the spectra, which is incomplete; see][]{Croom2004}.
This query provided \NQZraw\ sources.
After applying the fit quality criteria to this large \mgii\ 2QZ sample, it comprises \NQZ\ sources.
Generally, spectra obtained as part of the 2QZ and 2SLAQ AAO-based surveys are not flux calibrated.
Previous studies used single-band magnitudes, in either the ${\rm B}_{\rm J}$ or the $g$ bands, to derive monochromatic luminosities, assuming all spectra follow a uniform UV-to-optical SED of $f_{\rm \nu}\sim\nu^{-0.5}$ \cite[e.g.,][]{Croom2004,Richards2005_2SLAQ,Fine2008}.
Here we employed a more robust flux calibration scheme, which relies on all the photometric data available in the 2QZ and 2SLAQ catalogs. This procedure is discussed in Appendix~\ref{app_AAO_fcal}.

In order to expand the luminosity range of the \hbXmg\ sample, we preformed a separate query of \textit{all} $0.35<z<0.55$ 2QZ sources, ignoring their class and \texttt{ZQUAL}. 
For the selection of \mgXciv\ sources we focused on sources at $1.5<z<1.8$ (within the larger \mgii\ sample)
These choices were made in order to include sources which might be flagged as problematic due to the proximity of the \hb\ line  to the telluric features near 7000\AA\ or to the limit of the observed spectral range, and to avoid these problems from affecting the \civ\ line.
The \hbXmg\ sources which passed the fit quality criteria were manually inspected, to omit any non-QSO objects or otherwise unreliable spectra. 
These procedures resulted in merely \NQZhbmg\ sources with reliable \hb\ fits and \NQZmgc\ sources with reliable \civ\ fits.

To summarize, we have \NQZmg\ 2QZ sources in our final sample, \NQZhbmg\ of which have both \hb\ and \mgii, and \NQZmgc\ sources with both \mgii\ and \civ.
These sources span the magnitude range $\bj \sim 18-20.85$, which corresponds to the luminosity range $44\lesssim\lLbol\lesssim46$ at $z\simeq1.5$ (see \S\ref{subsec_bol_corr}).
The 2QZ sample thus reaches about a factor $3.75$ deeper than the SDSS sample.

\subsection{2SLAQ}
\label{subsec_sample_2slaq}

Sources from the 2SLAQ survey were selected by using the public 2SLAQ QSO catalog, compiled and presented in \cite{Croom2008}, and further analyzed in \cite{Fine2010_CIV}. Our query focused on $0.5<z<1.8$ sources which are classified as ``QSO'', and resulted in \NSLAQraw\ sources.
Here too, BALQSOs are largely omitted, due to their separate classification. 
Low-redshift 2SLAQ sources suffer from the same limitations affecting the 2QZ low-redshift sources. 
We preformed a separate search to locate \hbXmg\ candidates at $0.4<z<0.6$. 
Our careful manual inspection resulted in only \NSLAQhbmg\ sources with reasonable fits to both the \hb\ and \mgii\ lines. 
We chose \textit{not} to include any of these sources in the present analysis, due to their poor S/N and small number.
Sources with $1.5<z<1.65$ in the large \mgii\ sample have both \mgii\ and \civ. 
We have \NSLAQmgc\ such sources, out of a total of \NSLAQ.
These sources span the magnitude range $g\sim18-22$, which corresponds to $44\lesssim\lLbol\lesssim46$ at $z\simeq1.5$.
This is almost an order of magnitude deeper than SDSS, and about a factor of 2.5  deeper than 2QZ.

\subsection{Additional Samples}
\label{subsec_sample_small}

There are many smaller samples in the literature with measurements of more than one of the three emission lines discussed here. 
We choose to focus on samples for which the relevant line measurements are either preformed by procedures very similar to ours and publicly tabulated, or on samples for which such measurements could be preformed on publicly available spectra. 
In particular, we used the samples of 
BL05 (\NB04 sources),
N07 (and S04; \NN07 sources), 
\citet[][hereafter Sh07 hereafter; \NSh07 sources]{Shang2007},
\citet[][hereafter Sul07; \NSul07 sources]{Sulentic2007}, 
\citet[][hereafter Mc08; \NMc08 sources]{McGill2008}, 
\citet[][hereafter D09; \ND09 sources]{Dietrich2009_Hb_z2},
\citet[][hereafter M09; \NM09 sources]{Marziani2009}, 
and
T11 (\NT11 sources).
Basic information on these samples is given in Table~\ref{tab_samples}, while Appendix~\ref{app_samples} provides more details regarding the acquisition and contents of the samples. 
In particular, Appendix~\ref{app_samples} details which line measurements we used cases where sources appeared in more than one sample.

These small samples probe a wide range of redshift and luminosity and most of them have higher S/N than those typically obtained within the large SDSS, 2QZ and 2SLAQ surveys. 
In many of these samples the different lines were \textit{not} observed simultaneously.
Thus, it is possible that line and/or continuum variability contributes to the scatter in the relationships discussed in this work.

\section{Line and Continuum Fitting}
\label{sec_fitting}

We have developed a set of line fitting procedures to derive the physical parameters related to the \hb, \mgii\ and \civ\ lines.
These procedures are similar in many ways to those described in previous publications (e.g., S04, N07, NT07, S08, F08). 
In particular, the \hb\ and \mgii\ fitting procedures follow those used in NT07 and T11, respectively.
In short, a linear continuum model is fitted to the flux densities in two 20\AA-wide ``bands'' on both side of the relevant emission line. 
The bands are centered around 
4435 \& 5110\AA\ for \hb,
2655 \& 3020\AA\ for \mgii\ and
1450 \& 1690\AA\ for \civ.
For the \hb\ and \mgii\ lines, we subtract a template of \feii\ and \feiii\ emission lines. 
This is done by choosing the best-fit broadened and shifted template based on data published by \cite{BG92} and \cite{Vestergaard2004}. 
The grid spans a broad range both in the width and (velocity) shift of the iron features.
In the case of \mgii, we add flux in the wavelength range under the emission line itself (see Appendix~\ref{app_mgii})
The main emission lines are modeled by a combination of two broad and one narrow Gaussians. 
In the cases of \mgii\ and \civ\ this model is duplicated for each of the doublet components.
In Appendix~\ref{app_line_fitting} we discuss several minor but important technical issues concerning the fitting of the lines, including: 
(1) the ranges of all relevant parameters;
(2) the narrow ($\ltsim1000\,\kms$) components of all lines; 
(3) the improved iron template for the \mgii\ spectral complex and its scaling; 
(4) the difference between our fitted \Lthree\ and the ``real'' power-law continuum,   
and (5) the emission features surrounding the \civ\ line.

All fitting procedures were tested on several hundred spectra, selected from all samples used here. For these, we individually inspected the fitted spectra. 
In particular, we verified that the simplifications involved in the \civ\ fitting procedure (Appendix~\ref{app_civ}) are justified.
After fitting the SDSS, 2QZ and 2SLAQ samples, we filtered out low-quality fits by imposing the criteria $\chi^2<5$ and $R^2>0.2$.\footnotemark
\footnotetext{$R^2$ is the coefficient of determination, related to the amount of variance in the data that is accounted for by the model.}
Both these quantities were calculated over a range of $14000\,\kms$ centered on the peaks of the relevant emission lines, and adjusted for degrees of freedom. This choice is similar to the one we made in our previous analysis of large samples (NT07).
The smaller samples were treated manually, securing high-quality fitting results for all sources where adequate spectral coverage of the various lines was available.
We stress that even when two of the three main lines are observed in the same spectrum (i.e. for the \hbXmg\ and \mgXciv\ sub-samples), the fitting procedures are ran separately for each emission line complex. 
Thus, the parameters of the different models are fully independent, even when it is possible to assume otherwise.

The line fitting procedures provide monochromatic luminosities (\Luv, \Lthree\ and/or \Lop) and emission lines widths.
For the latter, we calculated the FWHM, the line dispersion $\sigma$ \cite[following][]{Peterson2004} and the inter-percentile velocity \cite[IPV; following][]{Croom2004}.
We estimated \RBLR\ for all the sources for which the ``\hb\ method'' is applicable, by:
\begin{equation}
\RBLR\left(\hb\right) = 538.2 \left (\frac{L_{5100}}{10^{46}\,\ergs}\right )^{0.65} \,\, \ld \,  .
\label{eq:R_L5100_original}
\end{equation}
This is an updated version of the correlation found by \cite{Kaspi2005}, taking into account the improved measurements presented in \cite{Bentz2009_RL_host}, but fitting the RM measurements of only the sources with $\Lop\gtsim10^{44}\,\ergs$, typical of the luminosities of high-redshift sources.
\mbh(\hb) is derived by assuming $f=1$ \cite[][and references therein]{Netzer_Marziani2010} and is given by: 
\begin{equation}
  \mbh\left(\hb\right)=1.05\times 10^8  
			\left(\frac{L_{5100}}{10^{46}\,\ergs}\right)^{0.65} 
			\left[\frac{\fwhb}{10^3\,\kms}\right]^2 \,\, \Msun \,\, .
\label{eq:M_Hb}
\end{equation}
We have also experimented with flatter \RBLR-\Lop\ relations, of $\RBLR\propto \Lop^{0.5}$, as suggested by some studies \cite[e.g.][]{Bentz2009_RL_host}. We found, as in \cite{Kaspi2005}, that these have little effect on our main results, provided that the constant in Eq.~\ref{eq:R_L5100_original} is adjusted accordingly.
In particular, for the median luminosity of the SDSS \hbXmg\ sub-sample ($\Lop=6\times10^{44}\,\ergs$) the \RBLR-\Lop\ relation of \cite{Bentz2009_RL_host} would have resulted in \RBLR\ estimates that are larger by 0.03 dex than the ones obtained by Eq.~\ref{eq:R_L5100_original}. 
For the \hbXciv\ and \mgXciv\ sub-samples we assumed the reverberation-based relation of \cite{Kaspi2007}: 
\begin{equation}
\RBLR\left(\civ\right) = 107.2 \left (\frac{\Luv}{10^{46}\,\ergs}\right )^{0.55} \,\, \ld \,\, .
\label{eq:R_L1450_K07}
\end{equation}
%
We also experimented with the relation of \cite{Vester_Peterson2006}, $\RBLR\propto\Luv^{0.53}$, and verified that none of the results that follow depend on this choice.

\section{Luminosities and Bolometric Corrections}
\label{sec_L_bol_corr}

\subsection{Monochromatic Luminosities and SED Shapes}
\label{subsec_L_SED}

\begin{figure}
\includegraphics[width=0.45\textwidth]{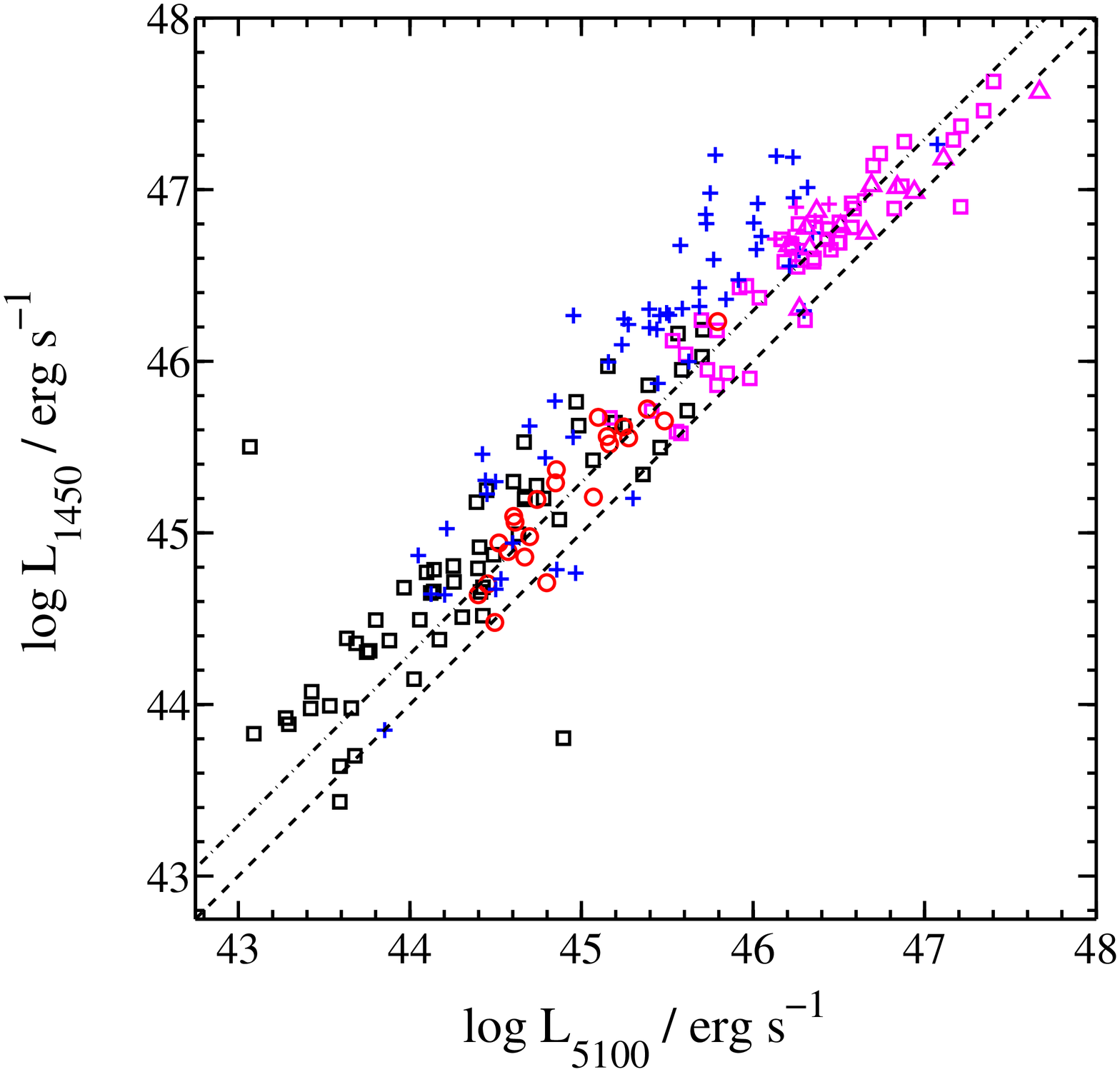}
\caption{
A comparison between \Luv\ and \Lop\ for for the different \hbXciv\ samples:
BL05 (black squares), N07+S04 (magenta squares), Sh07 (red circles), Sul07 (blue crosses) D09 (magenta crosses) and M09 (magenta triangles; see Table~\ref{tab_samples} for references and details).
The dashed line represents the 1:1 relation, 
and the dot-dashed line represents $\Luv=1.97\,\Lop$, the median value found for the N07+S04 sample, and typical for other high-luminosity sources.
A power-law continuum with $\alpha_{\nu}=-0.5$ \citep[e.g.][]{VandenBerk2001} would imply $\Luv=1.88\,\Lop$.
}
\label{fig:L1450_vs_L5100}
\end{figure}

The UV-optical continuum emission of type-I AGN, between $\sim$1300 and 5000\AA\ is well described by $f_{\nu}\propto \nu^{-0.5}$ (e.g., VdB01).
The typical luminosity scaling values in our SDSS sample are $\left<\Lthree/\Lop\right>=1.62$ and  $\left<\Luv/\Lthree\right>=1.39$, with standard deviations of 0.12 and 0.14 dex, respectively. 
Assuming the average conversion of \Lthree\ to the ``real'' continuum (see Appendix~\ref{app_mgii}), these ratios become $\left<\Lthree/\Lop\right>=1.41$ and $\left<\Luv/\Lthree\right>=1.59$, respectively.
For comparison, the ratios implied by an $\alpha_{\nu}=-0.5$ power law are 1.30 and 1.44, respectively.
Figure~\ref{fig:L1450_vs_L5100} shows that the smaller \hbXciv\ sub-samples generally follow the scalings of $\alpha_{\nu}\simeq-0.5$. 
However, the ratio of $\Luv/\Lop\simeq1.95$ provides a somewhat better scaling for these sources, for which both \Lop\ and \Luv\ are directly observed. 
This ratio is consistent with the VdB01 result, which is based on a composite of sources for which either \Lop\ or \Luv\ were observed.
In particular, all the PG quasars that are part of the reverberation-mapped sample of \cite{Kaspi2000} follow these luminosity scalings (these are part of the BL05, Sh07 and Sul07 samples).
Thus, we do not expect that SED differences would play a major role in scaling any relation that is consistent with the RM experiments.
This is a crucial point in single-epoch \mbh\ determinations and is further discussed in \S\ref{sec_civ_prob}.

We note that 87\% of the sources in the SDSS \mgXciv\ sub-sample have $\Luv>\Lthree$, and 93\% of the sources in the smaller \mgXciv\ sub-samples (Sh07, BL05, N07+S04, M09 \& D09) have $\Luv>\Lop$ (i.e., $\auvo>-1$).
In contrast, half of the sources (6/12) presented by \cite{Assef2011} have $\Luv\lesssim\Lop$. 
Such extreme UV-optical SEDs may represent AGN with intrinsically different properties or, perhaps, are affected by a higher-than-typical reddening.
These issues are further discussed in \S\ref{sec_civ_prob}.

\subsection{Bolometric Corrections}
\label{subsec_bol_corr}
To determine the bolometric luminosity (\Lbol) one has to assign a bolometric correction factor,  $\fbolwv=\Lbol/\lambda L_{\lambda}$.
Here we focus on \fbolthree.
Several earlier studies assumed a constant \fbolthree, e.g., $\sim5.9$ \cite[][]{Elvis1994}, or $\sim5.15$         
\cite[][]{Richards2006_SED}.
There are two problems with this approach:
{\bf (1)} The constant \fbolthree\ was based on the total, {\it observed} X-ray to mid-IR (MIR) SED of AGN. 
As such, it includes double-counting of part of the AGN radiation (the MIR flux originates from re-processing of the UV-optical radiation). This results in overestimation of \Lbol\ and thus \fbol\ \cite[e.g.,][]{Marconi2004}. 
{\bf (2)} The shape of the SED is known to be luminosity-dependent. 
This dependence is most significant at the X-ray regime, and perhaps also at UV wavelengths \cite[e.g.,][]{Vignali2003}. 
The general trend is of a \textit{decreasing} $L_{\rm X}$ with increasing $L_{\rm UV-opt}$. 
Thus, \fbol\ should decrease with increasing UV luminosity.
\cite{Marconi2004} provides a luminosity-dependent prescription for estimating \fbol(4400\AA) that addresses these two issues. 
The prescription can be modified to other wavelengths, by assuming a certain UV-optical SED.

\begin{figure}
\includegraphics[width=0.45\textwidth]{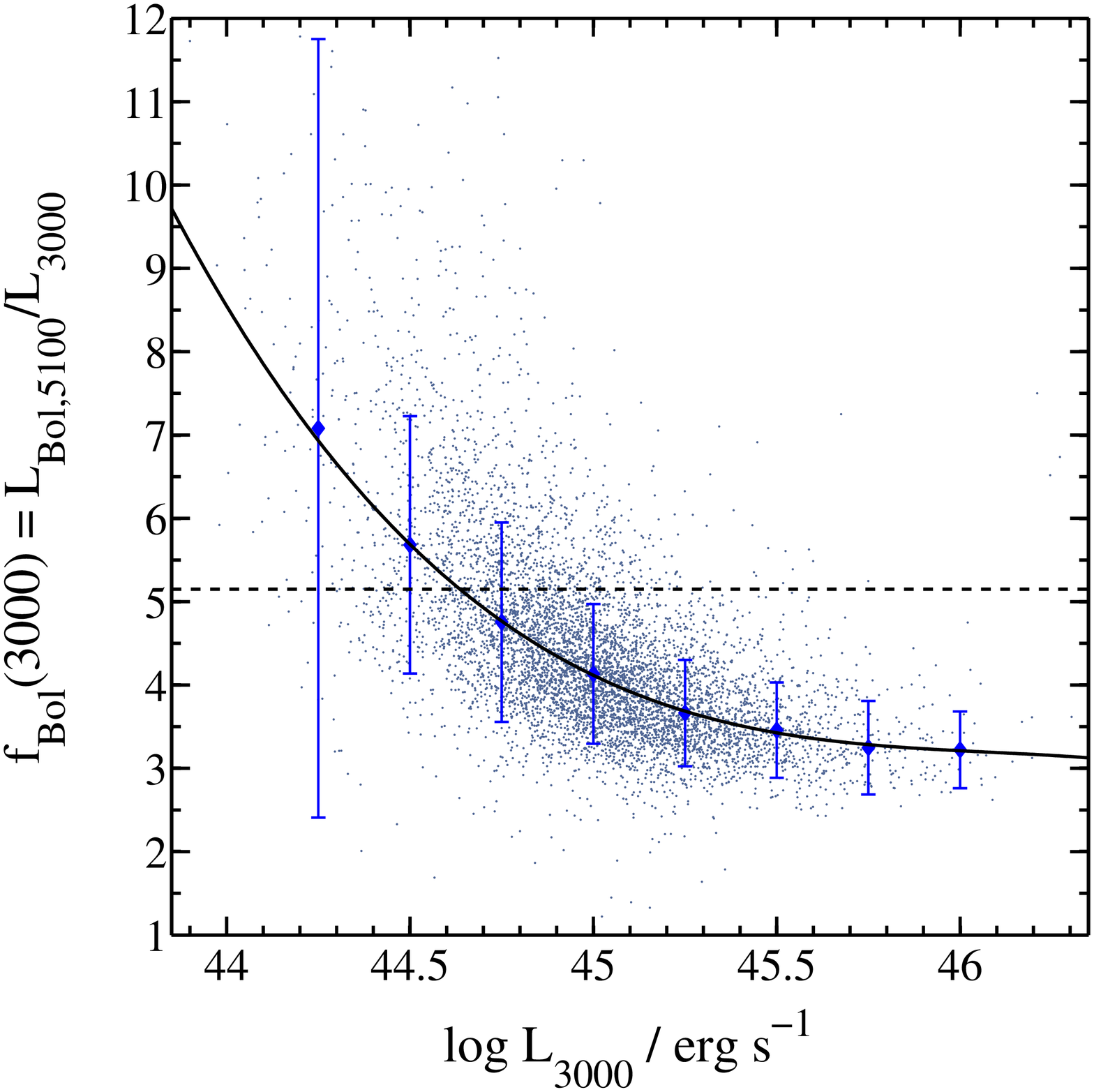}
\caption{
The relation between the empirically derived bolometric corrections ($\fbolthree\equiv\Lbol/\Lthree$) and \Lthree, for the SDSS \hbXmg\ sub-sample.
Large symbols and error bars represent the binned data and the corresponding scatter (standard deviations).
The solid black line represents the best-fit cubic polynomial to the binned data, given in Eq.~\ref{eq:fbol_uv_poly_1}.
The dashed horizontal line represents $\fbolthree=5.15$, the value used in several earlier studies 
\citep[e.g.,][]{Richards2006_SED}.
}
\label{fig:fbol_L3000_1}
\end{figure}

We derived a new prescription for \fbolthree\ by using the SDSS \hbXmg\ sub-sample.
First, we converted the measured \Lop\ of each source to \Lbol\ using the prescription of   
\cite{Marconi2004} and assuming $f_{\nu}\propto \nu^{-0.5}$. 
This provides, for each of the \NSDSShbmg\ sources, \Lthree\ and \Lbol/\Lthree, 
which are presented in Figure~\ref{fig:fbol_L3000_1}.
The derived bolometric corrections are systematically lower than the aforementioned fixed values.
For example, the typical correction is just 3.4 for sources with $\lLthree\simeq3\times10^{45}\,\ergs$, which is the median luminosity of the SDSS sources at $z\simeq1.2$. 
This is a factor of 1.5 lower than the value of 5.15 used in several other studies of \lledd\ at $z\sim1-2$ \cite[e.g.,][]{Fine2008}.
Fig.~\ref{fig:fbol_L3000_1} further reveals that, despite the large scatter at low \Lthree, there is a clear trend of decreasing \fbolthree\ with increasing \Lthree, as expected.
To quantify this trend, we grouped the data in bins of 0.2 dex in \Lthree\ and assumed that the error on \fbolthree\ is $\sigma_{\rm i}/\sqrt{N_{\rm i}}$, for the {\it i}-th bin in \Lthree.
An ordinary least squares (OLS) fit to the binned data points gives:
\begin{eqnarray} 
\label{eq:fbol_uv_poly_1}
\fbolthree =  4.12 -2.13 {\cal L}_{3000,45} \\ \nonumber
 + 1.76 {\cal L}_{3000,45}^2 -0.54 {\cal L}_{3000,45}^3  \, ,
\end{eqnarray}
%
%
where ${\cal L}_{3000,45}\equiv \log\left(\Lthree/10^{45}\,\ergs\right)$.
As Fig.~\ref{fig:fbol_L3000_1} clearly shows, this relation predicts relatively large bolometric corrections for low-luminosity sources, to the extreme of $\fbolthree\gtsim9.5$ for $\Lthree\lesssim 10^{44}\,\ergs$. 
Similarly high values were reported in the past, for a small minority of sources \cite[e.g., ][]{Richards2006_SED}. 
We suspect that the high values predicted by Eq.~\ref{eq:fbol_uv_poly_1} are the result of unrealistic extrapolation of the \cite{Marconi2004} $L_{\rm B}-\Lbol$ relation towards low luminosities, where the number of measured points is small. 
Moreover, we note that low-luminosity sources may suffer from more significant host-galaxy contamination, which results in a systematic over-estimation of \Lop\ and \Lbol.
We thus caution that the bolometric corrections we provide here should not be used for sources with  $\Lthree\lesssim 10^{44}\,\ergs$, where the emission from the host galaxy is comparable to that of the AGN and in particular in cases where the spectroscopic aperture affects the determination of the host emission \citep[see, e.g., the analysis in][]{Stern2012}.

\begin{figure}
\includegraphics[width=0.45\textwidth]{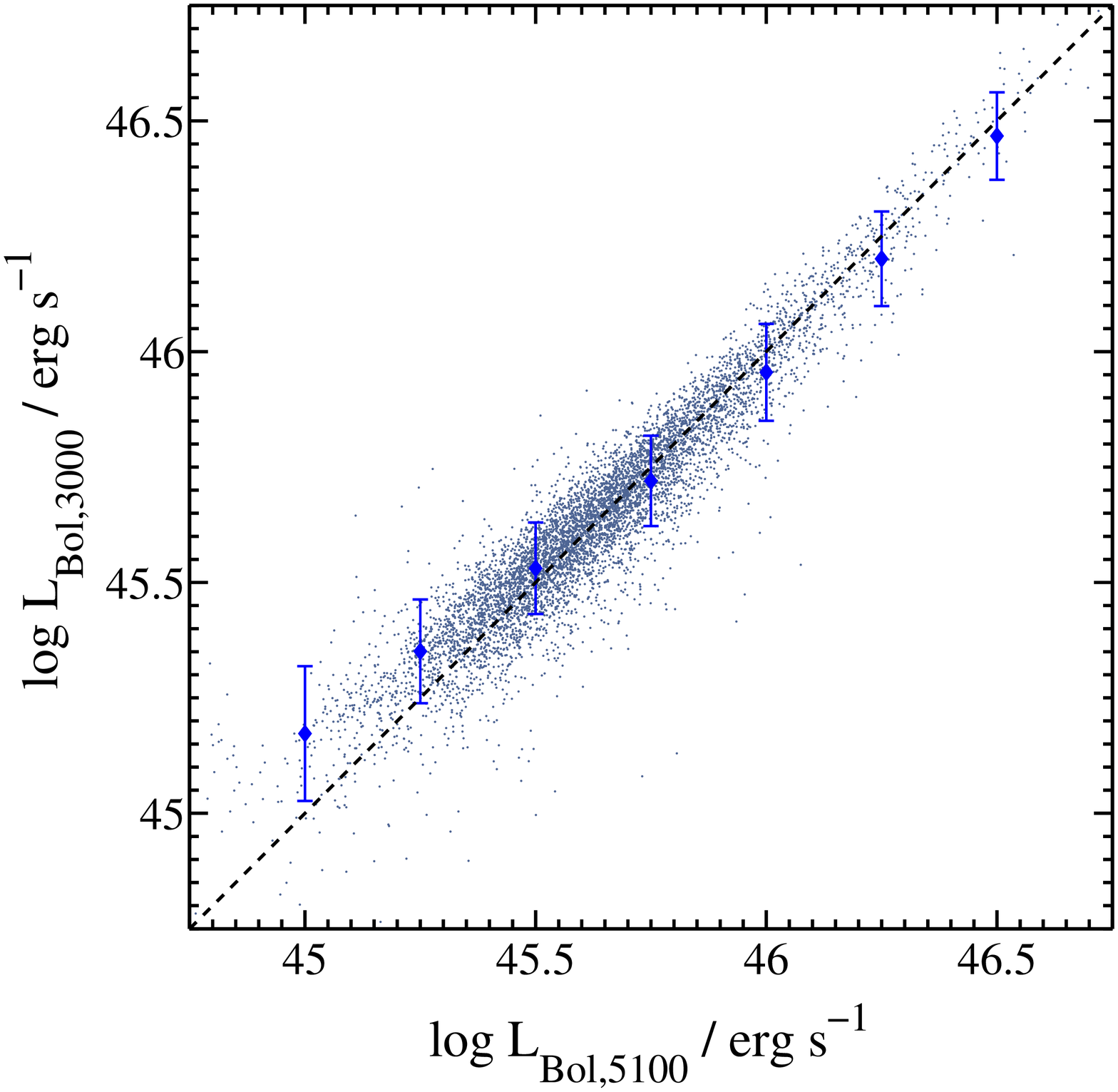}
\caption{
A comparison of estimates of \Lbol\ based on \Lthree\ and \Lop, for the SDSS \hbXmg\ sample.
Large symbols and error bars represent the binned data and the corresponding scatter (standard deviations).
The dashed line represents the the 1:1 relation.
}
\label{fig:Lbol_L3000_vs_L5100}
\end{figure}

Figure~\ref{fig:Lbol_L3000_vs_L5100} compares the bolometric luminosities derived from \Lthree\ by using Eq.~\ref{eq:fbol_uv_poly_1} to those calculated from \Lop. 
The two methods provide consistent estimates of \Lbol\ with a scatter of less than 0.09 dex (standard deviation of residuals).
The small scatter is probably dominated by the range of UV-to-optical slopes of individual sources.

We finally note that the real uncertainties on such estimates of \Lbol\ are actually governed by the range of global SED variations between sources, as well as the assumed physical (or empirical) model for the UV SED.
For example, the assumed exponent of the X-ray model and the $L_{\rm X}-L_{\rm UV}$ relation may amount up to $\sim$0.2 dex in $L_{\rm X}$, and thus in the calculated \Lbol\ \cite[see, e.g.,][]{Vignali2003,Bianchi2008}.
Two very recent studies further demonstrated these complications.
The study by \cite{Runnoe2012} showed that even the uniform \cite{Elvis1994} and \cite{Richards2006_SED} SEDs may provide \fbolthree\ as low as $\sim3$, given that the integration is limited to $\lambda<1\,\mic$ (thus neglecting re-processed emission. 
However, the best-fit trends of \cite{Runnoe2012} predict $\fbolthree\simeq5.3$ for sources with $\Lthree\simeq10^{45}$, consistent with the commonly used value of 5.15.
\cite{Jin2012a} used an accretion disk fitting method that results in a much higher (unobserved) far-UV luminosity for a given optical and/or near-UV luminosity.
Naturally, this model produced very high bolometric corrections, with $\fbolopt\gtsim12$ (and as high as $\sim$20-30) for several sources with $\Lop \simeq3\times10^{44}\,\ergs$, compared to the \cite{Marconi2004} prediction of $\fbolopt\simeq7$.
This completely new approach to the estimate of \Lbol\ in AGN will not be further discussed in the present paper.
Instead, we advice the usage of the corrections given by Eq.~\ref{eq:fbol_uv_poly_1}, which supplement the optical corrections of \cite{Marconi2004}.

\section{Virial \mbh\ Estimates: Basic Considerations}
\label{sec_mbh_general}

A main goal of this paper is to present a critical evaluation of the various ways to measure \mbh\ by using the RM-based ``virial'' method. 
It is therefore important to review the basic premise of the method and the justifications for its use. 

Four critical points should be considered:

\begin{enumerate}

\item
The emissivity weighted radius of the BLR, \RBLR, is known from direct RM-measurements almost exclusively for only the \hb\ and \civ\ emission lines, and to a much lesser extent for \mgii\ (see \S\ref{sec_intro}).
The expressions chosen for the present work are given in Eqs.~\ref{eq:R_L5100_original} and \ref{eq:R_L1450_K07}.
They depend on the measured \Lop\ and \Luv, and involve the assumption that the derived luminosities require no reddening-related, or other, corrections. 
The slope $\alpha$ of these correlations ($\RBLR \propto L^{\alpha}$) is empirically determined to be in the range 0.5--0.7, by a simple regression analysis of the observational results. 
It is not known whether the fundamental dependence is on \Lop, \Luv, the ionizing luminosity or perhaps \Lbol, although there are theoretical justifications for all these cases.
It is also not clear whether the slope itself depends on luminosity \cite[i.e. whether the same relation holds for all luminosities, see, e.g.,][]{Bentz2009_RL_host,Netzer_Marziani2010}.
Thus, the approach adopted here is to use all quantities as measured.

\item
The mean ratio of the RM-based \hb\ and \civ\ radii, assuming the $\RBLR-L$ relations, is about 3.7.
The number depends on the measured lags and the mean luminosities of the objects in the RM samples used to derive the relations, and may significantly vary for individual sources (see, e.g., the range of ratios in \citealt{Peterson2004}).
The mean ratio of the two involved luminosities in the two RM samples is $\Luv/\Lop\sim 1.9$. 
This is similar to the mean ratio in several much larger samples where the two wavelength regions are observed (e.g., S04, BL05, N07; see \S\ref{subsec_L_SED}) and is also similar to the SDSS-based composite spectrum of VdB01. The working assumption, therefore, is that this ratio represents the population of un-reddened type-I AGN well. 
Given this, the virial method cannot be applied directly to sources where $\Luv/\Lop$ deviates significantly from this typical value, since in such sources \RBLR\ may not scale with the luminosity in the same way as in the two RM samples used to derive the equations.  
For example, $\Luv/\Lop<<1.9$ may indicate significant continuum reddening, which will result in a systematic underestimation of \RBLR, if one uses Eq.~\ref{eq:R_L1450_K07} or similar relations.
Correcting for such reddening must be performed prior to the estimation of \RBLR\ or \mbh.

\item
The virial motion of the BLR, combined with the adoption of line FWHMs as
the gas velocity indicator, and the assumption that the global geometry of the \hb\ and \civ\ parts of the BLR are similar (e.g. two spheres of different radii) lead to the  prediction that
$\fwciv/\fwhb\simeq \sqrt{3.7}$. This is confirmed in a small number of intermediate luminosity objects showing the expected $\Luv/\Lop$ \cite[e.g.,][]{Peterson2004}. 
Given this, the simplest way to proceed to measure \mbh\ in large samples, which lack any additional information regarding the $\Luv/\Lop$ or $\fwciv/\fwhb$ ratios, is to assume the same is true for all sources.

\item
The best value of the geometrical factor $f$ in the mass equation (Eq.~\ref{eq:MBH_virial}) is 1.0. This is an average value obtained from the comparison of RM-based ``virial products'' and the $\mbh-\sigs$ relation, for about 30 low redshift type-I AGN \cite[e.g.,][]{Onken2004,Woo2010_LAMP_Msig,Graham2011}. None of these sources show a large deviation from this value. 
These calibrations rely predominantly on \hb, while \civ-related observables contribute to the estimation of virial products in only 5 sources. 
Unfortunately, there are no direct estimates of $f$ that are based {\it solely} on \civ.

Given points (ii) and (iii) above,
the value of $f$ used in \mbh\ estimates must be the same for the two emission lines. This is not meant to imply that certain sources cannot have two different virialized regions, for \hb\ and \civ, with different geometries and values of $f$. It only means that the method is based on certain samples with certain properties and hence should not be applied to objects with different properties. 
In objects where $f$ is not directly measured and the line width ratio deviate significantly from the above, e.g. objects with 
$\fwhb>\fwciv$, at least one of the lines should not be used as a \mbh\ indicator within the framework of the virial method. Since estimates of $f$ based on \fwhb\ are known to be correct in most of the \hb-RM sample, we prefer to adopt in such cases the assumption that \fwhb\ provides the more reliable mass indicator. 
\end{enumerate}

The above considerations suggest that a single epoch mass determination based on \hb\ and \Lop\ is a reliable mass estimate provided there is no significant continuum reddening. 
In case the amount of reddening is known, it should be taken into account prior to the application of the method.
In the absence of direct mass calibration based on lines other than \hb, there are only two alternatives: 
use theoretical conjectures, or a-priori knowledge about the line width. 
The first can be used for the \mgii\ line that is thought to originate from the same part of the BLR as \hb. 
If $\fwmg\simeq\fwhb$, then the line can be used for estimating \mbh. 
The second can be used for the \civ\ line since $\RBLR(\civ)/\RBLR(\hb)$ is known. 
As explained, the requirement is 
$\fwhb/\fwciv \simeq \sqrt{\RBLR\left(\civ\right)/\RBLR\left(\hb\right)}$. 
If this ratio is indeed found in the majority of objects with measured
\fwhb\ and \fwciv, then a single epoch mass estimate based on \civ\ can be safely used.
In the following sections we use such considerations to assess the validity of the use of the \mgii\ and \civ\ lines as mass indicators in type-I AGN.

\section{Estimating \mbh\ with \MgII}
\label{sec_mbh_mgii}

As discussed in \S\ref{sec_mbh_general}, the fact that there is no RM-based determination of \RBLR(\mgii) means that the only way to obtain a \mgii-based estimator for \mbh\ is to calibrate it against \mbh(\hb), based on the the assumption that $\RBLR(\mgii)=\RBLR(\hb)$ (verified by photoionization calculations).
We therefore have to show
that \Lthree\ can be used to estimate \RBLR(\hb), and that $\fwmg\simeq\fwhb$.
In what follows, we discuss these relations separately, and evaluate the ability of the (combined) virial product to reproduce \mbh(\hb).

\subsection{Using \Lthree\ to Estimate \RBLR}
\label{subsec_L3000}

\begin{figure*}
\centering
\includegraphics[width=0.47\textwidth]{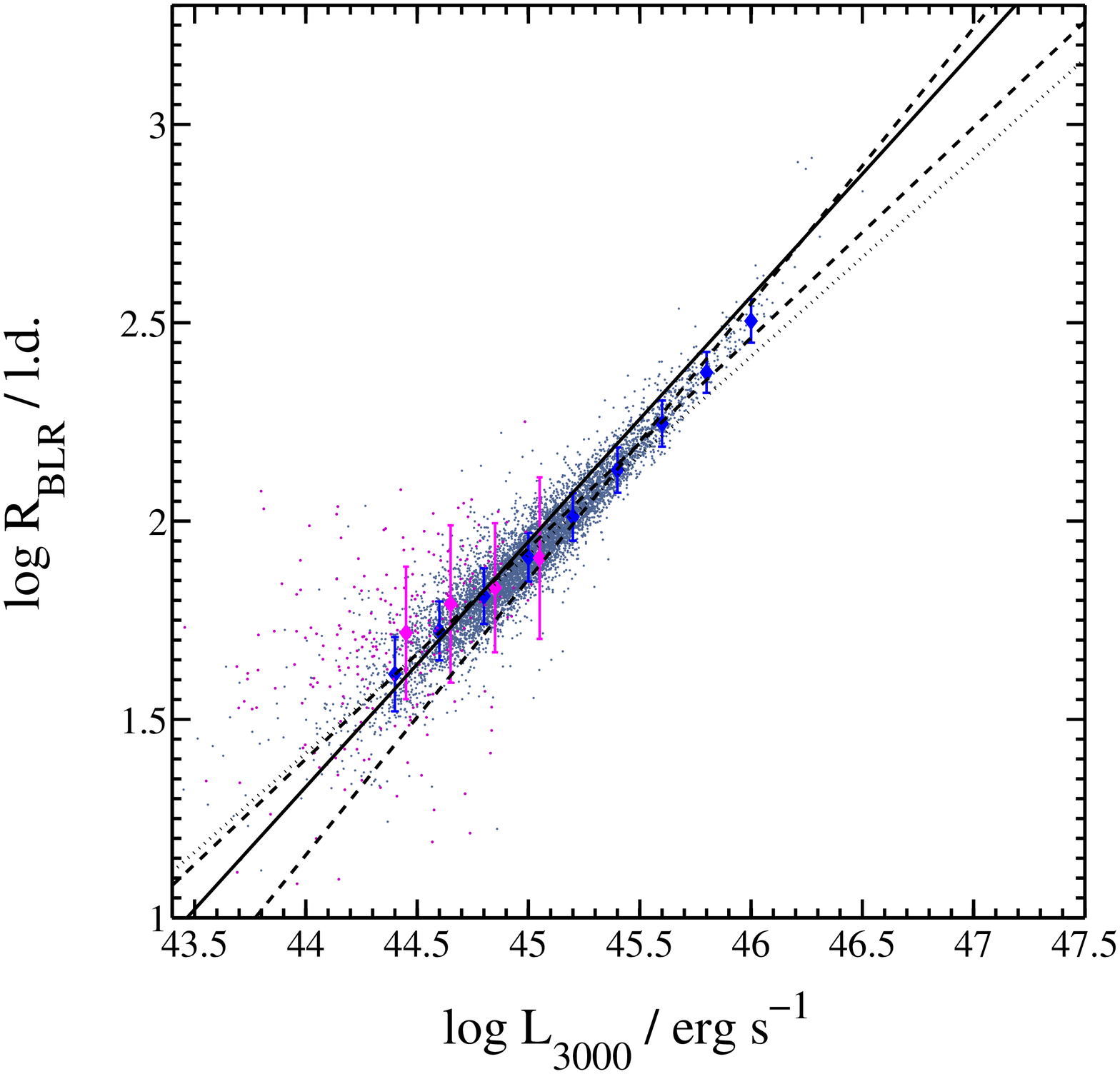}
\includegraphics[width=0.47\textwidth]{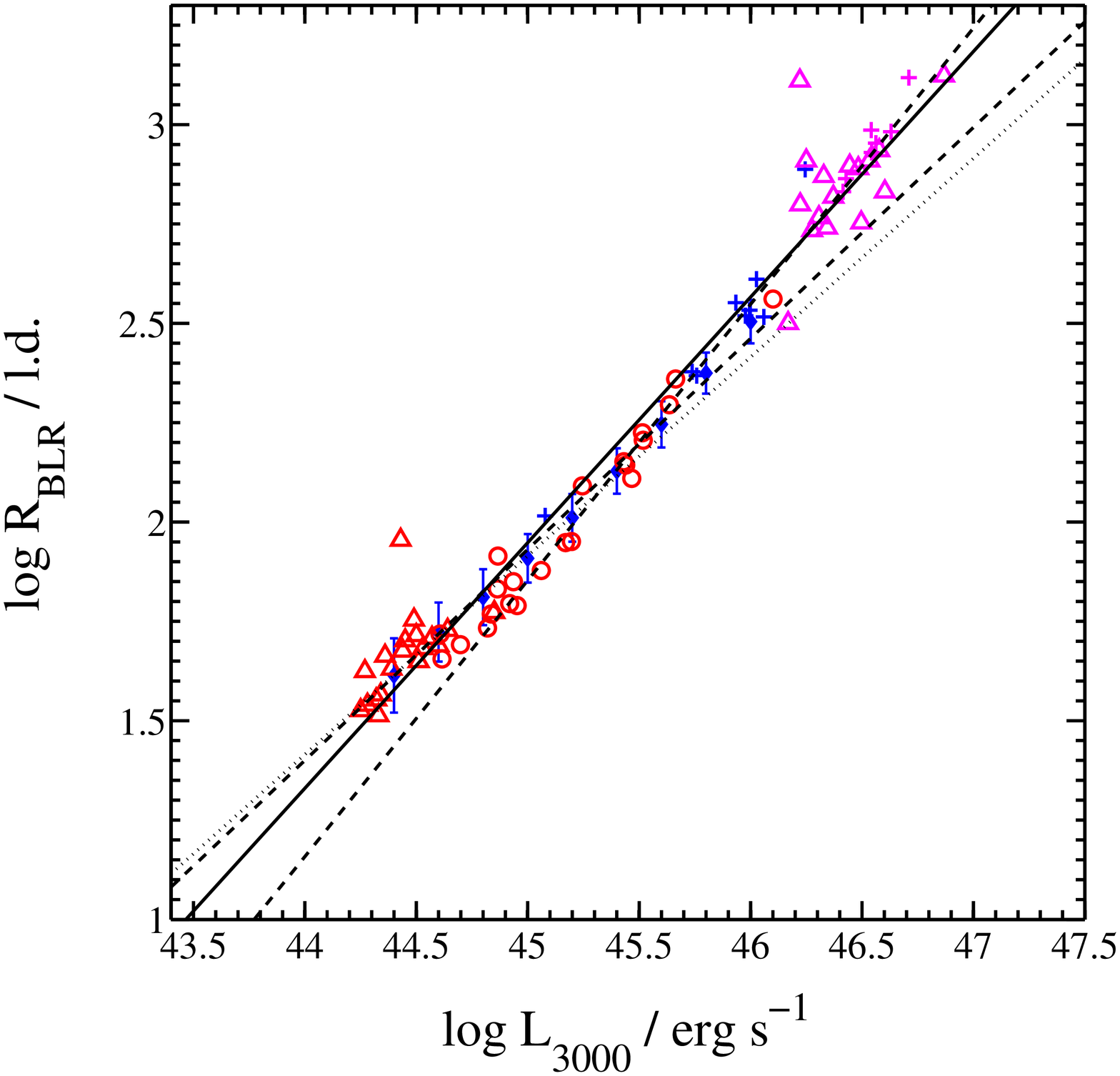}
\caption{
The relations between \RBLR, estimated from \Lop, and \Lthree, for the different \hbXmg\ sub-samples under study. 
\emph{Left - large samples:} blue and magenta points represent the SDSS and 2QZ \hbXmg\ sub-samples, respectively.
Larger symbols and error bars represent the binned data and the corresponding scatter (standard deviations).
\emph{Right - small samples:} the Sh07 (red circles), Mc08 (red triangles), D09 (magenta crosses) and M09 (magenta triangles) samples. The binned SDSS dataset is also included (blue diamonds and error bars).
In both panels, the solid lines represent our best-fit relations (Eq.~\ref{eq:R_L3000_final}).
Other lines provide reference for alternative relations:
dashed lines represent our best-fit relations for either high- or low-luminosity sources (slopes of 0.695 and 0.531, respectively), while the dotted line represents a relation with a slope of 0.5 (normalized to match the median \RBLR\ of the SDSS \hbXmg\ sub-sample at $\Lthree=10^{45}\,\ergs$). 
See the text and Table~\ref{tab_linear_fits} for more details.
}
\label{fig:RBLR_L3000}
\end{figure*}

In Figure \ref{fig:RBLR_L3000} we present the relation between the calculated values of \RBLR(\hb) and \Lthree, for all the sources in the different \hbXmg\ sub-samples. 
There is a clear and highly significant correlation between these two quantities. 
Since \RBLR\ is calculated directly from \Lop, this relation reflects the narrow distribution of UV-optical continuum slopes.
To quantify the $\RBLR-\Lthree$ relation, we bin the SDSS and 2QZ \hbXmg\ sub-samples (separately) in bins of 0.2 dex in \Lthree. The uncertainties on each binned data-point are assumed to be the standard deviations of the values included in the respective bin. The typical uncertainty is of 0.13 dex . 
This is a conservative choice, which attempts to account for the entire scatter in the data.\footnotemark
\footnotetext{An alternative choice would have been to estimate uncertainties as $\sigma/\sqrt{N}$. Due to the large number of SDSS sources in our sample, this would have decreased the uncertainties to below 0.01 dex, which is unrealistic.}
All other \hbXmg\ sub-samples remain un-binned. For these, we assume uncertainties of 0.1 dex on \RBLR\ and 0.05 dex on \Lthree. These choices reflect the absolute flux calibration uncertainties and the uncertainties related to the continuum and line fitting processes.
Since the uncertainties in both axes are comparable, we use the \texttt{BCES} \cite[][]{Akritas1996} and \texttt{FITEXY} \cite[][]{Press2002_numrec} fitting methods, both designed to also take into account the scatter in the data. 
All the \texttt{BCES} correlations are tested by a bootstrapping procedure with 1000 realizations of the data under study. We used the more sophisticated version of the \texttt{FITEXY} method, as presented by \cite{Tremaine2002}, where the error uncertainties on the data are scaled in
order to account for the scatter. Thus, all our \texttt{FITEXY} correlations resulted in
$\chi^{2}_{\nu}\simeq1$. 
The best-fit linear relations (also shown in Fig.~\ref{fig:RBLR_L3000}), parametrized as
\begin{equation}
\log \RBLR =\alpha\, \log\left(\frac{L_{3000}}{10^{44}\,\ergs}\right)\,+\beta \,\, ,
\label{eq:R_L3000_general}
\end{equation}
resulted in $\alpha\simeq0.615$ and $\beta\simeq1.33$ for both the \texttt{BCES} and \texttt{FITEXY} methods, 
The exact values and associated uncertainties are given in Table~\ref{tab_linear_fits}.
The standard deviation of the residuals is about 0.1 dex.
We note that some of the few low luminosity sources in Fig.~\ref{fig:RBLR_L3000} ($z<0.5$ sources from the Sh07 \& Mc08 samples) appear to lie above our best-fit relation. 
This might be due to the contamination of their optical spectra by host light, which would cause their \Lop\ (and hence \RBLR) to be slightly overestimated, although effect should be very small ($<0.05$ dex; see, e.g., Sh07 and \citet{Bentz2009_RL_host}).
In addition, the few extremely high luminosity sources ($z>1.5$ sources from the D09 \& M09 samples) also lie slightly above the best-fit relation of the combined dataset.
Thus, the slope of the $\RBLR-\Lthree$ relation may be somewhat shallower or steeper, in the low- or high-luminosity regimes, respectively.
We tested these scenarios by re-fitting the data after omitting data points either above $\Lthree=10^{46}\,\ergs$ or below $\Lthree=10^{45}\,\ergs$, that is keeping most of the (binned) SDSS data but omitting the extreme sources from the ``small'' samples, on either side of the luminosity range.
The results of this analysis are also presented in Table~\ref{tab_linear_fits}.
In particular, we find that for sources with $\Lthree\geq10^{45}\,\ergs$ the best-fit slope is $\alpha\simeq0.7$, while for sources with $\Lthree\leq10^{46}\,\ergs$ it is $\alpha\simeq0.5$.
Table~\ref{tab_linear_fits} also lists best-fit parameters for other choices of sub-samples.
In most of these cases, the derived slopes for the $\RBLR-\Lthree$ relations are between those reported by \cite{McLure2002} ($\alpha=0.47$) and by MD04 ($\alpha=0.62$).
The intercepts are also very similar to those of \cite{McLure2002}, and the intercept derived from the entire dataset (1.33) differs from the one derived in MD04 by only 0.06 dex, in the sense that our best-fit relation predicts higher \RBLR\ values for a given value of \Lthree.
The MD04 study relation was derived using only sources with $\Lthree > 10^{44}\,\ergs$, as is the case for most of the sources used here. 
Therefore, the differences between the MD04 relation and our results are not due to different luminosity regimes.


\begin{table*}
  \caption{Coefficients for \RBLR-$L$ Correlations}
  \label{tab_linear_fits}
  \begin{center}
    \begin{tabular}{lcccccccc} \hline \hline
    & 
  \multicolumn{4}{c}{BCES bisector} &
  \multicolumn{3}{c}{FITEXY} \\

   Luminosity \& sub-samples used &
   $\alpha$ &
   $\beta$ &
   $\sigma$ $^{\rm a}$ &
   $M_0$ $^{\rm b}$ &
   $\alpha$ &
   $\beta$ &
   $\sigma$ $^{\rm a}$ 
\\
\hline
$x=\log\left(\Lthree/10^{44}\,\ergs\right)$ &  &  &  &  &  &   \\
\hline
 SDSS  & 
 $0.482\pm0.004$  & $1.440\pm0.004$ & $0.066$ & $7.81$ & $0.531\pm0.004$ & $1.388\pm0.004$ & $0.064$ \\
 SDSS+2QZ & 
 $0.473\pm0.005$  & $1.450\pm0.005$ & $0.077$ & $7.98$ & $0.508\pm0.003$ & $1.415\pm0.004$ & $0.076$ \\
 ``Small'' $^{\rm b}$ & 
 $0.619\pm0.014$  & $1.340\pm0.022$ & $0.098$ & $5.61$ & $0.620\pm0.015$ & $1.334\pm0.025$ & $0.098$ \\
 SDSS (binned) + ``Small'' $^{\rm c}$ & 
 $0.618\pm0.014$  & $1.330\pm0.020$ & $0.094$ & $5.62$ & $0.619\pm0.014$ & $1.333\pm0.023$ & $0.094$ \\
 \dots\dots\dots $\Lthree\leq10^{46}\,\ergs$& 
 $0.531\pm0.015$  & $1.400\pm0.019$ & $0.071$ & $6.93$ & $0.544\pm0.025$ & $1.389\pm0.029$ & $0.071$ \\
 \dots\dots\dots $\Lthree\geq10^{45}\,\ergs$ & 
 $0.695\pm0.022$  & $1.170\pm0.037$ & $0.087$ & $4.67$ & $0.701\pm0.029$ & $1.158\pm0.059$ & $0.087$ \\
 SDSS+2QZ (binned) + ``Small'' $^{\rm c}$ & 
 $0.615\pm0.014$  & $1.340\pm0.019$ & $0.093$ & $5.66$ & $0.618\pm0.014$ & $1.330\pm0.023$ & $0.093$ \\
  &  &  &  &  & &  & \\
\hline
$x=\log\left(\Lmg/10^{42}\,\ergs\right)$ &  &  &  &  &  &  & \\
\hline
 SDSS & 
 $0.499\pm0.006$  & $1.38\pm0.007$ & $0.124$ &  & $0.450\pm0.005$ & $1.428\pm0.005$ & $0.123$ \\
 SDSS+2QZ & 
 $0.508\pm0.007$  & $1.36\pm0.008$ & $0.125$ &  & $0.440\pm0.005$ & $1.436\pm0.005$ & $0.123$ \\
 ``Small'' $^{\rm d}$ & 
 $0.713\pm0.034$  & $1.13\pm0.056$ & $0.151$ &  & $0.692\pm0.036$ & $1.166\pm0.066$ & $0.151$ \\
 SDSS (binned) + ``Small'' $^{\rm d}$ & 
 $0.684\pm0.032$  & $1.18\pm0.051$ & $0.146$ &  & $0.677\pm0.031$ & $1.190\pm0.057$ & $0.146$ \\
   
\hline
  \multicolumn{8}{l}{$^{a}$ Standard deviation of residuals, in dex.}  \\
  \multicolumn{8}{l}{$^{b}$ The scaling factor associated with the final \mbh\ estimator, i.e. the equivalents of the factor 5.6 in Eq.~\ref{eq:M_L3000_final}}  \\
  \multicolumn{8}{l}{$^{c}$ The Sh07, Sul07, Mc08, M09 and D09 \hbXmg\ sub-samples.}  \\
  \multicolumn{8}{l}{$^{d}$ The Sh07, Sul07 and M09 \hbXmg\ sub-samples.}  \\
    \end{tabular}
  \end{center}
\end{table*}


%
As Table~\ref{tab_linear_fits} and Fig.~\ref{fig:RBLR_L3000} demonstrate, the range of slopes and intercepts that describe the $\RBLR-\Lthree$ relation is relatively broad, and probably somewhat luminosity-dependent.
This situation is present in virtually all previous attempts to calibrate $\RBLR-L$ relations \cite[see discussion in, e.g.][]{Kaspi2000,Kaspi2005,Vester_Peterson2006}.
Notwithstanding this issue, in what follows we chose to estimate \RBLR\ by using the relation:
\begin{equation}
\RBLR\left(\Lthree\right) = 21.38 \left(\frac{\Lthree}{10^{44}\,\ergs}\right)^{0.62}\,\, \ld \, .
\label{eq:R_L3000_final}
\end{equation}
We caution that this relation may not be suitable for low-luminosity sources ($\Lthree<10^{45}\,\ergs$), where a shallower slope should be used.

\subsection{Using \Lmg\ to Estimate \RBLR}
\label{subsec_L_mgii}

\begin{figure}
\centering
\includegraphics[width=0.47\textwidth]{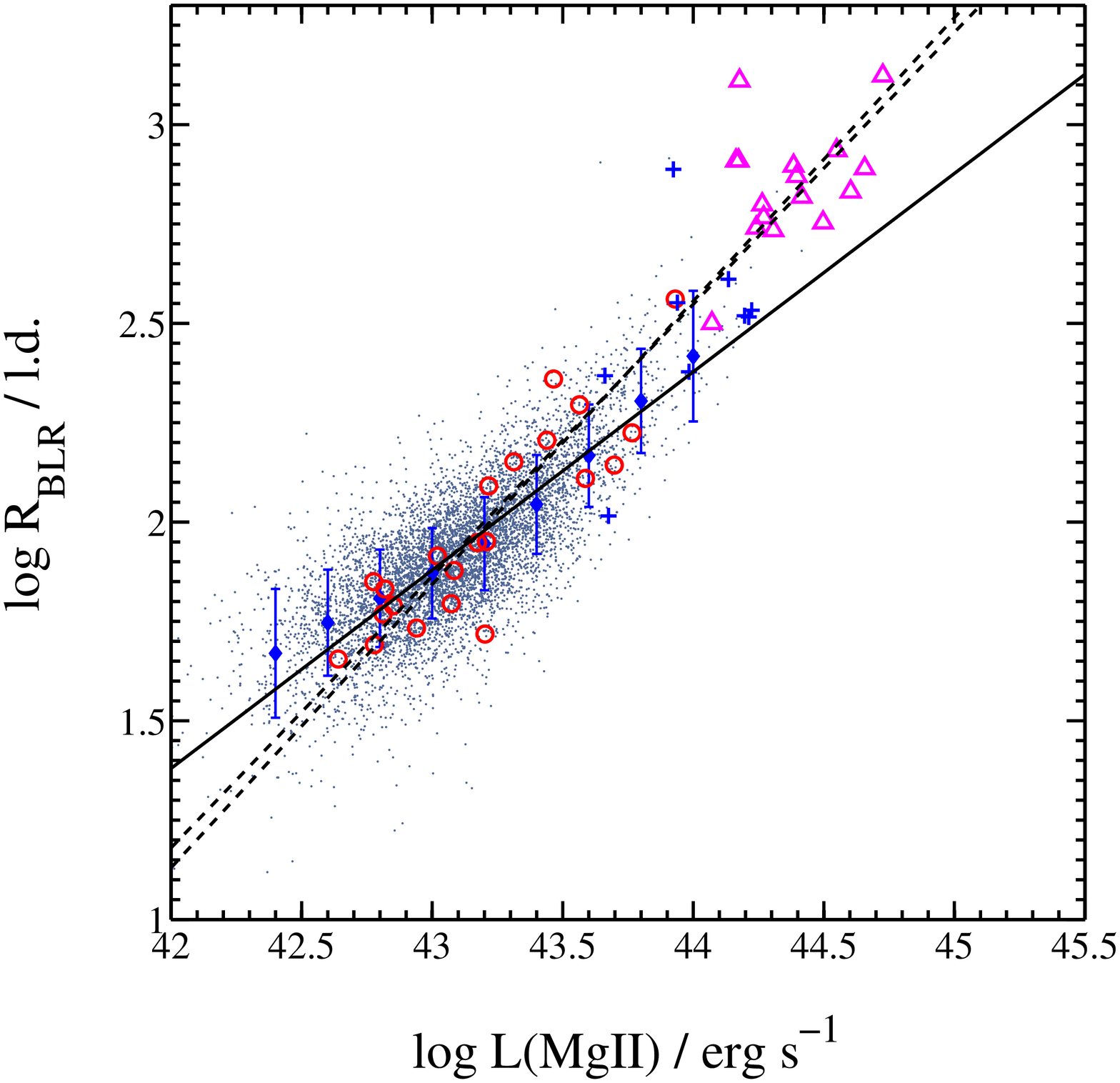}
\caption{
The relation between \RBLR, estimated from \Lop, and \Lmg, for several \hbXmg\ sub-samples under study. 
Symbols are identical to Fig.~\ref{fig:RBLR_L3000}. 
The solid black line represent the best-fit relation for the SDSS \hbXmg\ sub-sample, which follows $\RBLR\propto\Lmg^{0.5}$. 
The dashed lines alternative relations, derived by including the small, high-luminosity and high-redshift sample of M09. These relations follow $\RBLR\propto\Lmg^{0.7}$. 
See the text and Table~\ref{tab_linear_fits} for more details.
}
\label{fig:RBLR_Lmg}
\end{figure}

The line luminosity \Lmg\ can also be used to estimate \RBLR. 
Such an approach may be used to overcome the difficulties in determining \Lthree\ in NIR spectra of high-redshift sources.\footnotemark
\footnotetext{Several previous studies have calibrated $\RBLR-L$ relations for the \hb\ and \Halpha\ line luminosities \cite[e.g.,][]{Greene_Ho_Ha_2005,Kaspi2005}.}
The relation between \RBLR, as determined from \Lop, and \Lmg, is presented in Figure~\ref{fig:RBLR_Lmg}. 
The scatter in this relation is larger than that of the \RBLR-\Lthree\ relation. 
In particular, the 2QZ \hbXmg\ sub-sample shows considerable scatter and almost no correlation between \RBLR\ and \Lmg, probably due to the low-quality of the data at this extreme redshift range (see \S\ref{subsec_sample_2qz}).
Fitting the data with a linear relation of the form 
\begin{equation}
\log \RBLR =\alpha\, \log\left(\frac{\Lmg}{10^{42}\,\ergs}\right)\,+\beta\,\,
\label{eq:R_Lmg_general}
\end{equation}
gives the values of $\alpha$ and $\beta$ presented in Table~\ref{tab_linear_fits}. We get $\alpha\left[\Lmg\right]\simeq0.5$ for the SDSS and 2QZ \hbXmg\ sub-samples, but a considerably steeper slope (or a higher intercept) in the case we include the high-luminosity sources from the ``small'' samples. 
The scatter between the resulting (best-fit) estimates of \RBLR\ and those based on \Lop\ (Eq.~\ref{eq:R_L5100_original}), for the SDSS and 2QZ sub-samples, is about 0.13 dex, only slightly higher than that of the \RBLR-\Lthree\ relations.
Despite the advantages of using this \RBLR-\Lmg\ relation, we draw attention to the significant differences in the best-fit parameters that were derived from the different sub-samples, as well as by the two fitting methods, and caution that this relation is not as robust as the \RBLR-\Lthree\ one.
In addition, an accurate determination of \Lmg\ requires a reasonable determination of the \feii\ and \feiii\ features adjacent to the \mgii\ line, and thus still depends on the determination of the continuum.

\subsection{The Width of the \MgII\ Line}
\label{subsec_fwhm_mgii}

Our measure of \fwmg\ is different from the ones used in several earlier studies that measured the width of the \textit{total} (doublet) profile.
At large widths ($\fwmg\gtsim4000\,\kms$), the two widths are basically identical, 
since the width of the line is much larger than the separation between the two components.
For relatively narrow lines ($\fwmg\lesssim4000\,\kms$) the two measures differ significantly, and  $\fwhm\left(\mgii,{\rm single}\right) < \fwhm\left(\mgii,{\rm total}\right)$.
For example, the typical profiles with (total) $\fwmg\simeq2000\,\kms$ in our SDSS sample correspond to a single-component width of only $\sim1500\,\kms$. 
This implies that \mbh\ can be systematically \textit{overestimated} by a factor of $\sim1.8$ for such narrow-line objects.
This issue is crucial for studies of sources with high accretion rates.
To correct earlier results that used the entire profile, we suggest the following simple relation, which is based on a fit to the SDSS data:  
\begin{eqnarray}
\label{eq:FW_Mg_s_d}
\fwhm\left(\mgii,{\rm single}\right) = 1.01\times \fwhm\left(\mgii,{\rm total}\right) \\ \nonumber
- 304 \, \kms .
\end{eqnarray}
We used this prescription to correct the relevant tabulated values of \fwmg\ published in earlier papers (see Table~\ref{tab_samples}).

\begin{figure*}
\centering
\includegraphics[width=0.47\textwidth]{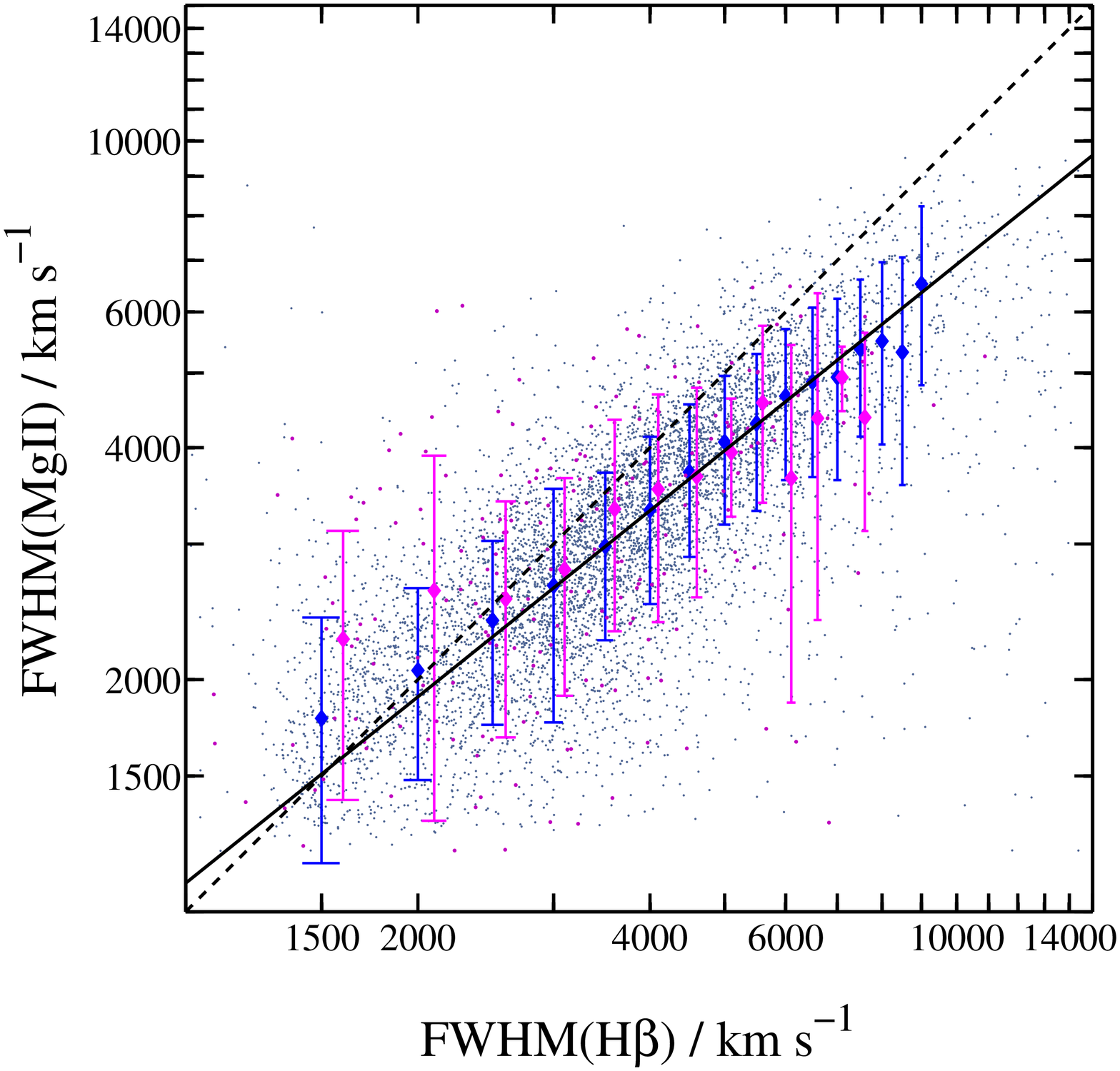}
\includegraphics[width=0.47\textwidth]{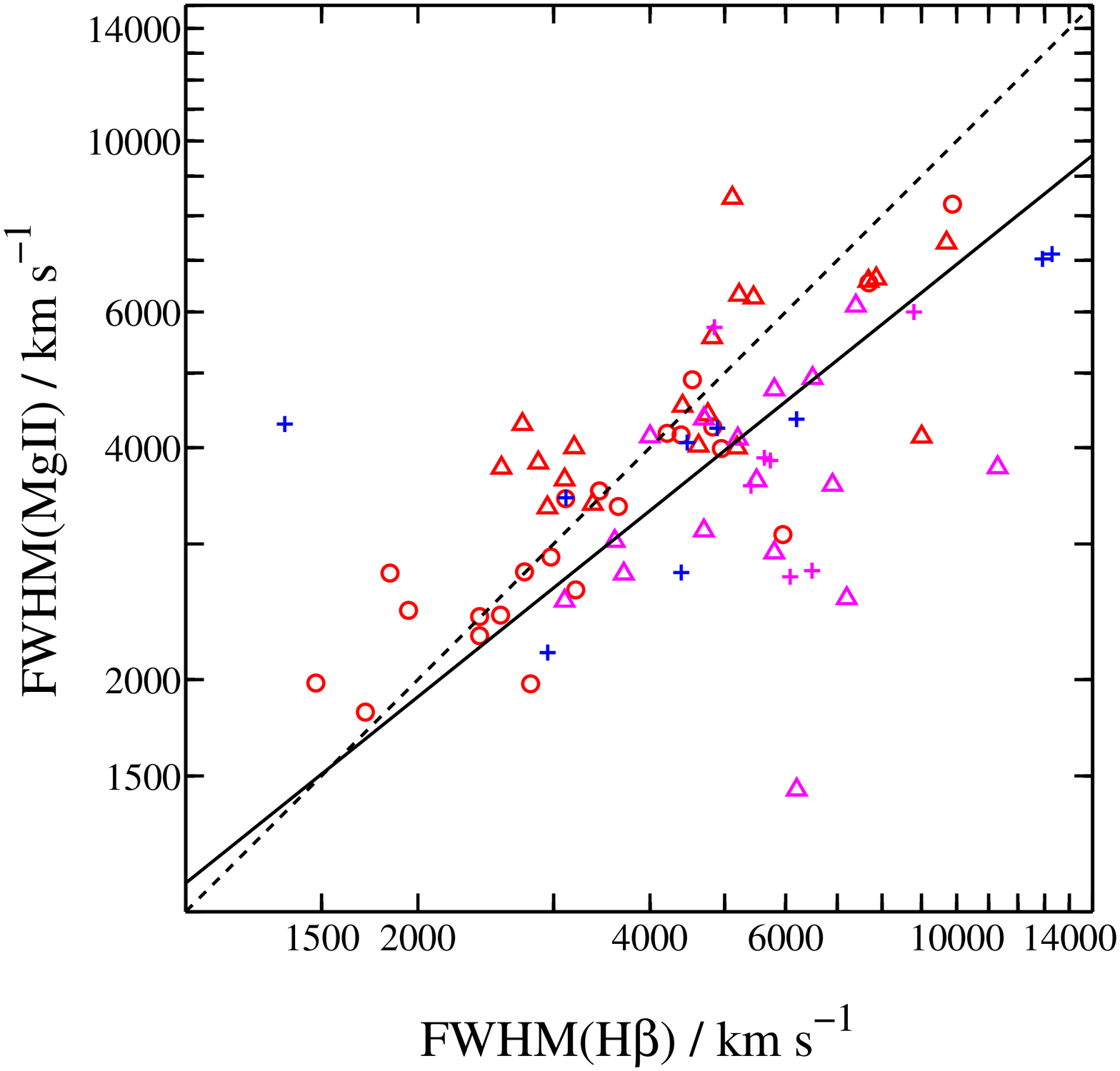}
\caption{
A comparison of \fwmg\ and \fwhb, for the large (\emph{left}) and small (\emph{right}) \hbXmg\ sub-samples.
In both panels, symbols are identical to Fig.~\ref{fig:RBLR_L3000}, 
and the dashed lines represent the 1:1 relations.
Solid lines represent the best-fit power-law relation, given in Eq.~\ref{eq:FW_Mg_vs_Hb}.
}
\label{fig:FW_Mg_FW_Hb}
\end{figure*}

Several studies showed that \fwmg\ is very similar to \fwhb\ \cite[e.g.][and references therein]{Shen_dr5_cat_2008}. This justifies the use of \fwmg\ as a tracer of virial BLR cloud motion.
Indeed, the distribution of $\fwmg/\fwhb$ for our SDSS \hbXmg\ sub-sample peaks at -0.05 dex and the standard deviation is 0.15 dex.
A similar comparison of the IPV line widths results in an almost identical distribution.
In order to further test this issue, we present in Figure~\ref{fig:FW_Mg_FW_Hb} a direct comparison between \fwmg\ and \fwhb, for \textit{all} the \hbXmg\ sub-samples. 
As expected, there is a strong correlation between the widths of the two lines.
However, 
in most of the cases with $\fwhb\gtsim4000\,\kms$, the \mgii\ line is narrower. This trend is also reflected in the binned data shown in Fig.~\ref{fig:FW_Mg_FW_Hb}. 
For example, for sources with $\fwhb\simeq9000\,\kms$, the typical value for the \mgii\ line is merely $\fwmg\simeq6000\,\kms$.
We find that the best-fit relation between these line widths is
\begin{eqnarray}
\label{eq:FW_Mg_vs_Hb}
\log\, \fwmg = \left(0.803 \pm 0.007\right) \log \fwhb \, \\ \nonumber
 +\, \left(0.628 \pm 0.027\right) \, ,
\end{eqnarray}
based on the BCES bisector. This relation is also shown in Fig.~\ref{fig:FW_Mg_FW_Hb}.
Our result is in excellent agreement, both in terms of slope and intercept, with that of \cite{Wang_MgII_2009}, which is based on a subset of our SDSS \hbXmg\ sub-sample (about 10\% of the sources; see \S\ref{subsec_sample_sdss}). 
The general trend we find between \fwmg\ and \fwhb\ is captured, in essence, by the statistical correction factors suggested by \cite{Onken_Koll2008}, which are useful for large samples.
However, it is not at all clear whether the fit reflects a real, global trend. 
The alternative is that $\fwmg\simeq\fwhb$ up to $\sim6000\,\kms$, beyond which there are some differences in the mean location of the strongest line emitting gas.

\subsection{Determination of $\mbh\left(\mgii\right)$}
\label{subsec_f_L3000_mgii}

\begin{figure}
\includegraphics[width=0.47\textwidth]{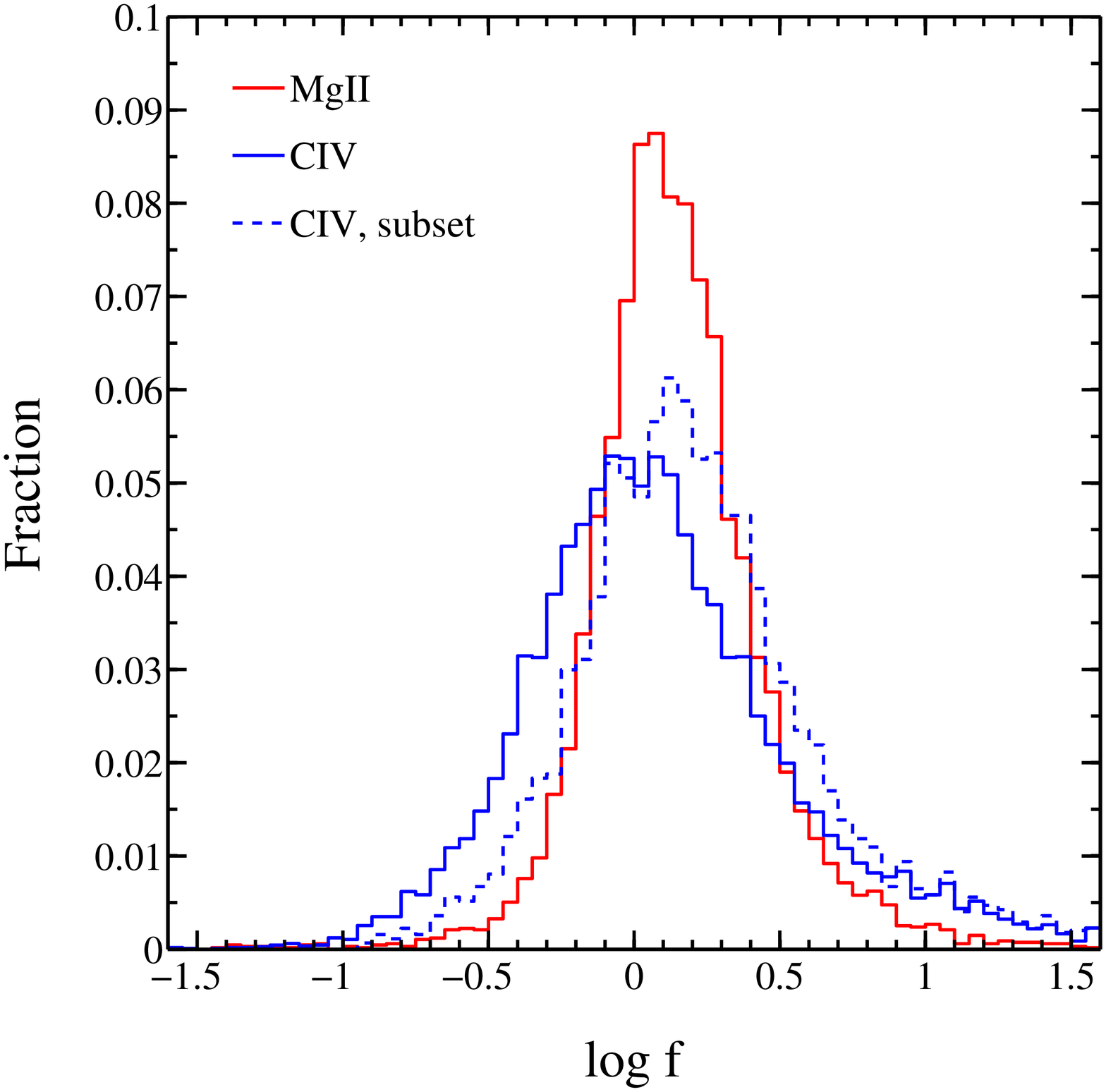}
\caption{
The distributions of empirically derived normalization factors $\left(f\right)$, for different emission lines and sub-samples. 
The solid red line represents the distribution of \fmg\ (see \S\ref{subsec_f_L3000_mgii}), for the SDSS \hbXmg\ sub-sample.
The solid blue line represents the distribution of \fciv\ for the entire SDSS \mgXciv\ sub-sample, while the dashed blue line represents the subset of sources which show $|\voff(\civ)|<500\,\,\kms$.
}
\label{fig:f_mg_civ_hist}
\end{figure}

For each source in the \hbXmg\ sub-samples we calculated an ``empirical scaling factor'':
\begin{equation}
  \mumg \equiv \frac{1}{G} \, \RBLR\left(\Lthree\right)\,\fwmg^{2} \, ,
\label{eq:f_factor_hbXmg}
\end{equation}
where $\RBLR\left(\Lthree\right)$ is calculated through  Eq.~\ref{eq:R_L3000_final}. 
In Figure~\ref{fig:f_mg_civ_hist} we show the distribution of the relevant normalization factor, defined as $\fmg=\mbh\left(\hb\right)/\mumg$, for the SDSS \hbXmg\ sub-sample. 
The distribution has a clear peak at a median (mean) value is $\fmg=1.33$ (1.42), and a standard deviation of 0.32 dex. This is in agreement with the expected value of $\fmg=1$.
For comparison, the \cite{McLure_Dunlop2004} \mbh\ estimator was derived assuming $\fmg=0.89$.\footnotemark
\footnotetext{The MD04 derivation assumes $f=1$ (for $\mbh\left[\hb\right]$), but also introduces an offset of about -0.05 dex to minimize the differences between the resulting \mbh(\mgii) estimates and the reverberation results.
Our analysis considers \fmg\ to be the factor required to correct for such an offset.}
We further investigated whether \fmg\ depends on other observables, such as source luminosity, SED shape or EW(\mgii). However, we find no significant correlations of this type. 
The only exception, a marginal anti-correlation with \lledd\ (see \S\ref{subsec_bol_corr}) is most probably driven by the strong dependence of both \lledd\ and \fmg\ on \fwhb.
In addition, the small scatter in the $\RBLR-\Lthree$ relation, in comparison with that of the $\fwhb-\fwmg$ relation, suggests that the range of derived \fmg\ values is driven solely by the scatter in $\fwhb/\fwmg$. 
This is supported by the fact that we find no correlation between \fmg\ and $\Lthree/\Lop$ while the significant correlation with $\fwhb/\fwmg$ (not shown here) tightly follows a power-law with the expected slope of about 2.

Next, we combine our best-fit $\RBLR-\Lthree$ relation (Eq.~\ref{eq:R_L3000_final}) with \fwmg\ and the median value of \fmg\ to obtain the final form of the \mgii-based \mbh\ estimator,
\begin{equation}
  \mbh=5.6\times10^6\left
[\frac{L_{3000}}{10^{44}\,\ergs}\right]^{0.62} \left[\frac
    {{\rm FWHM}(\mgii)}{10^3 \,\kms\,}\right]^{2} \,\,
\Msun \ .
\label{eq:M_L3000_final}
\end{equation}
This estimator differs from the one given in MD04 in its overall normalization, which is higher by a factor of about 1.75 than the MD04 one. 
Thus, we find that the MD04 \mbh\ estimator causes an \textit{underestimation} of \mbh\ by $\sim0.25$ dex.
This result is consistent with the findings of \cite{Shen_dr7_cat_2011}, where the slope of the $\RBLR-\Lthree$ relation was forced to be 0.62.
Naturally, the choice of a different $\RBLR-\Lthree$ relation, according to the parameters listed in Table~\ref{tab_linear_fits}, would result in subtle luminosity-dependent differences between our \mbh\ estimates and those of MD04. 
We repeated the above steps for several other choices of parameters ($\alpha$ and $\beta$ in Eq.~\ref{eq:R_L3000_general}), and list in Table~\ref{tab_linear_fits} the resulting scaling of the final \mbh\ estimator (i.e. the equivalents of 5.6 in Eq.~\ref{eq:M_L3000_final} above).\footnotemark
\footnotetext{For clarity, we calculated these factors only for the parameters derived using the \texttt{BCES} method.}
Thus, one can use a different \Lthree-based virial \mbh\ estimator, fully described by $\alpha$ and $M_0$ in Table~\ref{tab_linear_fits}, according to the required luminosity range (see discussion in \S\ref{subsec_L3000}).

\begin{figure*}
\centering
\includegraphics[width=0.47\textwidth]{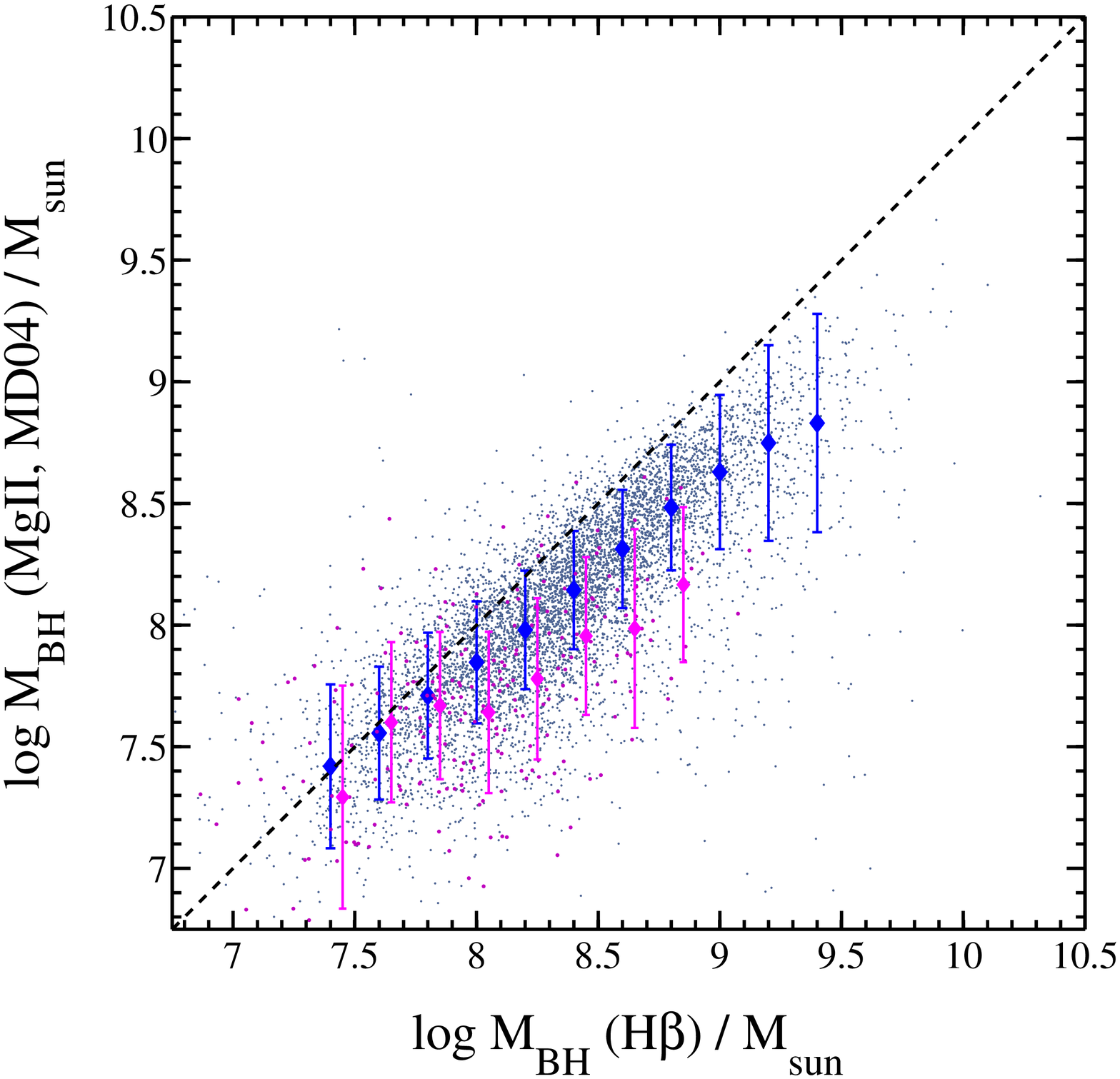}
\includegraphics[width=0.47\textwidth]{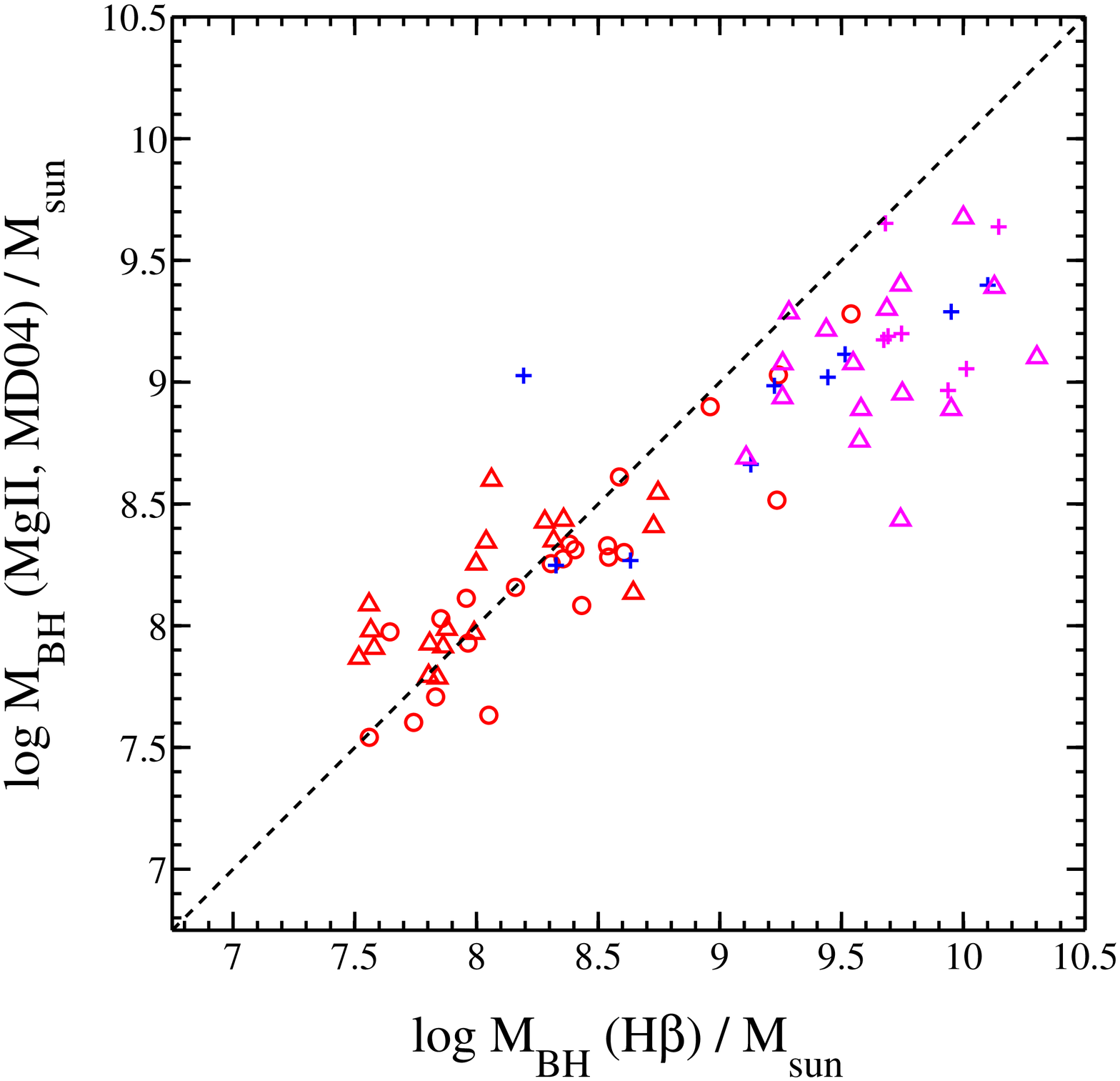}
\caption{
A comparison of estimates of \mbh\ based on the \hb\ line and the 
\citet{McLure_Dunlop2004} prescription, for the large (\emph{left}) and small (\emph{right}) \hbXmg\ sub-samples.
In both panels, symbols are identical to Fig.~\ref{fig:RBLR_L3000},
and the dashed lines represent the the 1:1 relations.
}
\label{fig:MBH_MD04_Hb}
\end{figure*}

\begin{figure*}
\centering
\includegraphics[width=0.47\textwidth]{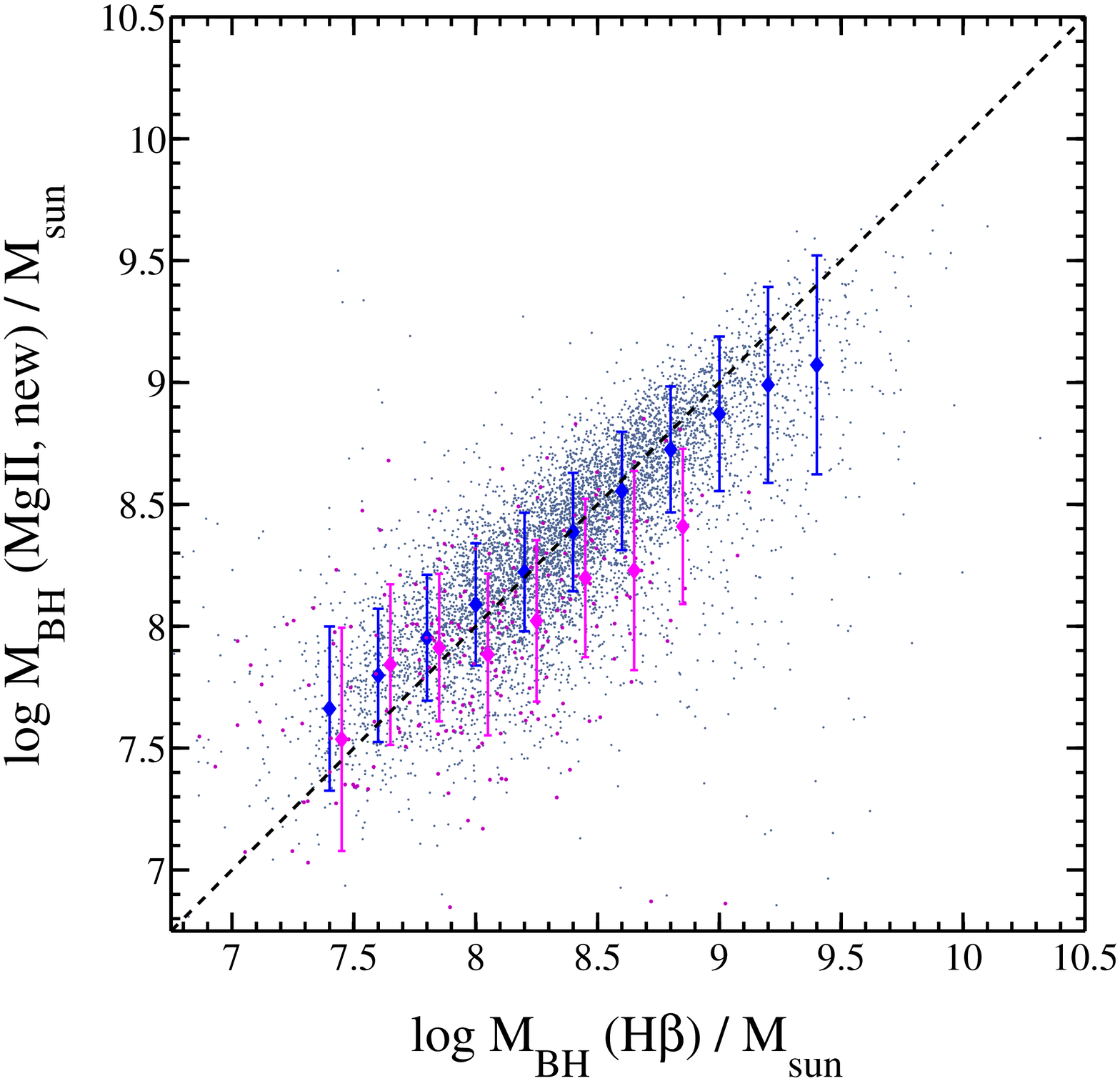}
\includegraphics[width=0.47\textwidth]{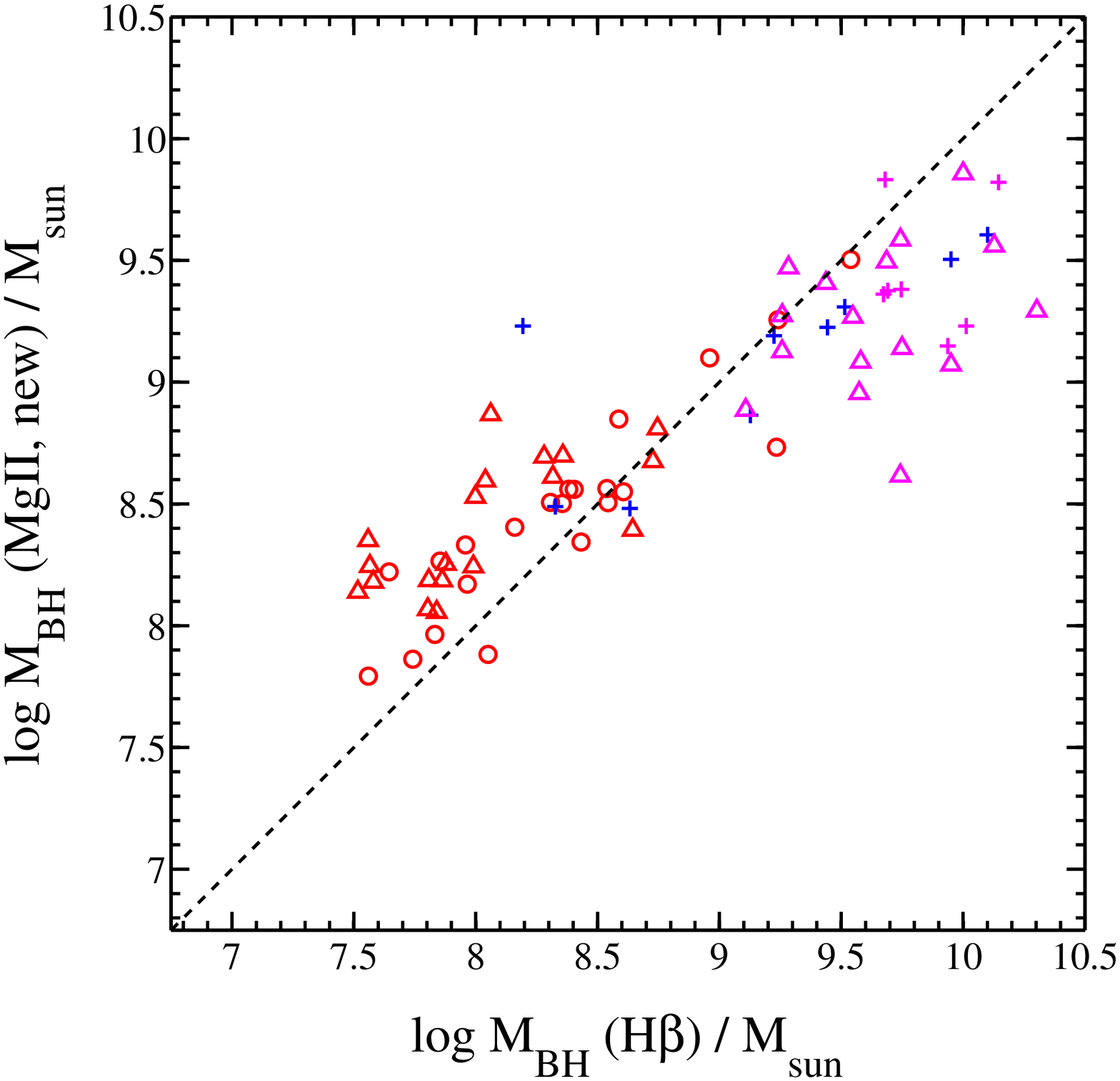}
\caption{
A comparison of \hb-based estimates of \mbh, and the new calibration of a \mgii-based \mbh\ estimator,
for the large (\emph{left}) and small (\emph{right}) \hbXmg\ sub-samples.
The \hb- and \mgii-based \mbh\ estimates are obtained through Eqs.~\ref{eq:M_Hb} and \ref{eq:M_L3000_final}, respectively.
In both panels, symbols are identical to Fig.~\ref{fig:RBLR_L3000}, 
and the dashed lines represent the the 1:1 relations.
}
\label{fig:MBH_final_vs_Hb}
\end{figure*}

To evaluate the improvement we obtained in estimating \mbh(\mgii), we compare in Figure~\ref{fig:MBH_MD04_Hb} the masses obtained with the \hb\ (following Eq.~\ref{eq:M_Hb}) and the MD04 methods, for all the \hbXmg\ sub-samples.
Clearly, there is a systematic offset between the two estimators, in the sense that \mbh(\mgii,MD04) is typically lower than \mbh(\hb). 
The median offset within the SDSS \hbXmg\ sub-sample is, indeed, $\sim0.25$ dex.
In Figure~\ref{fig:MBH_final_vs_Hb} we preform a similar comparison, but this time using the \textit{new} \mgii-based \mbh\ estimates (Eq.~\ref{eq:M_L3000_final}).
The systematic shift seen in Fig.~\ref{fig:MBH_MD04_Hb}, particularly at the high mass end, has completely disappeared. 
The median difference between the two estimates is negligible, although the scatter (standard deviation of residuals) remains about 0.32 dex.

Several studies suggested to use \mbh\ estimators where the exponent of the velocity term differs from 2 \cite[e.g.,][]{Greene_Ho_Ha_2005,Wang_MgII_2009}. 
Like any additional degree of freedom, 
it is expected that this approach may reduce the scatter between \hb- and \mgii-based estimates of \mbh. 
The usage of this empirical approach abandons the fundamental assumption of virialized BLR dynamics, which is the basis for all mass determinations considered here. 
For the sake of completeness, we derived a relation by combining Eqs~\ref{eq:R_L3000_final} and \ref{eq:FW_Mg_vs_Hb} and minimizing the systematic offset with respect to the the \hb-based \mbh\ estimators. This process resulted in the relation 
$\mbh=3.62\times10^6\left
 [\frac{L_{3000}}{10^{44}\,\ergs}\right]^{0.62} 
 \left[\frac{{\rm FWHM}(\mgii)}{10^3 \,\kms\,}\right]^{2.5} \,\,
 \Msun$ .
The scatter between this estimator and the one based on \hb, for the SDSS \hbXmg\ sub-sample, is of 0.34 dex, almost identical to the one obtained following Eq.~\ref{eq:M_L3000_final} above. 
Since this relation departs from the simple virial assumption, we chose not to use in what follows.

\mbh\ can also be obtained by using the \RBLR-\Lmg\ relation presented in \S\ref{subsec_L_mgii}. 
We have repeated the steps described above, assuming $\left(\RBLR/\rm{l.d.}\right)=24 \left( \Lmg/10^{42}\,\ergs\right)^{0.5}$ (see Table~\ref{tab_linear_fits}), and obtained 
\begin{equation}
  \mbh=6.79\times10^6\left
  [\frac{\Lmg}{10^{42}\,\ergs}\right]^{0.5} \left[\frac
  {{\rm FWHM}(\mgii)}{10^3 \,\kms\,}\right]^{2} \,\,
\Msun \ .
\label{eq:M_Lmg_final}
\end{equation}
\mbh\ estimates based on this relation are also consistent with those based on \hb, and the scatter for the SDSS \hbXmg\ sub-sample is of 0.33 dex, indistinguishable from the one achieved by using \Lthree.
We note that the \mgii\ line presents a clear ``Baldwin effect'', i.e. an anti-correlation between EW(\mgii) and \Lthree\ \cite[e.g.][BL05]{Baldwin1977_BE}. 
Therefore, the usage of \Lmg\ probably incorporates some other, yet-unknown properties of the BLR.

%

We tested the consistency of our improved \Lthree-based determinations of \lledd\ with those based on \Lop\ (and \hb).
For this, we calculated \lledd\ for the SDSS \hbXmg\ sub-sample based on the two available approaches: either using the bolometric corrections given by Eq.~\ref{eq:fbol_uv_poly_1} and \mbh\ estimates given by Eq.~\ref{eq:M_L3000_final}, or using the bolometric corrections of \cite{Marconi2004} and \mbh\ estimates given by Eq.~\ref{eq:M_Hb}.
In all cases we assume $L_{\rm Edd}=1.5\times10^{38} \left(\mbh/\Msun\right)\,\ergs$, appropriate for solar metallicity gas. 
The comparison of the two estimates of \lledd\ clearly shows that the two agree well, with a scatter of about 0.31 dex, and negligible systematic difference.
On the other hand, using the bolometric correction of \citet[][5.15]{Richards2006_SED} and the MD04 estimates of \mbh\ results in a \textit{systematic overestimation} of \lledd\ by a factor of about 2.2.

\section{\CIV-based estimates of \mbh}
\label{sec_civ_prob}

As discussed in \S\ref{sec_mbh_general}, under the virial assumption, 
and the known ratio of \fwciv/\fwhb\ in a small number of sources with RM-based \RBLR\ measurements, it is possible to use a reliable estimator for the size of the \civ-emitting region (Eq.~\ref{eq:R_L1450_K07}) to calibrate a \civ-based estimator for \mbh, by combining \RBLR(\Luv), \fwciv\ and a known $f$-factor ($f=1$ in our case).
Since for the samples where \RBLR(\civ) is directly measured it is \textit{smaller} than \RBLR(\hb), by a factor of $\sim$3.7, we expect $\left(\fwciv/\fwhb\right)\simeq\sqrt{3.7}=1.9$.
In what follows, we address the different ingredients of a virial \civ-based estimator, and show that for a large number of the type-I AGN studied here
the \civ\ measurements are \textit{not} consistent with the virial assumption.

\begin{figure*}
\centering
\includegraphics[width=0.49\textwidth]{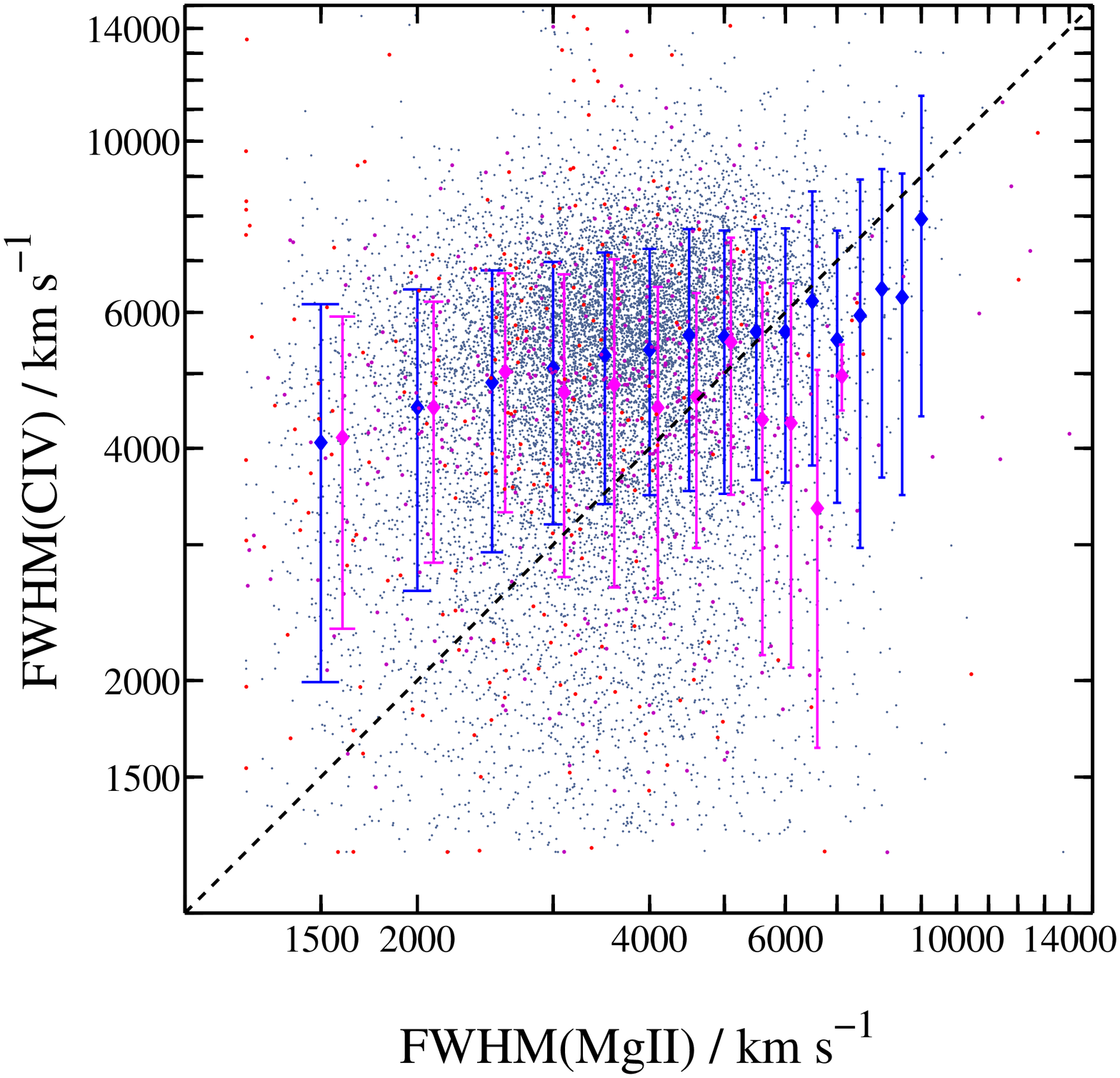}
\includegraphics[width=0.49\textwidth]{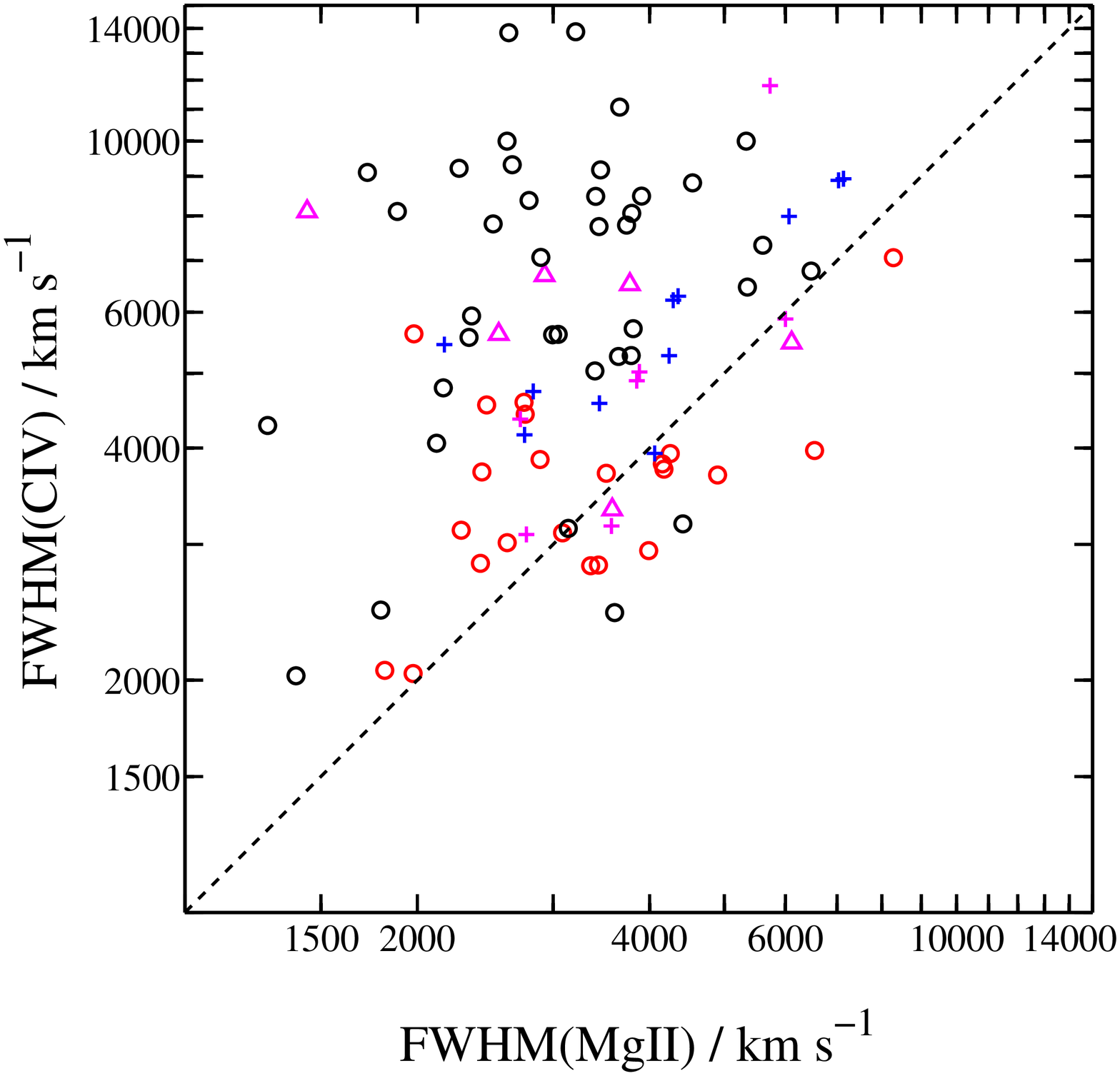}
\caption{
A comparison of \fwciv\ and \fwmg\ for the different \mgXciv\ sub-samples.
\emph{Left - large samples:} blue, magenta and red points represent the SDSS, 2QZ and 2SLAQ , respectively. 
Larger symbols and error bars represent the binned data and the corresponding scatter (standard deviations).
\emph{Right - small samples:} black circles represent the T11 sample (at \zfpe), while all other symbols are identical to previous figures.
In both panels, the dashed lines represent the the 1:1 relations.
}
\label{fig:FW_C4_vs_Mg}
\end{figure*}

\begin{figure}
\includegraphics[width=0.47\textwidth]{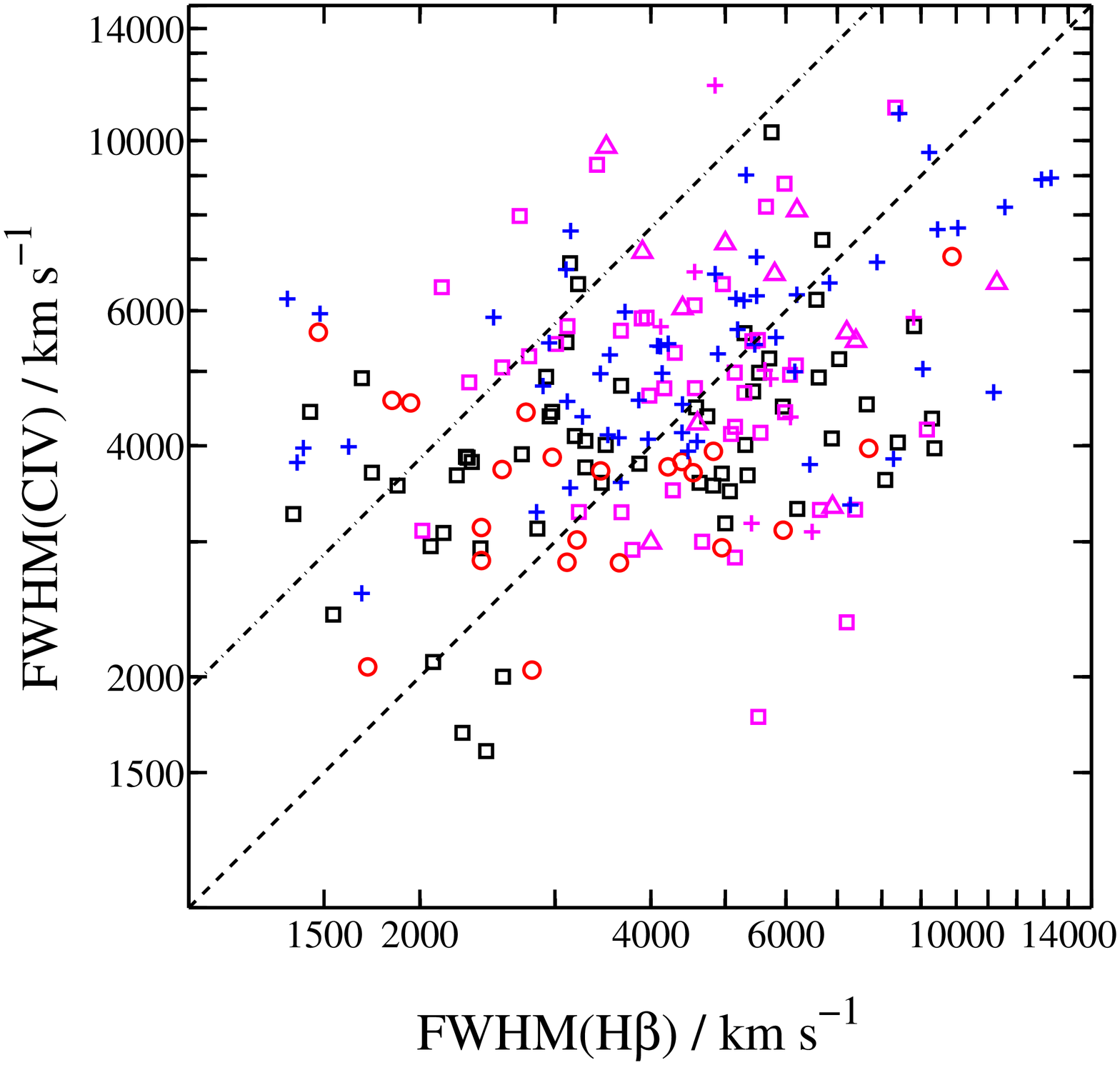}
\caption{
A comparison of \fwciv\ and \fwhb\ for all the \hbXciv\ sub-samples.
Symbols are identical to Fig.~\ref{fig:L1450_vs_L5100}.
The dashed line represents the 1:1 relation, while the dot-dashed line represents $\fwciv=\sqrt{3.7}\fwhb$, as predicted by the virial assumption (see \S\ref{sec_mbh_general} for details).
Note the large fraction of sources with $\fwciv \lesssim \fwhb$, 
in sharp contrast with the virial assumption, and the very large scatter.
}
\label{fig:FW_C4_vs_Hb}
\end{figure}

Figures~\ref{fig:FW_C4_vs_Mg} and \ref{fig:FW_C4_vs_Hb} compare \fwciv\ with \fwmg\ and \fwhb, respectively. 
The scatter in both figures is much larger than the scatter in the \mgii-\hb\ comparison diagram (Fig.~\ref{fig:FW_Mg_FW_Hb}), 
and the line widths do not follow each other.
In practical terms, for any observed (single) value of \fwciv, the corresponding values of \fwhb\ covers almost the entire range of $\sim2,000-10,000\,\kms$, in contrast with the value expected from the assumption of virialized motion and an identical $f$.
We verified that alternative measures of the line width, such as the IPV, do not reduce the scatter or present any more significant relations between the different lines.
Most importantly, a significant fraction of the sources under study ($\sim45\%$) exhibit $\fwciv \leq \fwhb$ (Fig.~\ref{fig:FW_C4_vs_Hb}), and 26\% of the sources in the SDSS, 2QZ \& 2SLAQ \mgXciv\ sub-samples have $\fwciv\leq\fwmg$ (Fig.~\ref{fig:FW_C4_vs_Mg}), in contrast to the expectations of the virial method.  
The large scatter, and the lack of any correlation between \fwciv\ and either \fwhb\ or \fwmg, were identified in several earlier studies of local \cite[e.g.][BL05]{Corbin1996}, intermediate- (e.g., S08, F08) and high-redshift (S04, N07, T11) samples.
We note that the high fraction of sources where \fwciv\ seems to defy the expectations is not due to a specific population of narrow-line sources (i.e., NLSy1s), and/or low-quality UV spectra, as was suggested by \cite{Vester_Peterson2006}.
In particular, we find that 55\% of the \hbXciv\ sources with $\fwhb > 3000\,\kms$ (i.e., 88 out of 155 broad-line sources) present $\fwciv \leq \fwhb$. 
In addition, 51\% of the sources in the N07+S04, D09 \& M09 samples (33 out of 65 sources), where the \civ\ line was measured in high-quality optical spectra, also show $\fwciv \leq \fwhb$.

\begin{figure*}
\begin{center}
\includegraphics[width=0.47\textwidth]{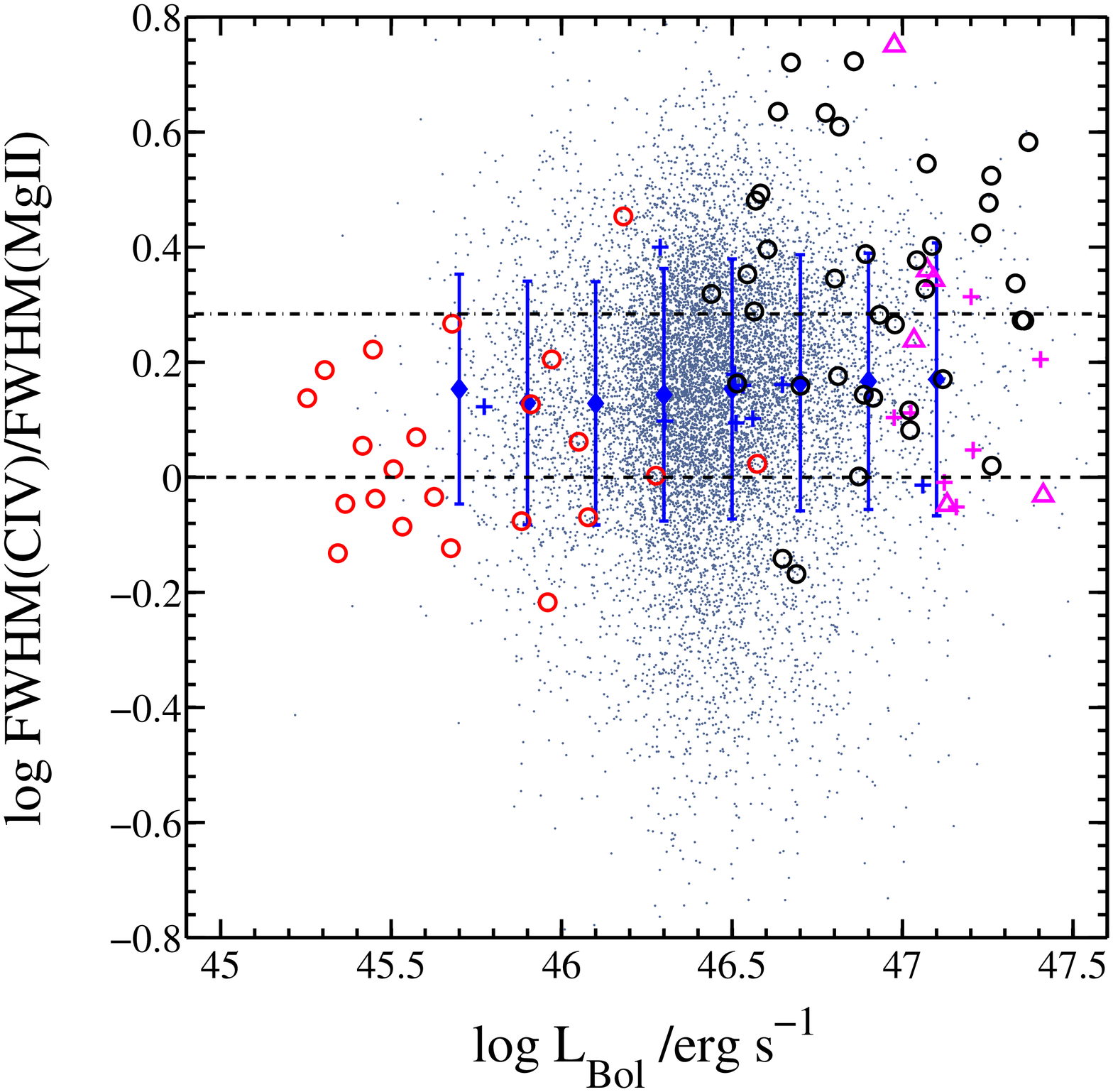}
\includegraphics[width=0.47\textwidth]{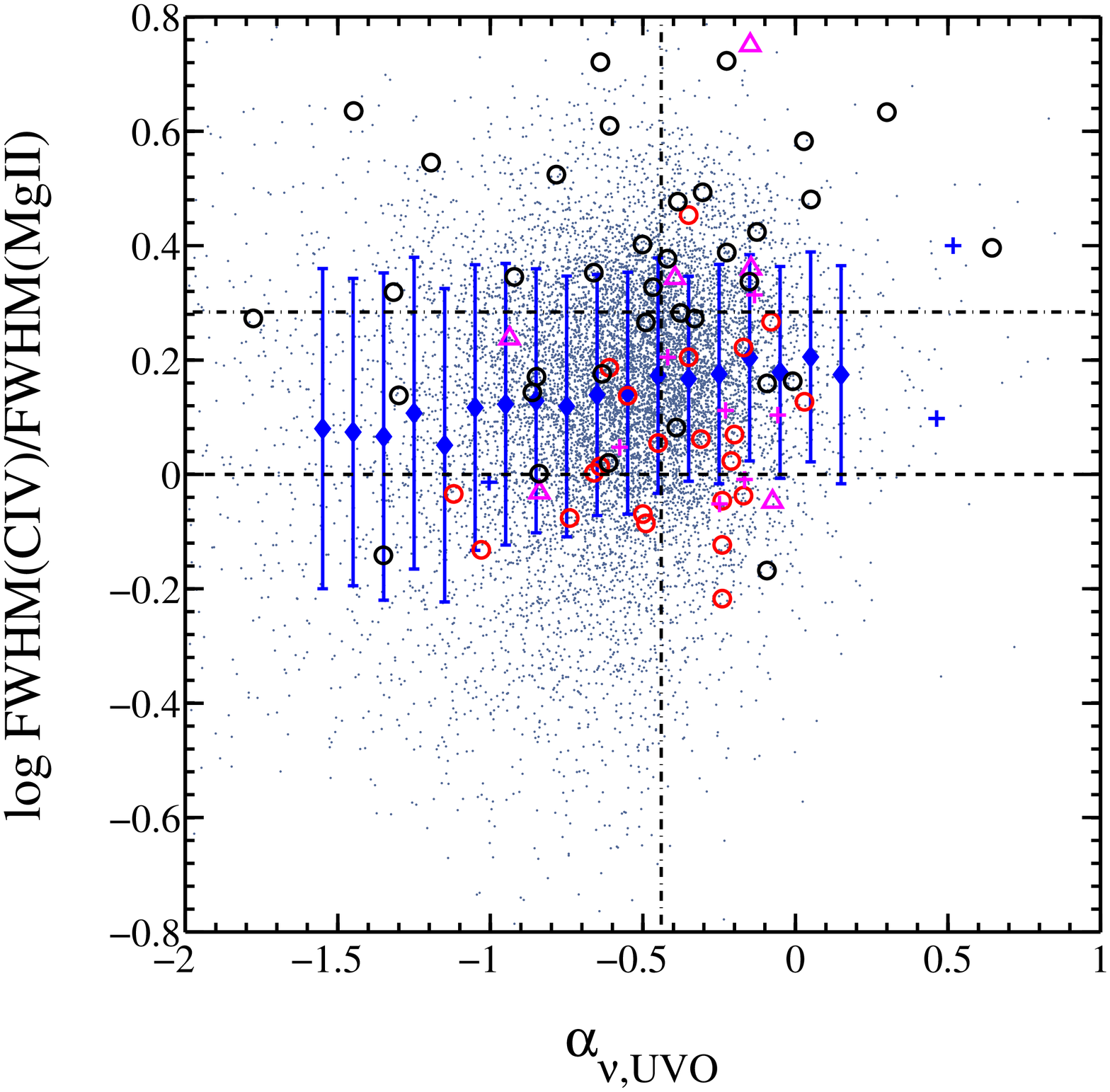}
\includegraphics[width=0.47\textwidth]{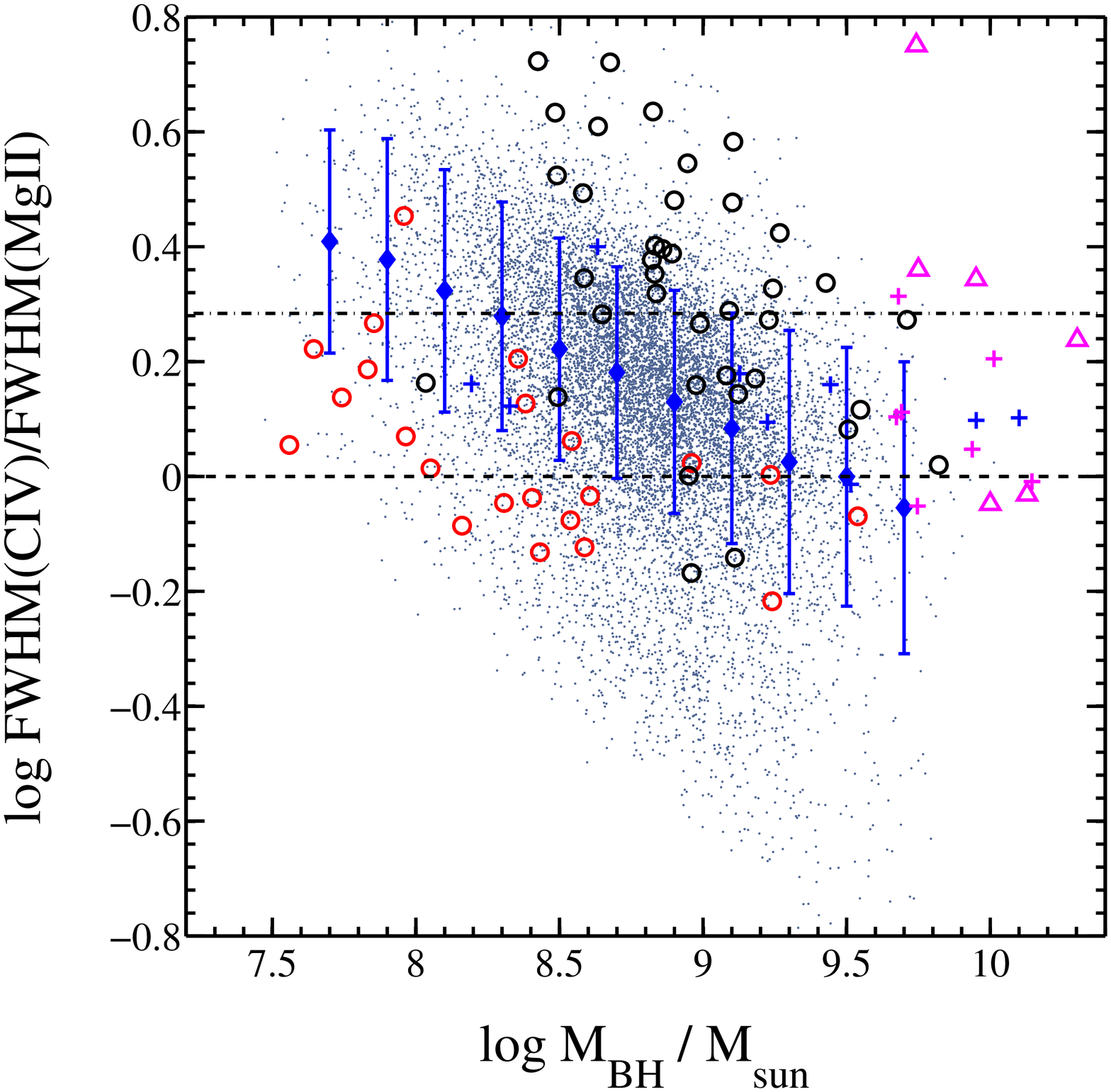}
\includegraphics[width=0.47\textwidth]{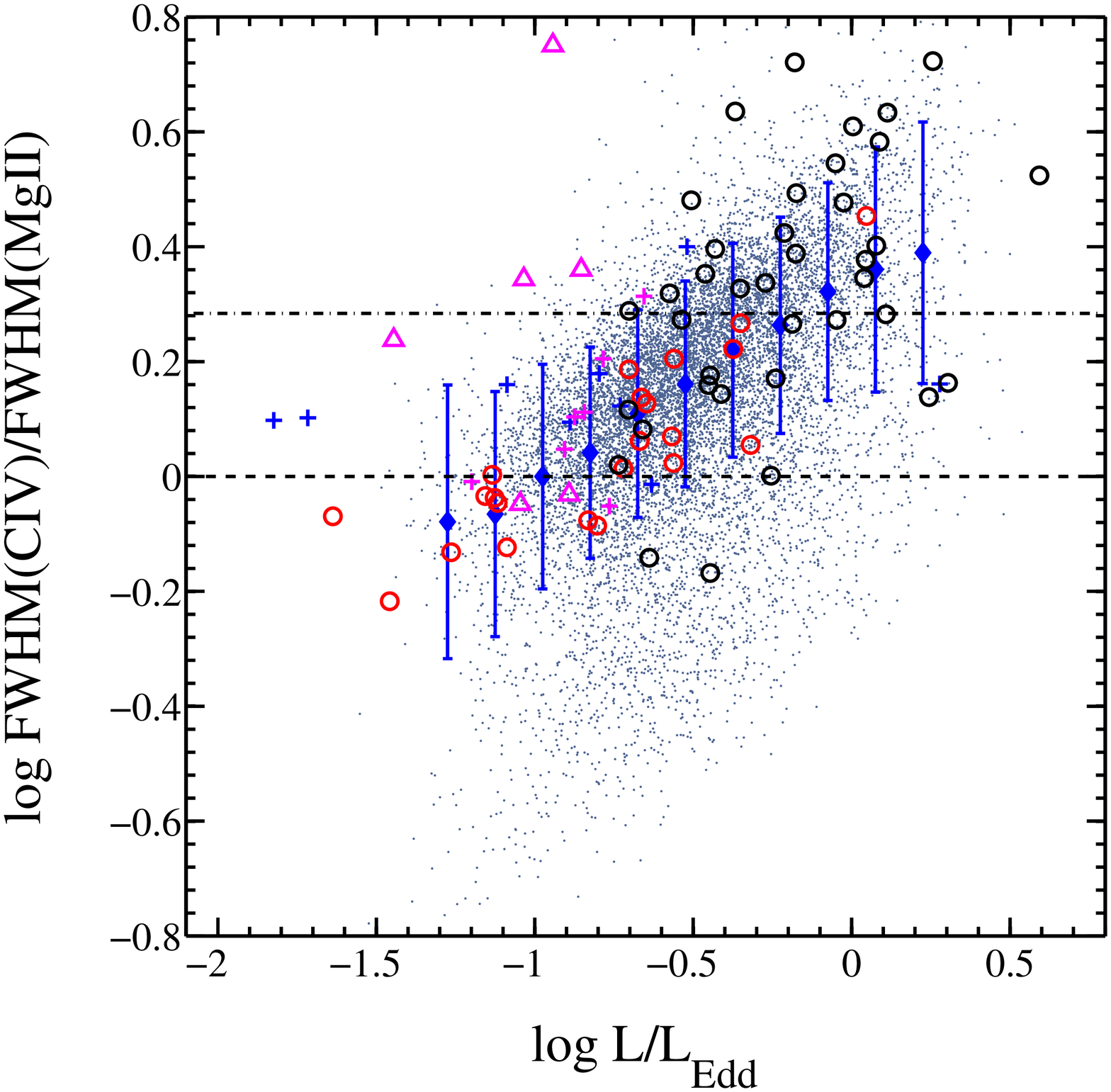}
\end{center}
\caption{
The relations between the difference in the \mgii\ and \civ\ line width, $\log\left(\fwciv/\fwmg\right)$, and several AGN properties (left to right, top to bottom): \Lbol, \auvo , \mbh\ and \lledd. 
In all panels, small blue dots represent the SDSS \mgXciv\ sub-sample, while larger blue diamond symbols and error bars represent the binned data and the corresponding scatter (standard deviations).
Symbols for several of the small samples are identical to previous figures.
The dashed lines represent $\fwciv=\fwmg$, while the dot-dashed lines represent $\fwciv=\sqrt{3.7}\fwmg$, as expected from the virial assumption.
Note the large scatter in all panels, as well as the lack of significant correlations among samples of different luminosities. 
The apparent trends within the SDSS \mgXciv sub-sample (in the lower panels) are due to its limited luminosity range and the dependence of \mbh\ and \lledd\ on \fwmg.
}
\label{fig:FW_diff_civ_mg}
\end{figure*}

To further understand the large scatter in Figs.~\ref{fig:FW_C4_vs_Mg} and \ref{fig:FW_C4_vs_Hb}, we tried to look for correlations between either $\fwciv/\fwmg$ or $\fwciv/\fwhb$ and other AGN properties.
The different panels in Figure~\ref{fig:FW_diff_civ_mg} compare $\fwciv/\fwmg$ with \Lbol, \mbh, \lledd, and the shape of the optical-UV SED, \auvo.
In this comparison, \Lbol, \mbh, \lledd\ and \auvo\ were calculated from \hb-related observables whenever possible (i.e. the M09, Mc08 and Sh07 samples), and using Eq.~\ref{eq:M_Hb} and the \cite{Marconi2004} bolometric corrections. 
In all other cases these quantities were calculated from the \mgii-related observables and Eqs.~\ref{eq:M_L3000_final} \& \ref{eq:fbol_uv_poly_1}.
The shape of the SED is calculated following 
$\auvo= -\left[\log\left(\Lop/\Luv\right)/\left(5100/1450\right)\right]-1$, or 
$\auvo= -\left[\log\left(\Lthree/\Luv\right)/\left(3000/1450\right)\right]-1$, for the \mgXciv\ sub-samples that lack \Lop\ measurements.
Most panels show considerable scatter, and a lack of any significant correlations.
Although the SDSS sub-sample shows some systematic trends of $\fwciv/\fwmg$ with \mbh\ and \lledd\ \cite[see also][]{Wang2011_BLR_CIV}, these trends completely disappear once the smaller samples are taken into account. 
These trends are thus a result of the very limited range in luminosity of the SDSS \mgXciv\ sub-sample, and the dependence of \mbh\ (and \lledd) on \fwmg.
In particular, we draw attention to the lack of a significant (anti-)correlation between $\fwciv/\fwmg$ and \auvo. Such an anti-correlation is expected due to the similar forms of the $\RBLR-L$ relations (Eqs.~\ref{eq:R_L5100_original} \& \ref{eq:R_L1450_K07}), the uniformity of UV-optical SEDs (\S\ref{subsec_L_SED}), and the virial assumption.
We will return to this point below.
The different panels in Fig.~\ref{fig:FW_diff_civ_mg} also demonstrate that the population of sources that do not comply with the basic expectations (i.e. sources with $\fwciv<\fwmg$) cannot be distinguished from the general population.

An alternative way to examine this issue is to drop the assumption that $f$ is the same for all lines, and consider empirical estimates of the $f$-factors associated with a \civ-based estimator of \mbh. 
This can be done by combining the $\RBLR-\Luv$ relation \cite[given by ][Eq.~\ref{eq:R_L1450_K07}]{Kaspi2007} and \fwciv, for all the \hbXciv\ and \mgXciv\ sub-samples, into a ``virial product'' of the form
\begin{equation}
 \muciv= 2.09 \times 10^8 \left [\frac{\Luv}{10^{46}\,\ergs}\right]^{0.55} 
 \left[\frac{\fwciv}{10^3\,\kms}\right]^2 \,\, \Msun\, .
 \label{eq:mu_civ}
\end{equation}
We can then calculate the \civ-related $f$-factors ($\fciv\equiv\mbh/\muciv$), where \mbh\ is determined either from \Lop\ \& \fwhb\ (Eq.~\ref{eq:M_Hb}) or \Lthree\ \& \fwmg\ (Eq.~\ref{eq:M_L3000_final}).
As mentioned above, we expect $\fciv=f\left(\hb\right)=1$. 

\begin{figure*}
\begin{center}
\includegraphics[width=0.47\textwidth]{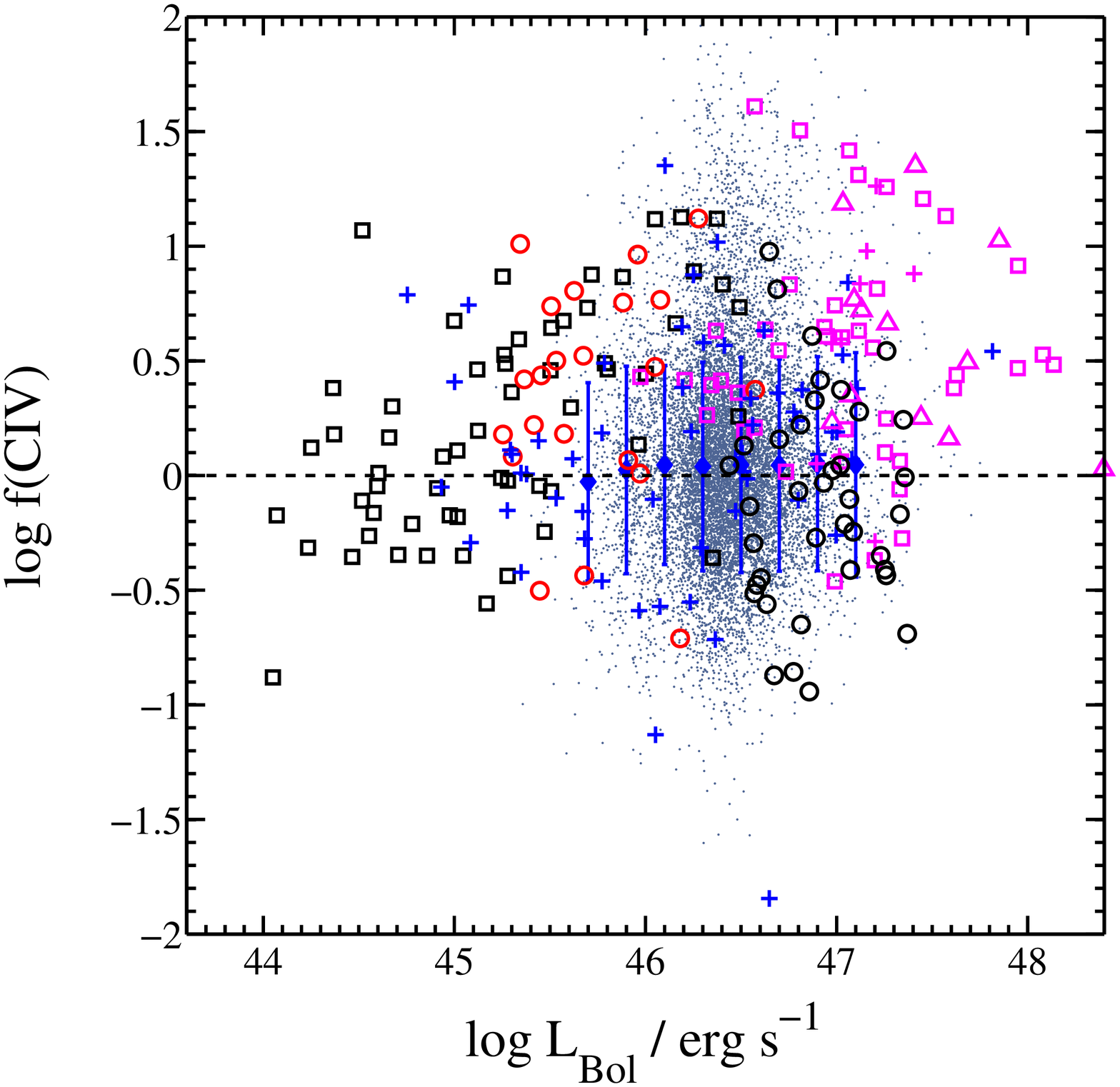}
\includegraphics[width=0.47\textwidth]{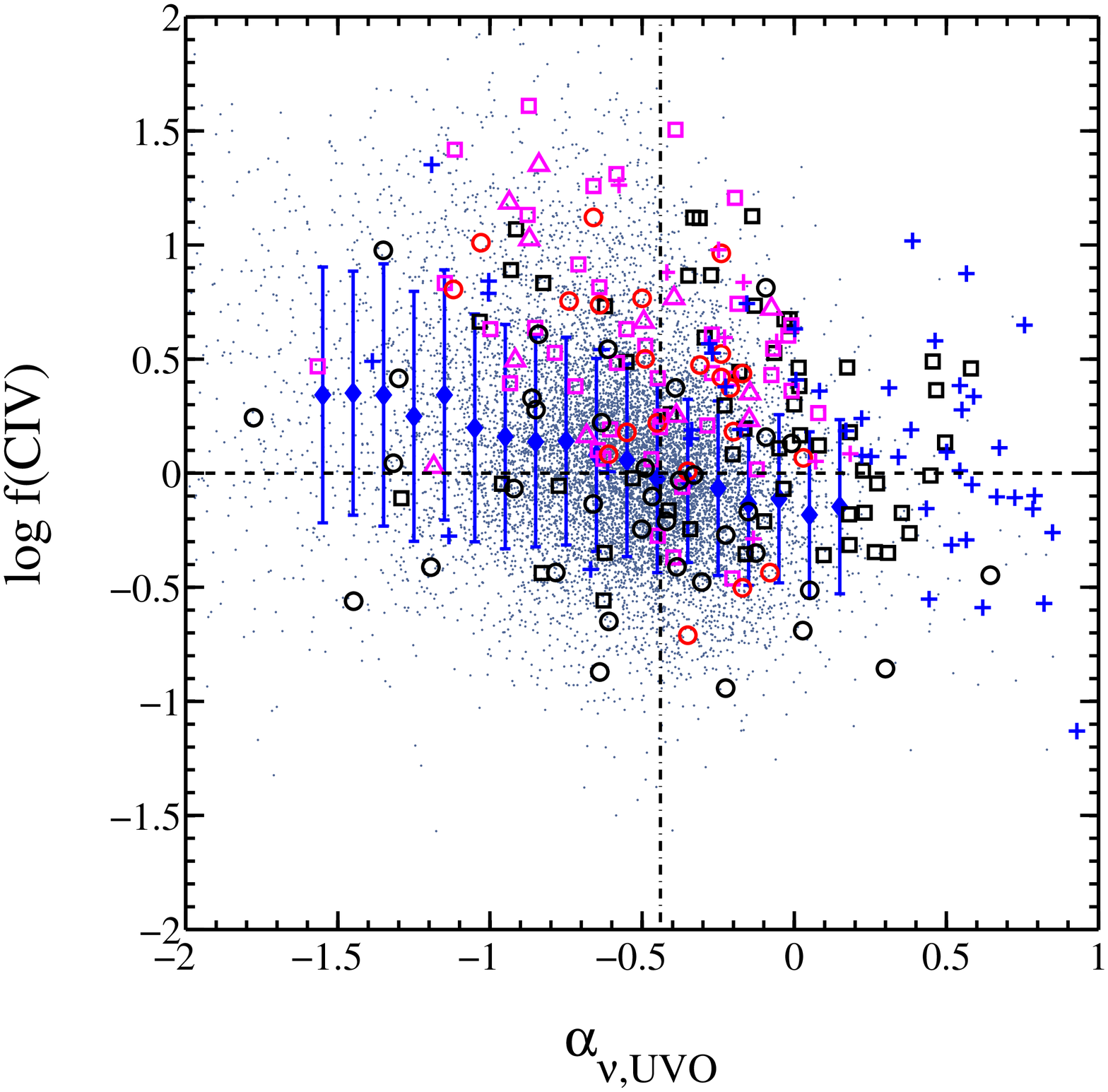}
\includegraphics[width=0.47\textwidth]{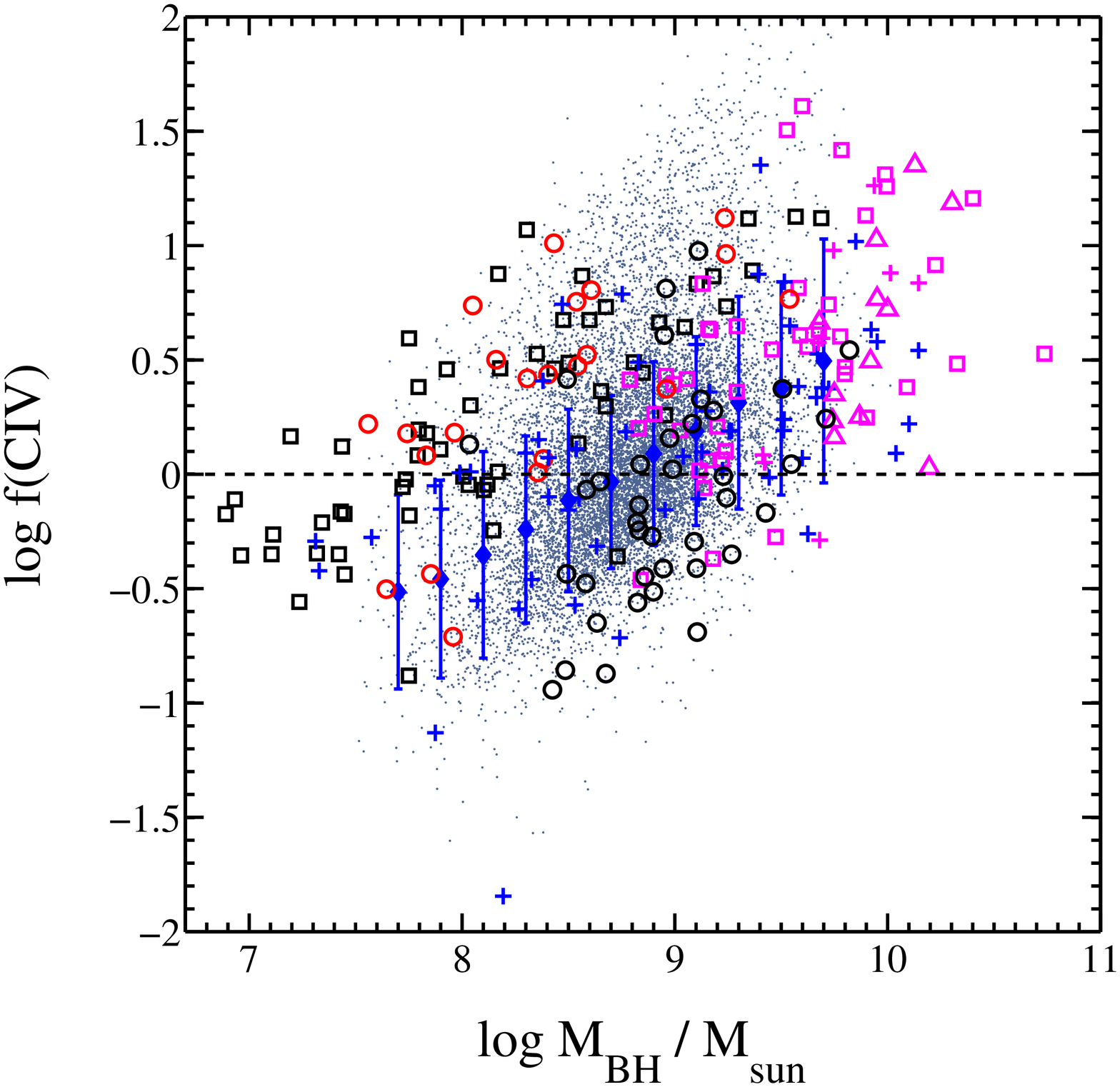}
\includegraphics[width=0.47\textwidth]{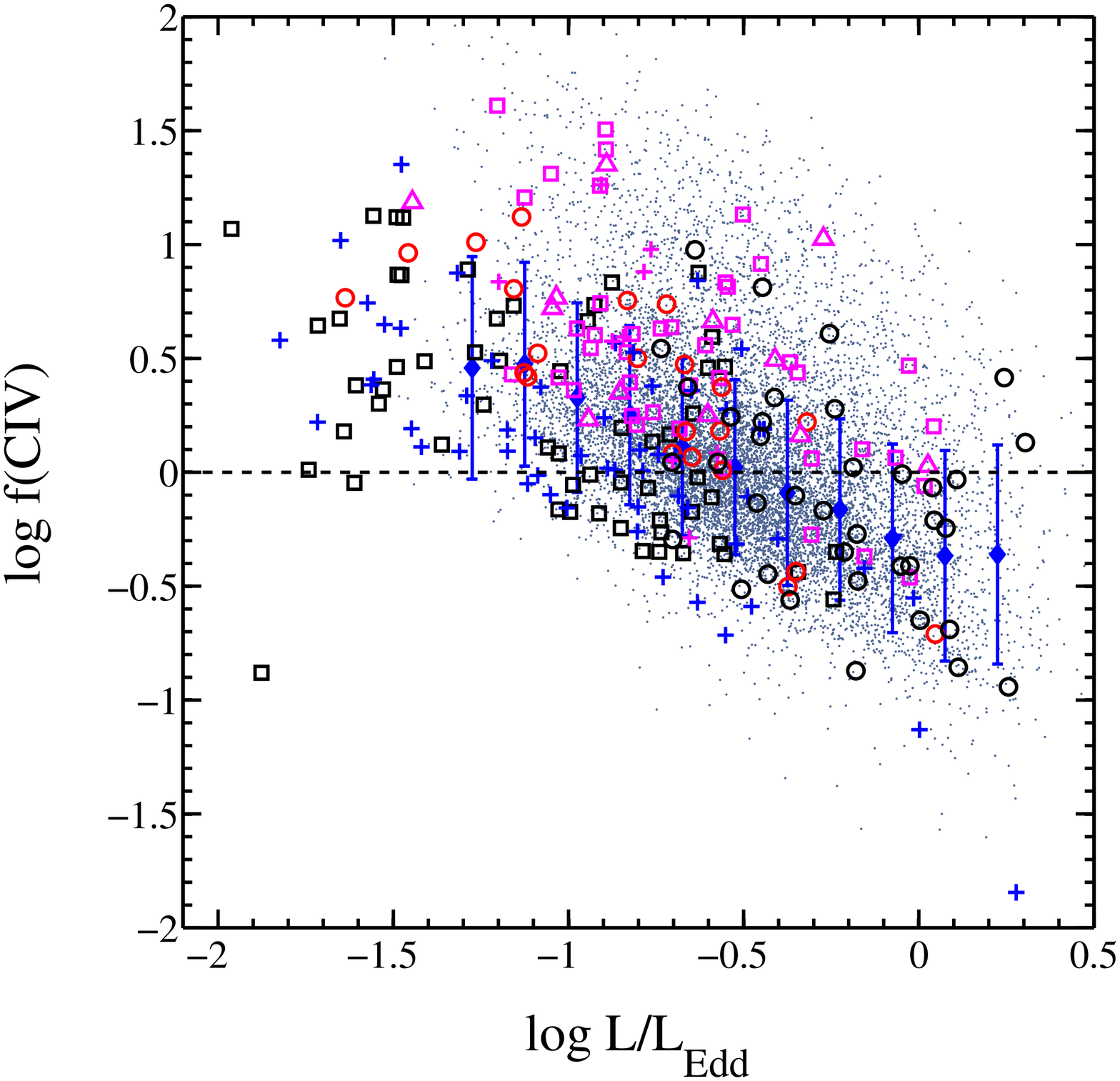}
\end{center}
\caption{
The relations between the scaling factor \fciv\ and several AGN properties (left to right, top to bottom): \Lbol, \auvo, \mbh\ and \lledd. 
Symbols are identical to previous figures.
The dashed lines mark $\fciv=1$, as expected from the virial assumption and $f\left(\hb\right)=1$ .
Note the large scatter in all panels, as well as the lack of significant correlations among samples of different luminosities.
Here too, the apparent trends within the SDSS \mgXciv\ sub-sample (in the lower panels) are due to its limited luminosity range and the dependence of \mbh\ and \lledd\ on \fwmg.
}
\label{fig:f_civ_props}
\end{figure*}

The distribution of \fciv\ for the SDSS \mgXciv\ sub-sample is shown in Figure~\ref{fig:f_mg_civ_hist}. 
It shows a very broad distribution with a peak at about $\fciv=1.1$, in agreement with the expected value. 
Thus, formally, the usage of \Luv\ and \fwciv\ can reproduce the correct \textit{typical} \mbh\ for a large sample of sources.
The recent study of \cite{Croom2011} showed that, indeed, for large samples with narrow ranges of redshift and luminosities, the typical (average) \mbh\ can be determined solely based on the distribution of luminosities, regardless of \fwhm.
However, the scatter in \fciv\ (the standard deviation), for the SDSS \mgXciv\ sub-sample, is about 0.46 dex. Moreover, 25\% of the sources show $\fciv>2.1$ and an additional 25\% show $\fciv<0.6$. 
This practically prohibits the usage of a single scaling factor to determine \mbh\ in individual sources. 
Since the scatter in $\Luv/\Lthree$ and $\Luv/\Lop$ is rather small (see \S\ref{subsec_L_SED}), and since the slopes in the different $\RBLR-L$ relations are similar, most of the scatter in \fciv\ is probably due to the scatter in $\fwciv/\fwmg$ and $\fwciv/\fwhb$.
We investigate this point further by plotting, in Figure~\ref{fig:f_civ_props}, the $f$-factor against several observed properties.
The large scatter, and lack of correlation among samples of different luminosities, is evident in all the panels, and in particular in the comparison with \auvo. 
In contrast, a comparison of \fciv\ with $\fwciv/\fwmg$ exhibits a very prominent correlation, which closely follows a $\fciv\propto\left(\fwciv/\fwmg\right)^2$ trend.
This clearly demonstrates that the main source for differences between the different virial \mbh\ estimators are the large and unsystematic differences between \fwciv\ and \fwmg\ (or other velocity measures).

The recent study by \cite{Assef2011} presented detailed measurements of \hb\ and \civ\ for 12 lensed QSOs at $z=1.5-3.6$, drawn from the CASTLES survey. These authors suggest that the differences between \mbh(\hb) and \mbh(\civ) \cite[estimated using the calibration of ][]{Vester_Peterson2006} are mainly driven by the shape of the UV-optical SED (i.e., \Luv/\Lop), and provide empirical correction terms to account for these differences.
As noted in \S\ref{subsec_L_SED}, the (small) sample of \cite{Assef2011} includes heavily reddened sources, with about a half of their sources having $\Lop>\Luv$ (corresponding to $\auvo<-1$). 
Less than 15\% of our SDSS sources show such red SEDs.
The corrections provided by \cite{Assef2011} thus minimize the effect of the SED shape, or reddening, on \mbh\ estimates.
However, as we showed above, the main driver for the discrepancy between \mbh(\hb) and \mbh(\civ) is most probably related to \fwciv, and not to the SED shape. 
Indeed, almost half of sources in the \cite{Assef2011} sample have $\fwciv\ltsim\fwhb$, in contrast with the virial assumption, and in agreement with the fraction of such sources in our \hbXciv\ sub-samples.
Moreover, \cite{Assef2011} report a large scatter between \fwciv\ and \fwhb, similarly to our findings. 
We also note that the usage of correction terms which are based on \Lop\ (or \fwhb) is impractical for large samples of $z\gtsim0.8$ sources. On the other hand, if the \hb\ line for such sources is available (through NIR spectroscopy), it can be used to determine \mbh\ directly (e.g., by using Eq.~\ref{eq:M_Hb}), and we see no point in using it solely to correct systematics in \mbh(\civ).

Figs.~\ref{fig:FW_diff_civ_mg} and \ref{fig:f_civ_props} demonstrate that the discrepancies associated with a \civ-based virial product cannot be accounted for even \textit{if} other basic AGN properties are known.
Moreover, some of these properties \textit{cannot} be reliably determined \textit{a-priori}, given a single spectrum that contains only the \civ\ line.\footnotemark
\footnotetext{While \Lbol\ can be determined from \Luv, and \auvo\ can be estimated given a broad enough spectral coverage, \mbh\ and thus \lledd\ depend on the ability of to construct a \civ-based estimator for \mbh.}
We therefore tested the possibility of correcting the discrepancy described above by using only \civ-related observables, namely the shift (\voff[\civ]) of the line center and the equivalent width of the line (EW[\civ]).
Previous studies suggested that the large blue-shifts often observed in \civ\ indicate that the emission originates from non-virialized gas motion \cite[e.g., BL05,][]{Richards2011_CIV}. 
Such a scenario may explain why a simple virial product (Eq.~\ref{eq:mu_civ}) fails to scale with \mbh.
A trend with EW(\civ) might be expected if, for example, there was a common origin for the difference in \fwhm\ and the well-known Baldwin Effect.

\begin{figure}
\includegraphics[width=0.47\textwidth]{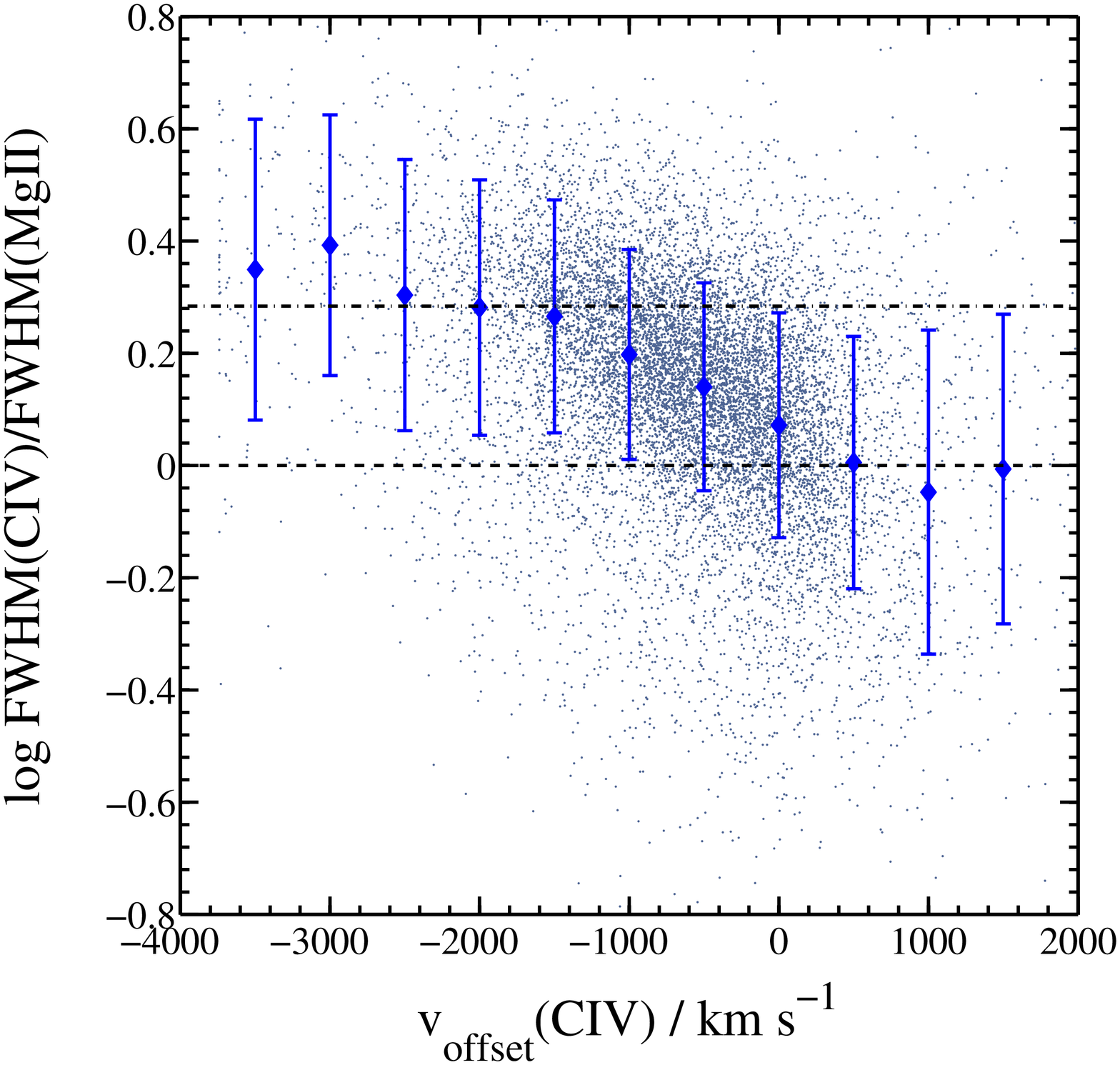}
\caption{
The relation between the difference in the \mgii\ and \civ\ line width, and the offset of the \civ\ line with respect to the \mgii\ line, for the SDSS \mgXciv\ sub-sample.
Larger symbols and error bars represent the binned data and the corresponding scatter (standard deviations).
The dashed line represents $\fwciv=\fwmg$, while the dot-dashed line represents $\fwciv=\sqrt{3.7}\fwmg$, as expected from the virial assumption.
}
\label{fig:FW_diff_civ_mg_vs_voff}
\end{figure}

We first verified that there is no clear trend of neither $\fwciv/\fwmg$ nor \fciv\ with EW(\civ).
In particular, the population of sources with $\fwciv<\fwmg$ cannot be distinguished from the rest of the sources based on their EW(\civ).
Next, We calculated the shifts of the \civ\ line relative to the \mgii\ for the \mgXciv\ sub-samples under study.
We assumed that the rest-frame center of the \mgii\ doublet is at 2799.11\AA, and that the \civ\ profile is centered at 1549.48\AA. 
Our fitting procedures calculate the observed line centers as the peaks of the entire best-fit BLR profile.
Figure~\ref{fig:FW_diff_civ_mg_vs_voff} presents the resulting \voff(\civ) against $\fwciv/\fwmg$. 
As is the case in Fig.~\ref{fig:FW_diff_civ_mg}, the scatter is very large and covers more than a factor of 5 in $\fwciv/\fwmg$. However, there appears to be a clear trend of an increasing $\fwciv/\fwmg$ with increasing \civ\ \textit{blue}-shift (negative velocities).

Figure~\ref{fig:FW_diff_civ_mg_vs_voff} draws attention to two particular types of sources which may, in principle, be used to overcome the problems associated with \fwciv.
First, \civ\ lines in sources with small offsets may be dominated by virialized gas. 
To test this, we selected a subset of about 4500 sources from the SDSS \mgXciv\ sub-sample that have 
$|\voff(\civ)|<500\,\kms$.
The distribution of \fciv\ for this subset is shown in Fig.~\ref{fig:f_mg_civ_hist}. 
Clearly, the large scatter in $\fwciv/\fwmg$, of more than 0.2 dex, propagated to the virial products of these sources. 
Moreover, the typical \fciv\ for this subset is 1.49 - considerably higher than the expected value (of unity). 
Second, sources with blueshifts of about $1500-2500\,\kms$ appear to match the expected value of $\fwciv/\fwmg \simeq 1.9$. Here too, the scatter in $\fwciv/\fwmg$ for the relevant sources is larger than 0.2 dex, and the scatter in \fciv\ is 0.5 dex. Moreover, there is no physical motivation for focusing on lines with such particular offset velocities, which are probably produced by non-virialized gas.
Thus, \civ\ cannot be used to precisely estimate \mbh\ even for these specific sources.
We also note the great difficulty in robustly determining \voff(\civ) in optical spectra of high redshift sources, where the systemic redshift determination heavily relies on a few UV lines, including \civ\ itself and the complex, often partially absorbed \Lya\ spectral region.

We conclude that the \civ\ line is an unreliable probe of the kinematics of the BLR gas. There is a significant population of type-I AGN, indistinguishable from the general population, for which the width of the \civ\ line contradicts the basic virial expectation. 
A single-epoch \mbh\ estimator which relies on the width of the \civ\ line provides results that deviate by $\pm$0.46 dex from the more reliable \hb-based estimator. 
This scatter is due to \fciv\ only and was estimated assuming a constant $f\left(\hb\right)$ and negligible uncertainties in the $\RBLR-L$ relations and in the measurements
of all relevant observables. The uncertainty in individual \civ-based estimates of \mbh\ can be considerably larger (see \citet{Woo2010_LAMP_Msig} for the case of $f\left[\hb\right]$).

\section{Discussion and Conclusions}
\label{sec_diss_con}

The reliable estimators of \mbh\ and \lledd\ obtained in the present work should enable us to examine the observed distributions of these quantities, and their evolution over cosmic epochs. 
However, since both \mbh\ and \lledd\ depend on the source luminosity, their observed distributions, at any given redshift, would suffer from selection effects due to the flux limit of the respective survey.
The treatment of this issue requires the application of dedicated statistical methods \cite[e.g.,][NT07]{Kelly2010}, 
which is beyond the scope of this paper.
We thus only briefly present here the preliminary results that concern the {\it upper} envelope of the \mbh\ and \lledd\ distributions, 
and defer the full analysis of the observed SMBH evolution to a forthcoming paper.

We apply the \hb- and \mgii-based prescriptions to all the samples listed in Table~\ref{tab_samples}.
For sources with reliable \hb\ measurements (mainly SDSS sources at $z\leq0.75$), the bolometric luminosities are estimated using the \cite{Marconi2004} corrections and \mbh\ is calculated through Eq.~\ref{eq:M_Hb}.
For sources with reliable \mgii\ measurements (mainly SDSS sources at $0.75\ltsim z\ltsim2$)
we use the bolometric corrections given by Eq.~\ref{eq:fbol_uv_poly_1}, while \mbh\ is calculated by Eq.~\ref{eq:M_L3000_final}.
In all cases where more than one line is observed, we prefer \hb-based measurements on other measurements.

\begin{figure*}
\centering
\includegraphics[width=0.90\textwidth]{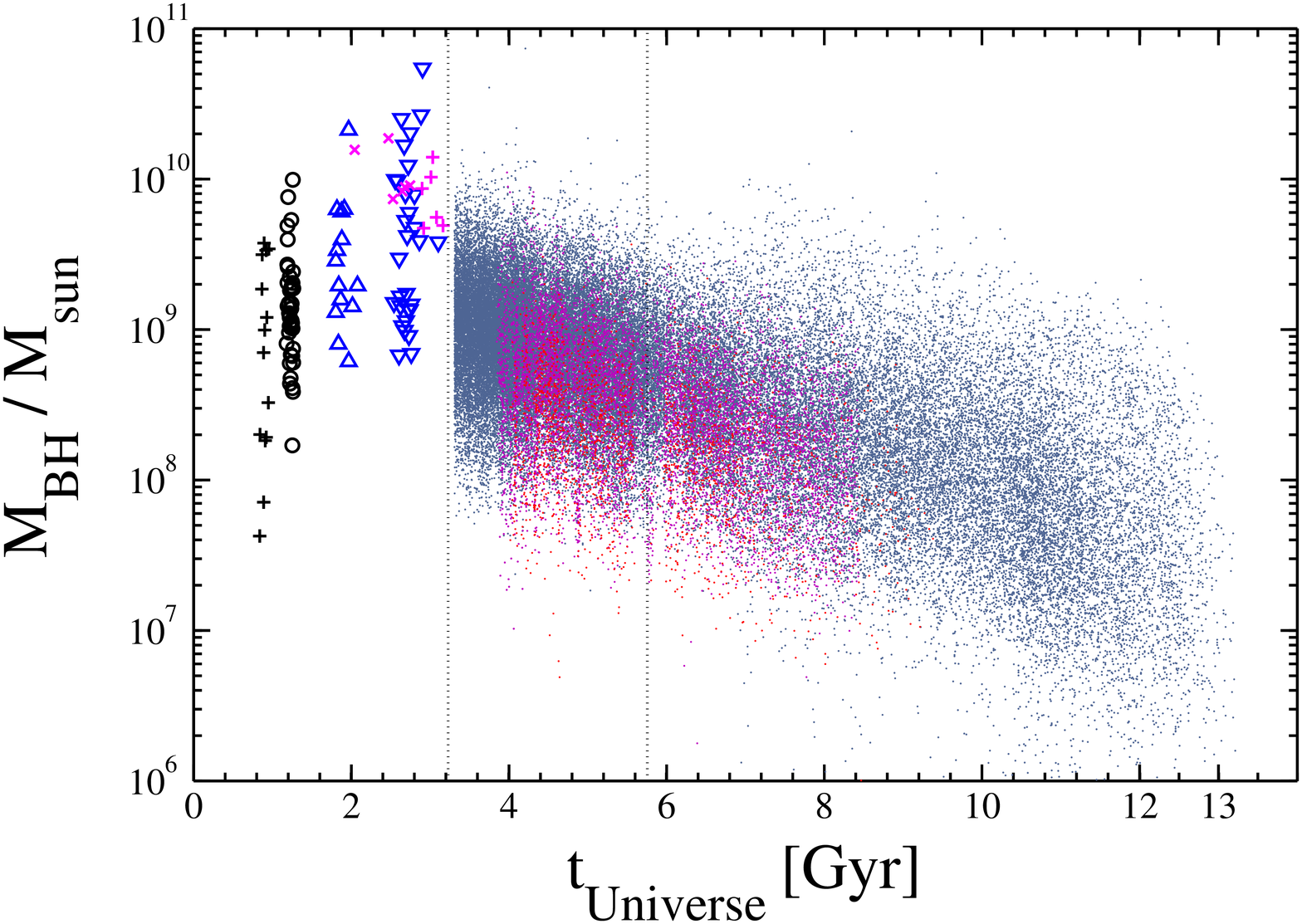}
\caption{
The evolution of \mbh\ along the age of the Universe for several of the samples studied here:
the SDSS, 2QZ and 2SLAQ samples at $z<2$ (small blue, magenta and red points, respectively); 
the combined N07+S04 sample at \znetprev\ (blue triangles);
the $z>2$ sources from the M09 (magenta x-signs) and D09 (magenta crosses) samples; 
the T11 sources at \zfpe\ (black circles); 
 and a sample of \zsix\ sources, taken from the studies of \citet{Kurk2007} and \citet[black crosses]{Willott2010}.
 The \mbh\ estimates are based on \hb\ (for the N07+S04, D09, M09 and $z<0.75$ SDSS sources) or \mgii\ (for the T11, \zsix\ and $z>0.75$ SDSS sources), following Eqs.~\ref{eq:M_Hb} or \ref{eq:M_L3000_final}, respectively.
 Dotted vertical lines mark $z=1$ and $2$.
}
\label{fig:MBH_t_all}
\end{figure*}

Figure~\ref{fig:MBH_t_all} describes the evolution of \mbh\ with the age of the Universe, for the SDSS, 2QZ \& 2SLAQ samples (at $z\ltsim2$), as well as several of the $z>2$ samples (N07+S04, M09, D09 and T11). 
For completeness, we also include a small sample of \zsix\ sources, taken from the studies of \cite{Kurk2007} and \cite{Willott2010}.
Fig.~\ref{fig:MBH_t_all} suggests that the most massive BHs grow at the fastest rates at $z\gtsim4.5$, reach their final masses ($\mbh\gtsim10^{10}\,\Msun$) before $z\sim2$, and remain mostly inactive thereafter (see discussion in T11).
These objects are probably the progenitors of the most massive {\it relic} SMBHs found in the centers of giant elliptical galaxies \cite[M87 and several other BCGs, see, e.g.][]{McConnell2011_MBH_10}. 
The most luminous (SDSS) AGN at $z\sim1$ are considerably less massive, and are probably the descendants of the less luminous AGN at $z\sim2$, which may be fainter that the 2SLAQ flux limit. 
Such a scenario should be tested critically, by comparing the number densities of the different populations of AGN, which trace a sequence of increasing \mbh\ with cosmic time.
Finally, Fig.~\ref{fig:MBH_t_all} indicates that the shut-down of accretion onto SMBHs, at $z\sim1-2$, cannot be associated with a certain ``maximal'' value of \mbh.

\begin{figure*}
\centering
\includegraphics[width=0.7\textwidth]{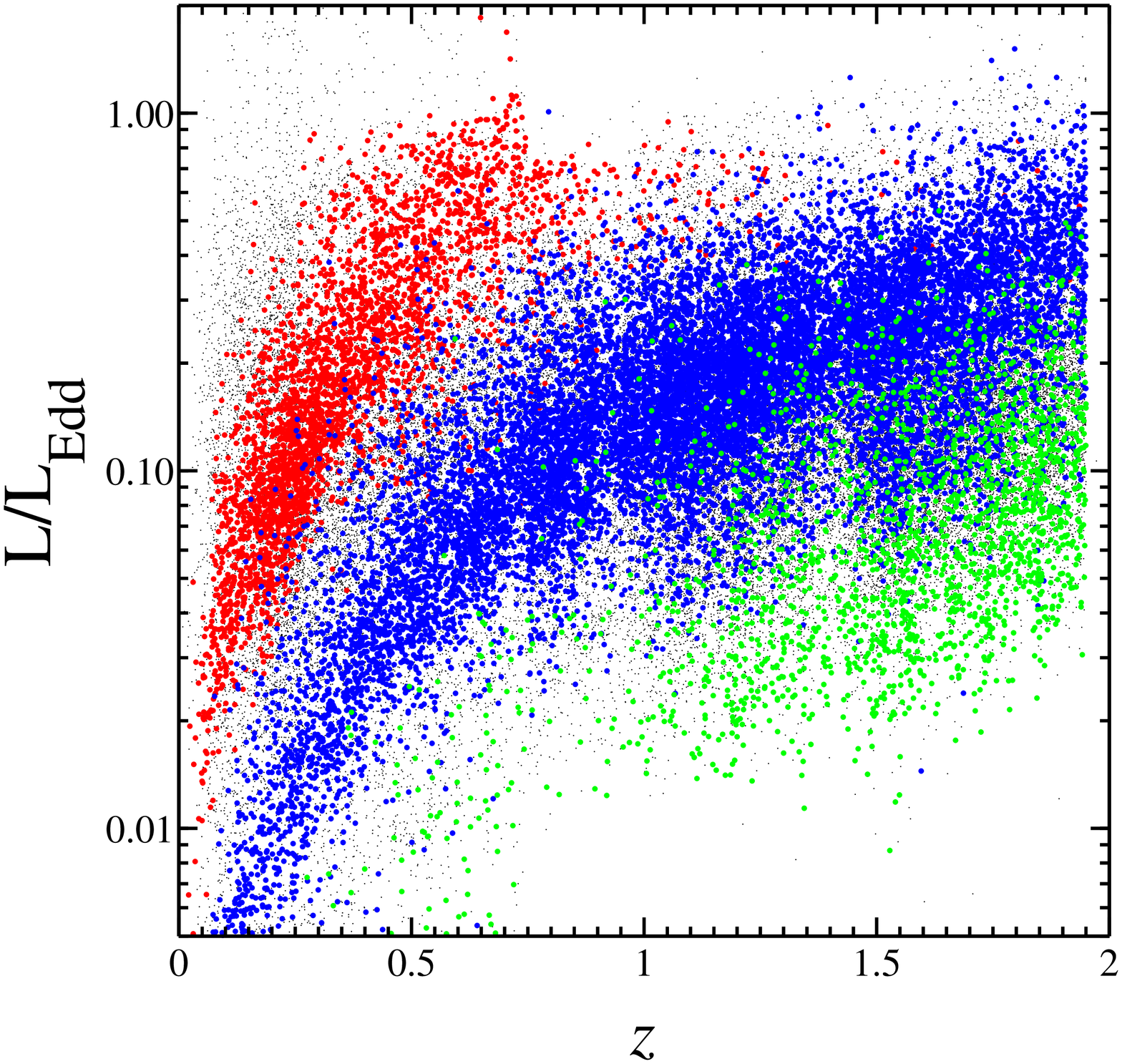}
\caption{
The evolution of \lledd\ with redshift, for the entire SDSS sample studied here. 
Red, blue and green points highlight sub-samples of sources, with \mbh\ is in the range of $10^{7.5-7.8}$, $10^{8.5-8.8}$ and $10^{9.5-9.8}\,\Msun$, respectively.
Note the lack of very massive BHs that accrete at high rates (i.e. sources with $\mbh > 3\times10^{9}\,\Msun$ and $\lledd>0.2$) at $z<2$.
}
\label{fig:Ledd_z_all}
\end{figure*}

Figure~\ref{fig:Ledd_z_all} shows the evolution of \lledd\ with redshift for the SDSS sample studied here.\footnotemark
\footnotetext{The 2QZ and 2SLAQ samples add little information in to this figure, and will be analyzed in a forthcoming paper.}
We draw attention to the smooth transition seen in Fig.~\ref{fig:Ledd_z_all} at $z=0.75$. This region marks the transition from \mbh\ and \lledd\ estimates that are based on \hb\ and \Lop, to those based on \mgii\ and \Lthree. 
As explained in \S\ref{subsec_f_L3000_mgii}, this transition wouldn't appear as smooth if these quantities were obtained using the methods of \cite{McLure_Dunlop2004} and \cite{Richards2006_SED}.
The steep rise of the upper envelope of the \lledd\ distribution with redshift (see also NT07) flattens at $z\sim1$, so that the highest-\lledd\ sources approach the expected limit of $\lledd\simeq1$.
We stress that the upper envelope of the \lledd\ distribution is not affected by selection effects related to the flux limits of the various samples.
For example, an un-obscured AGN at $z=1.5$, powered by a SMBH with $\mbh=10^9\,\Msun$ and accreting at the Eddington limit could have easily been observed within the SDSS, since its observed $i$-band magnitude would be $\sim$16.4.
Such a source would appear as a green point, at $z=1.5$ and $\lledd=1$ in Fig.~\ref{fig:Ledd_z_all} - a region in parameter space which is clearly dominated by sources with much lower \mbh. 
Thus, Fig.~\ref{fig:Ledd_z_all} suggests that the vast majority of very massive BHs, with $\mbh\gtsim10^9\,\Msun$, do not accrete close to their Eddington limit even at $z\sim2$, which is often considered as ``the epoch of peak Quasar activity''. Such SMBHs have probably experienced periods of faster mass accumulation at $z>3$ \cite[][T11]{Willott2010,DeRosa2011}.
This scenario is supported by the extreme rareness of AGNs with $\Lbol > 10^{47}$ at $z\sim1-2$ \cite[see, e.g.,][and references therein]{Croom2004,Hasinger2005,Richards2006_QLF,Hopkins2007_QLF}.
As mentioned above, the flux limits have a considerable influence on the faintest sources, and practically determine the low-end of the{\it observed} distributions we present, at all accessible redshifts.
Indeed, deeper surveys (e.g., zCOSMOS, VVDS) have revealed populations of $z\sim1-2$ AGN with $\lledd\simeq0.05$  
\cite[e.g.][]{Gavignaud2008,Trump2009b_MBH,Merloni2010}.

To conclude, in this first of two papers we presented a systematic study of the methods we use to measure \mbh\ and \lledd\ in type-I AGN at $z\gtsim1$.
This was done through a combination of larger samples, spanning a broader range in redshift and luminosity, in comparison to previous works. 
We focused on the observables associated with the \hb, \MgII\ and \CIV\ emission lines, 
as these are typically available for single epoch spectra of high-redshift sources.
Our main findings are:
\begin{enumerate}

\item 
We provide a luminosity-dependent prescription for obtaining \Lbol\ from \Lthree. 
The resulting \Lbol\ is consistent with the one obtained from \Lop\ and differs by a factor of $\sim1.5$ from previous estimates of this correction, for luminous sources.
 
\item 
The width of the \mgii\ line was shown to follow closely the width of the \hb\ line up to $\fwhm\simeq6000\,\kms$, beyond which the \mgii\ line width appears to saturate (\S\ref{subsec_fwhm_mgii}).
 
\item 
We obtained an improved $\RBLR-\Lthree$ relation, with a best-fit slope of 0.62, consistent with previous studies \cite[e.g.,][]{McLure_Dunlop2004}. We further find that the relation is probably luminosity-dependent and provide several alternative slopes and scalings.
Combining this with point (ii) above, we obtained an \mbh\ estimator that is highly consistent with the \hb-based one (Eq.~\ref{eq:M_L3000_final}). The scatter between the two is 0.32 dex (\S\ref{subsec_f_L3000_mgii}). 
This relation provides \mbh\ estimates that are systematically {\it higher}, by a factor of 1.75, than those of the \cite{McLure_Dunlop2004} \mbh\ estimator.

\item
The combination of points (i) and (iii) above produces \mgii-based estimates of \lledd\ that are systematically {\it lower}, by a factor of about 2.2, than those derived by previously published prescriptions. 

\item 
The width of the \civ\ line shows no correlation with either \hb\ or \mgii, and for most sources is substantially different from the what is expected from the virial assumption. The large scatter in the ratios of  line widths ($\sim$0.5 dex) does not correlate with any other observable AGN property, and thus cannot be corrected for (\S\ref{sec_civ_prob}). 
The problematic behavior of \fwciv\ dominates the associated virial products and practically prohibits any reliable measurements of \mbh\ using the \civ\ line.

\item 
We do not find a significant population of excessively massive BHs ($\mbh>10^9\,\Msun$) that accrete close to their Eddington limit. 
Such high mass BHs are observed as slowly accreting sources ($\lledd\sim0.05-0.1$), as early as $z\simeq2$ (Fig.~\ref{fig:Ledd_z_all}), and might be observable in yet deeper surveys.
 
\end{enumerate}
A full analysis of the evolutionary trends mentioned here, and others, will be presented in a forthcoming paper.

Two new relevant papers were published after the submission of the present paper to the journal. 
The study of \cite{Shen_Liu_2012} presented a new sample of 60 luminous SDSS AGN at $z\sim1.5-2.2$, for which the \civ, \mgii, \hb\ and \ha\ lines were measured, using high quality NIR spectroscopy. 
The conclusions of the \cite{Shen_Liu_2012} study regarding the problematic usage of \civ\ as an \mbh\ estimator are in excellent agreement with our findings. 
In particular, \cite{Shen_Liu_2012} find a large discrepancy between \fwciv\ and \fwhb, and almost half of their sources show $\fwhb>\fwciv$. 
The study of \cite{Ho2012} presented simultaneous rest-frame UV-to-optical spectra of 7 luminous AGN at $z\simeq1.5$, which include the \civ, \mgii, \CIII\ and \ha\ lines. 
All \mgii-based estimates of \mbh\ were shown to be consistent with those based on \ha, while \civ-based estimates of \mbh\ showed large discrepancies, of up to a factor of $\sim$5 (compared with \mbh[\ha]).
Thus, both the \cite{Shen_Liu_2012} and \cite{Ho2012} studies strengthen our conclusions, using additional high quality, high redshift samples. 
%

%


\section{Acknowledgments}

We thank the anonymous referee for his/her careful reading of the manuscript and detailed comments, which allowed us to improve the paper.
We thank Paola Marziani and Lutz Wisotzki for providing spectra and ancillary data for the HES sources; 
Gary Ferland for providing his Iron emission model; and Stephen Fine for providing 2SLAQ-related data and useful comments on the analysis of 2QZ and 2SLAQ sources.
We also thank Yue Shen for useful comments.
This study makes use of data from the SDSS (http://www.sdss.org/collaboration/credits.html). 
Funding for this work has been provided by the Israel Science
Foundation grant 364/07 and by the Jack Adler Chair for Extragalactic Astronomy.

 


\appendix

\section{Photometry-Based Flux Calibration for 2QZ and 2SLAQ Spectra}
\label{app_AAO_fcal}

\begin{figure*}
\centering
\includegraphics[width=0.49\textwidth]{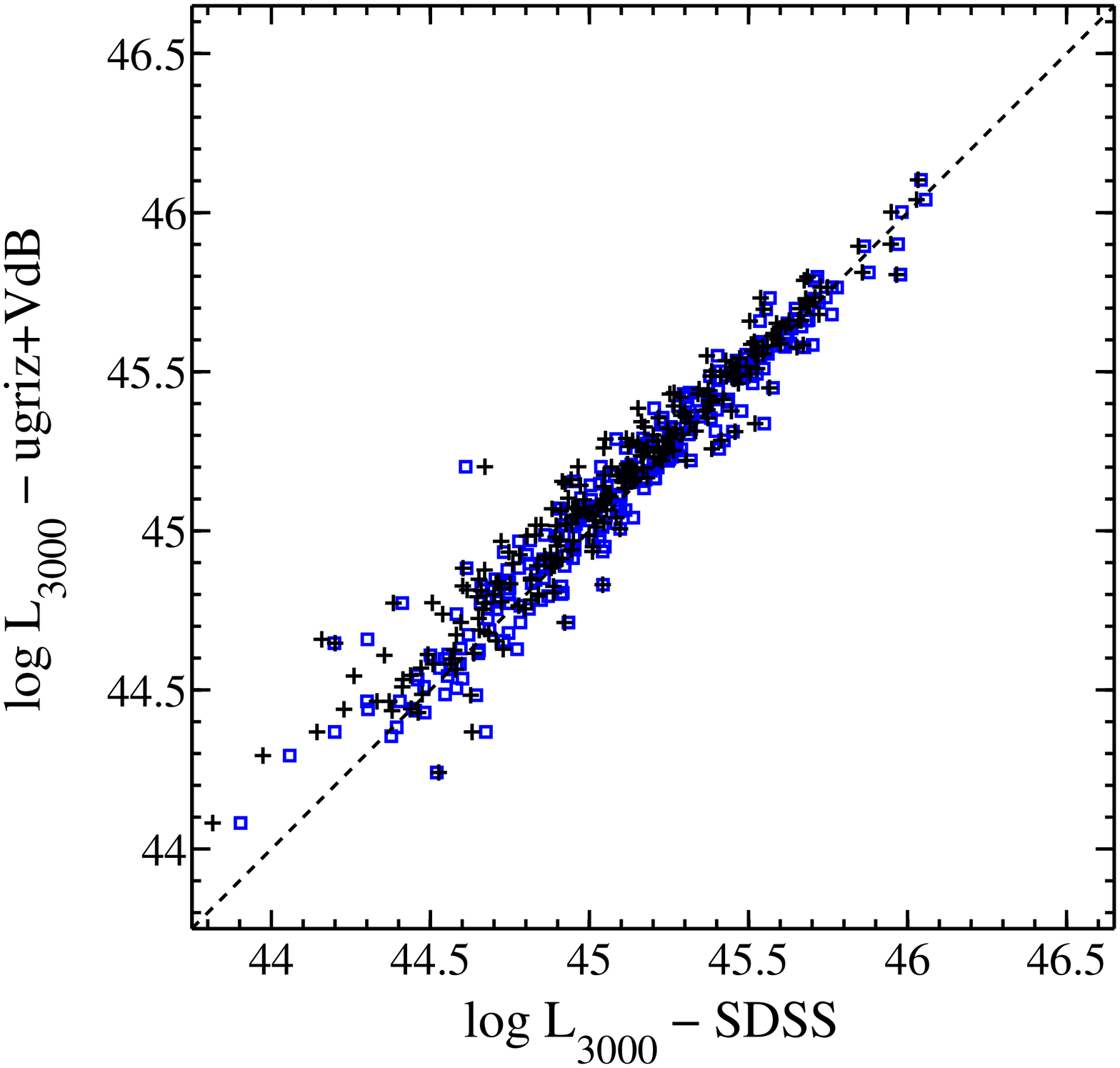}
\includegraphics[width=0.49\textwidth]{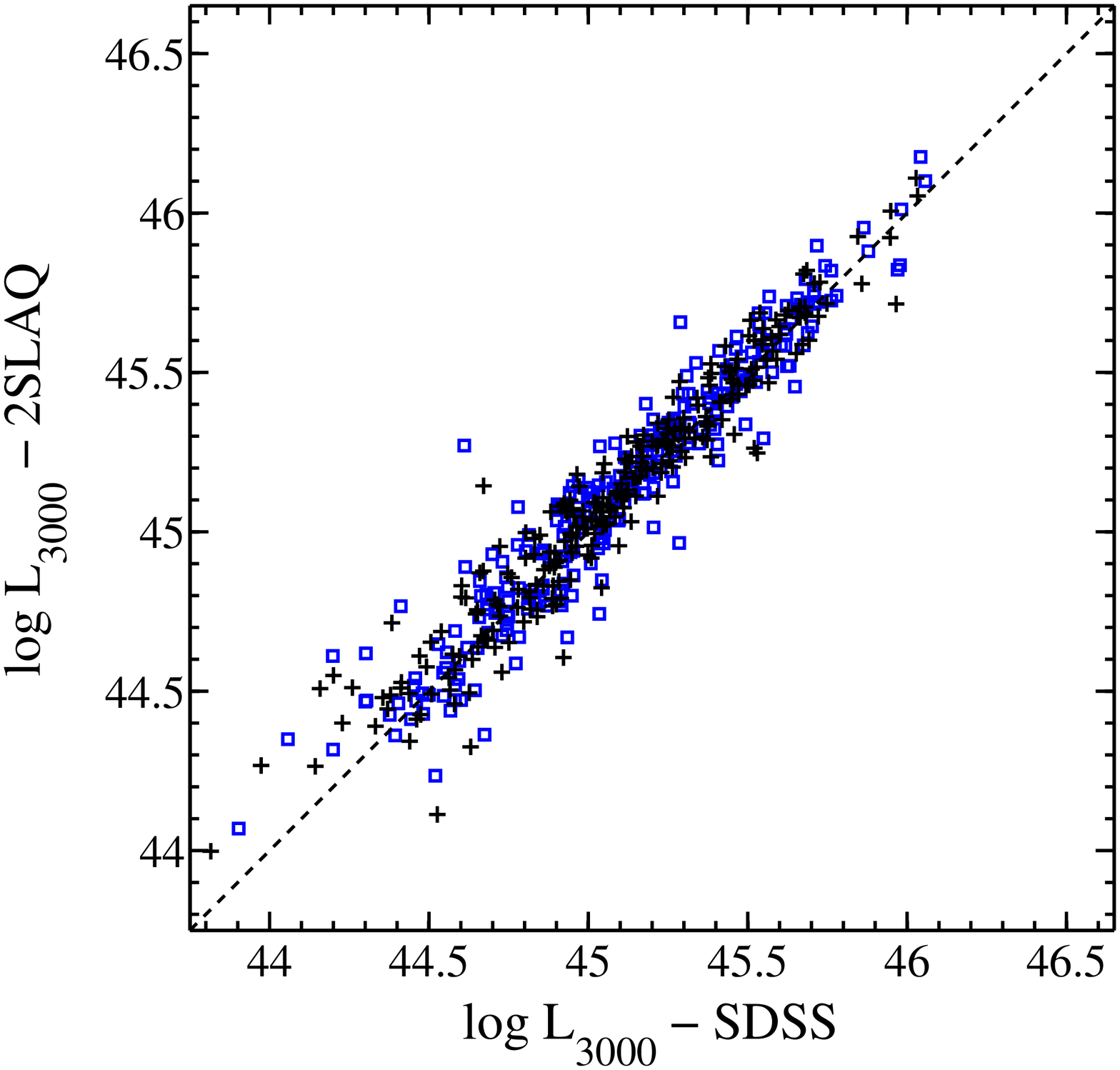}  
\caption{
Comparison of \Lthree, as measured from SDSS spectra and from our calibrated 2SLAQ spectra, for the \NSLAQXSDSS\ sources that have duplicate high-S/N spectroscopy of the \mgii\ spectral region.
\emph{Left:} calibration obtained by the nearest photometric band and assuming the SED follows the uniform composite spectrum of \citet{VandenBerk2001}.
\emph{Right:} calibration obtained by fitting a power law to all the available SDSS ($ugriz$) magnitudes. 
In both panels, blue squares indicate luminosities measured directly from the observed spectra, while black crosses represent the results of our \mgii-fitting procedure.
The dashed lines represent the 1:1 relations.
}
\label{fig:fcal_L3000_2SLAQ_vs_SDSS}
\end{figure*}

Spectra obtained as part of the 2QZ and 2SLAQ surveys, using the AAT, are not flux calibrated.
Previous studies used single-band magnitudes, in either the \bj- or $g$ bands, to derive monochromatic luminosities, assuming all spectra follow a uniform UV-to-optical SED of $f_{\rm \nu}\propto\nu^{-0.5}$ \cite[e.g.,][]{Croom2004,Richards2005_2SLAQ,Fine2008}.
Here we employed a more robust flux calibration scheme, which relies on all the photometric data available in the 2QZ and 2SLAQ catalogs.
All sources within the 2QZ catalog have tabulated $u$, \bj\ and $r$ band magnitudes, and all sources within the 2SLAQ catalog have the full set of $ugriz$ magnitudes.
The following two procedures were tested using a unique sample of sources, which have duplicate spectroscopy from both the SDSS and 2SLAQ surveys, kindly provided by S. Fine. 
Although the original sample had \NSLAQXSDSSraw\ sources, here we focus only on the \NSLAQXSDSS\ sources that passed basic quality criteria, and that have $z<1.3$, in order to avoid the effects of the telluric features on the estimation of \Lthree.
The (fully calibrated) SDSS spectra of these sources were fitted using our \mgii-fitting procedure (see \S\ref{sec_fitting}). 

First, we test how can the determination of monochromatic luminosities be improved if one uses the photometric band which is as close as possible to the (observed-frame) spectral region under question.
In the case of \Lthree, for sources at $z\simeq1$, 1.5 or 2, the relevant bands are $r$, $i$ or $z$, respectively. 
To test this method for the \NSLAQXSDSS\ sources, we shifted the composite of VdB01 to the appropriate redshift and then scaled to the observed SDSS (AB) magnitude in the nearest available band. 
We used the full composite, including emission lines, and the scaling was done through synthetic photometry, using the entire passband of the relevant filter.
Next, we measured the continuum flux density of the shifted and scaled template in a narrow band around (rest-frame) 3000\AA, similarly to the main \mgii\ fitting procedure.
The left panel in Figure~\ref{fig:fcal_L3000_2SLAQ_vs_SDSS} compares \Lthree\ obtained through this method to the one measured from the SDSS spectra.
The luminosity estimates are highly consistent with those derived from our \mgii\ fits of the SDSS spectra - 
, the mean difference is 0.06 dex and the scatter around the 1:1 relation is 0.11 dex.

Since the method described above assumes a uniform SED, it necessarily results in calibrated spectra where ratios of \lamLlam\ estimated at \textit{any} two wavelength bands remain constant. 
In such a situation, there is no justification for re-calibrating any specific $\RBLR-\lamLlam$ relation, since all slopes will remain those of the initial relation ($\RBLR-\Lop$ in our case), and the change in scaling factors would \textit{only} reflect the assumed (uniform) SED.
Thus, we devised an additional method that attempts to overcome this issue.

\begin{figure}
\includegraphics[width=0.47\textwidth]{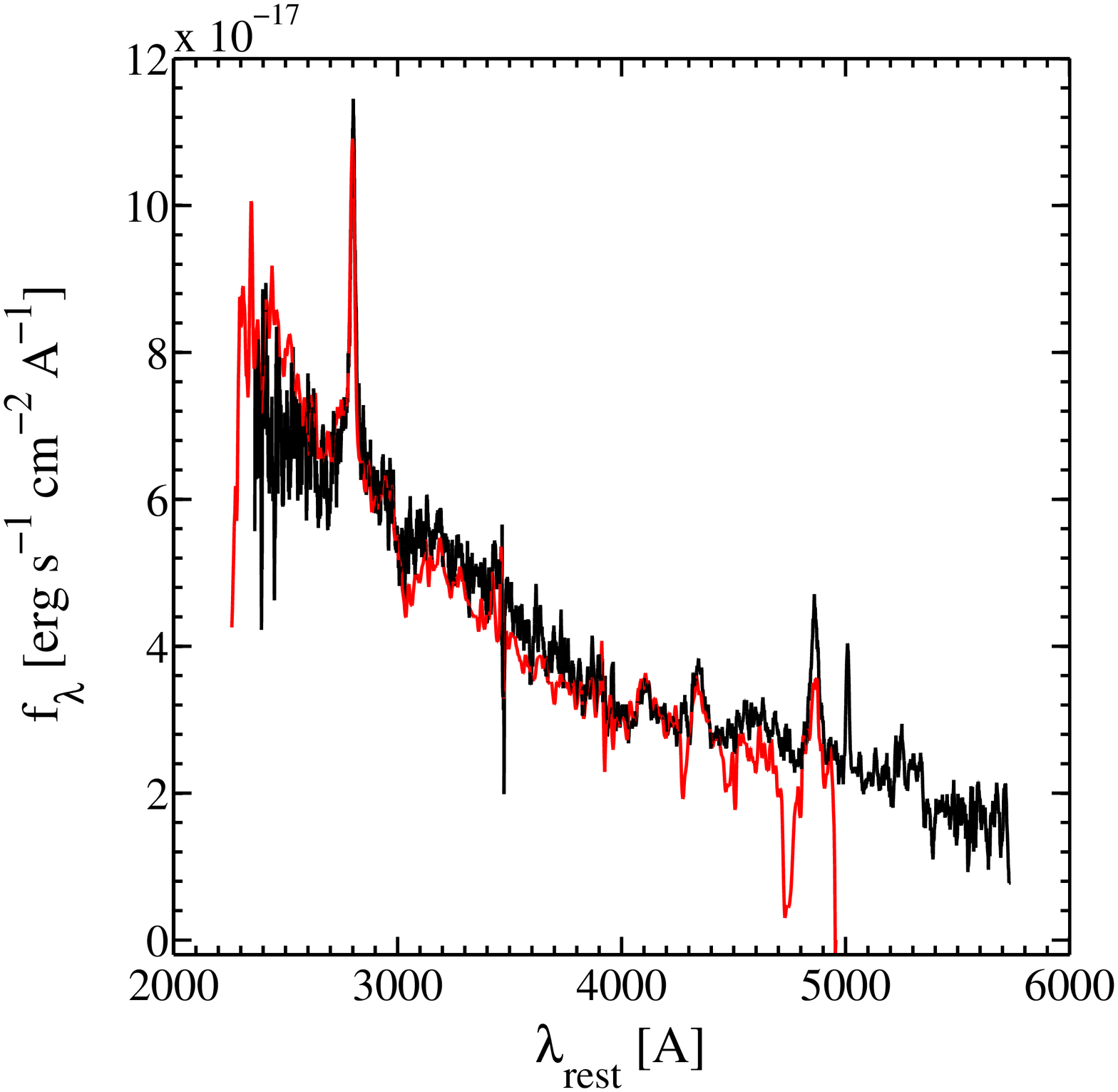}
\caption{
Example of a 2SLAQ spectrum, calibrated using our procedure, for object 
J003844.86-004404.5 ($z=0.6074$). 
The SDSS spectrum is shown in black (smoothed by a boxcar filter of 11 pixels). 
The 2SLAQ spectrum, calibrated using all the available SDSS photometry and a best-fit power law, is shown in red (smoothed over 9 pixels). 
}
\label{fig:fcal_L3000_pl_example}
\end{figure}

We assume that each AAO-observed source can be described as a combination of a power-law continuum and an emission (and absorption) line spectrum. 
The power-law component is provided by a grid of spectra of the form $f_{\nu}\propto\nu^{\alpha}$, 
with $\alpha$ varying between -2 and 0 in steps of 0.05.
Each uncalibrated spectrum is first smoothed using very broad boxcar averaging (251 pixels, i.e. $\sim$1080\AA).
Next, the input spectrum is divided by its smoothed version, to establish a (relative) emission line spectrum.
The emission line spectrum is then multiplied by each of the power-law spectra in our grid.
All such ``combined'' spectra are scaled to match the observed $r$ band magnitude of the source.
Finally, the best combined spectrum is chosen, such that it best matches the observed $g$ or \bj\ magnitudes, for sources from the 2SLAQ or 2QZ catalogs, respectively.
Figure~\ref{fig:fcal_L3000_pl_example} illustrates the resultant pseudo-calibrated spectrum of J003844.86-004404.5 ($z=0.6074$), observed as part of the 2SLAQ survey, compared with its SDSS spectrum.
The right panel of Figure~\ref{fig:fcal_L3000_2SLAQ_vs_SDSS} compares \Lthree\ obtained through the power-law fitting method to the one measured from the SDSS spectra. 
The mean difference between the estimates if now merely 0.03 dex, and the scatter around the 1:1 relation is 0.10 dex, indistinguishable from the scatter found for the ``uniform SED'' method outlined above. 

In conclusion, since the two methods provide consistent estimates of \Lthree, both can be used in future studies that rely on 2QZ and 2SLAQ spectra of type-I AGN. 
However, only the full power-law fitting method allows for a non-uniform UV-optical continuum SED, 
and thus we adopt it in this paper.

\section{Additional Information on Small Samples}
\label{app_samples}

The following provides additional information on the literature samples used in this work. 

\begin{itemize}

\item \NB04 PG quasars at $0.25\lesssim z\lesssim 0.5$, from the study of \citet[][BL05 in the present paper]{Baskin2005}.
 For these sources, the \hb\ line was observed as part of the survey of \cite{BG92}, and fitted using the procedure detailed in \S\ref{sec_fitting}.
 From the original sample of 81 sources, we omit the \NSh07 sources that were also studied in \citet[][see below]{Shang2007}.
 The \civ\ line was observed as part of several \textit{HST} and \textit{IUE} campaigns, and the archival spectra were fitted by BL05. 
 We adopt their \civ\ measurements (Table 1 of BL05) but use our own \hb\ fits.

\item \NN07 sources at $z\simeq2.4$ and $z\simeq3.3$ from the study of \cite[][N07 in the present paper]{Netzer2007_MBH}, which incorporates also the sample presented in \cite[][S04 in the present paper]{Shemmer2004}. 
The \civ\ line was observed in a variety of optical surveys and the \hb\ line was observed in a dedicated \hkband\ campaign.
All line measurements for this sample were carried out using the procedures detailed in \S\ref{sec_fitting}.

\item \NSh07 PG quasars at $0.1\lesssim z \lesssim 0.4$ from the study of \cite[][Sh07 in the present paper]{Shang2007}. 
The \civ\ line was observed by the \textit{HST} and the \mgii\ and \hb\ lines were observed in a dedicated, almost-simultaneous ground-based optical campaign.
This dataset is exceptionally useful for our analysis, due to the high quality of the observations \& related calibrations, and the detailed measurements of \hb, \mgii\ \textit{and} \civ\ for all \NSh07 sources. 
We thus preferred the usage of these measurements over those reported in BL05 (see above), for all the Sh07 sources. 
We used the tabulated parameters of the Sh07 study for all lines, as well as the tabulated continuum parameters.
However, since the Sh07 study fits a power law continuum over the entire (rest-frame) UV-optical range, the tabulated parameters underestimate \Lthree\ as defined here (see \S\ref{sec_fitting}). We correct for this by applying the typical correction factor derived in \S\ref{sec_fitting}.

 \item \NSul07 sources at $z\lesssim0.3$ from the study of \cite[][Sul07 in the present paper]{Sulentic2007}. 
 This study presents a compilation of \civ\ measurements that are based on archival \textit{HST} spectra of 130 sources, many of which overlap with the BL05 or Sh07 samples. 
 We chose \NSul07 sources for which \hb\ was reliably measured, by applying our fitting procedure to spectra obtained either from the SDSS or from the catalog of \cite{Marziani2003_cat}. 
 We omitted all sources with $z<0.05$, to avoid severe host galaxy contamination.
 For the sake of consistency, we prefer our own fits over the \hb\ line measurements published within \cite{Marziani2003_cat}.
 For 11 sources, the SDSS spectra also cover the \mgii\ line. These sources are thus part of the small subset of sources with measurements of all three emission lines.
The SDSS counterparts are removed from the large SDSS sample.

 \item \NMc08 SDSS sources at $z\simeq0.36$ from the study of \cite[][Mc08 in the present paper]{McGill2008} for which the \mgii\ and \hb\ lines were observed in a dedicated Keck/LRIS campaign \cite[][]{Woo2006_Msig}. 
 The SDSS spectra of these sources were removed from our SDSS sample.

 \item \ND09 sources at $z\sim1-2.2$ from the study of \cite[][D09 in the present paper]{Dietrich2009_Hb_z2} for which the \civ\ (9 sources) and \mgii\ (7 sources) lines were observed as part of several optical surveys. 
 The \hb\ line was observed in a dedicated campaign in the $J$ and $H$ bands.

 \item \NM09 sources at $z\sim1-2.4$ from the study of \cite[][M09 in the present paper]{Marziani2009} for which the \civ\ line was observed as part of the Hamburg-ESO survey \cite[]{Wisotzki2000} and the \hb\ line was observed in the $sZ$, $J$ and $H$ bands.

\item \NT11 sources at $z\simeq4.8$ from the study of \cite[][T11 in the present paper]{Trakhtenbrot2011} for which the \civ\ line was observed as part of the SDSS and the \mgii\ lines were observed by a dedicated \hkband\ campaign.
 All line measurements for this sample were carried out using the procedures detailed in \S\ref{sec_fitting}.

\end{itemize}

\section{Line Fitting}
\label{app_line_fitting}

\subsection{The \hb\ complex}
\label{app_hb}

We fitted the \hb\ line complex in a very similar manner to the one described in NT07. 
The linear continuum is fitted between two bands around either 4435\AA\ or 4730\AA\ and 5110\AA. 
The continuum is subtracted from the observed spectrum and an initial fit of the narrow \oiii$\lambda\lambda 4959,5007$\AA\ lines is used in order to determine the systematic redshift and the velocity dispersion of the narrow line region (NLR) gas, so to assist with the full \hb-\oiii\ fit (see below). 
The two \oiii\ lines are forced to have the same systematic shift and width, in the range  $300<\fwhm<1200\,\kms$, and a flux ratio of 1:3.

The grid of \feii\ templates is constructed by broadening and shifting the original template of \cite{BG92}. 
The convolution is preformed with a single Gaussian profile with a width that increases in steps of $25\,\kms$, in the range of 1200 to 20000 \kms. 
The templates are shifted in steps of 69 \kms ($\sim$1\AA), in the range of -1000 to +1000 \kms.
Our \feii\ grid thus consists of 23343 different, normalized, templates.
The best-fit template is chosen through a standard $\chi^2$ minimization scheme, which focuses on the flux density in the 4400-4650\AA\ range and provides the best-fitting scaling factor. 
The scaled best-fit template is then subtracted from the spectra so that the continuum can be re-measured.
The process is repeated until we receive continuum- and iron-free spectra. 
Several of the samples considered here (N07; Sh07; M09; D09) were originally fitted by codes that neglected the possible shift of the \feii\ template. However, the differences in the main derived parameters (\Lop\ and \fwhb) are negligible \cite[see ][]{Hu2008a,Marziani2012}.

Finally, we fit an emission line model which consists of 5 Gaussians: two broad and one narrow components for the \hb\ line and a single narrow component for each of the \oiii\ lines. 
Each of the broad components is allowed to have a width in the range $1200<\fwhm<20000\,\kms$, and 
to be shifted by as much as 1000 \kms, relative to the redshift determined by the initial \oiii\ fit.
All three narrow components are forced to have the systemic shift and line widths.
These are further constrained to differ only by as much as 200 \kms (shift) and 30\% (width), from the parameters determined in the initial \oiii\ fit, but not to exceed the allowed range of 
$300<\fwhm<1200\,\kms$.
The flux ratio of the two \oiii\ features is again fixed at 1:3, while the ratio between the narrow \hb\ component and the \OIII\ line is free to take any value in the range of $0.05-1$ (i.e. the presence of {\it any} significant \oiii\ signal dictates the presence of a narrow \hb\ component).

\subsection{The \MgII\ complex}
\label{app_mgii}

%

We fit a linear pseudo-continuum is to the flux around 2655 and 3020\AA. 
This component completely ignores the Balmer continuum emission and thus does \textit{not} represent the ``real'' (accretion originated) underlying continuum emission.
The typical difference between the ``real'' \Fthree\ and the one we measure here is, however, small. 
For example, combining the composite spectrum of VdB01 and the best-fit UV-optical power law continuum ($f_{\nu}\propto\nu^{-0.44}$) implies that the measured \Fthree\ is larger by a factor of 1.16.
As a second test, we used the SDSS sample under study. The ``real'' 3000\AA\ continuum was estimated by fitting a power law between narrow bands at 2200 and 4200\AA.
This limits the test to sources with $0.75<z<1.2$. The observed \Fthree\ was measured by applying the full-scale \mgii-fitting procedure.
The resulting median factor for these 14532 sources is 1.15 with a standard deviation of 0.06 dex ($10\%$). 
We use this factor to convert the ``real'' \Lthree\ measurements of the Sh07 samples, to match our \Lthree\ measurements.
We use a \feii\ \& \feiii\ template made of the composite prepared by Vestergaard \& Wilkes (2001) and several additional \feii\ lines in the region not covered by the above composite, kindly provided by G.~Ferland (private communication). 
The new lines are similar to the ones presented in \citet[][Fig.~13]{Sigut_Pradhan2003} and in \citet[][Fig.~5]{Baldwin2004_Fe}. 
The \feii\ flux under the \mgii\ line is dominated by the red extension of the $\sim$2750\AA\ emission complex.
The additional flux tends to flatten in sources with very broad \feii\ lines, but the overall effect on the \mgii\ fitting is very small. We consider this template to be more reliable than the addition of constant flux under the \mgii\ line, as done by, e.g., \cite{Kurk2007} and \cite{Fine2008}.
Other studies, such as \cite{Salviander2007} and \cite{Shen_dr5_cat_2008}, use templates that are very similar to ours, following the models in \cite{Sigut_Pradhan2003}.
A visual inspection of the spectra suggests that the iron lines blueward and redward of \mgii\ must not share a common scale factor. 
Instead, our procedure finds the template which best fits the spectrum in the range of 2600-2700\AA, in terms of shift and width of the Iron features. 
The scaling in the 2900-3030\AA\ region is allowed to vary by up to 15\%. 
We then combine the two templates, characterized by the same shift and width, so that they join at 2810\AA.
Each of the two doublet profiles consists of two broad and one narrow emission Gaussians, and a single (narrow) absorption Gaussian. 
We have fixed the intensity ratio of the two broad \mgii\ doublet components to 1:1, suitable for optically-thick lines, while for the narrow components the ration is 1:2.
Since the narrow emission component of \mgii\ is difficult to disentangle, we have experimented with an alternative procedure in which the width of the narrow \mgii\ line was bound to match the
one measured from the \OIII\ line, in sources where both are observed (the SDSS \hbXmg\ sub-sample). 
This technique resulted in either a width which is similar to the one achieved by the free-fitting
version, or in a fit of inferior quality. 
Our preferred procedure is, therefore, to freely fit the narrow \mgii\ doublet components with the additional constraints of $300<\fwhm<1200\,\kms$, and a shift of no more than 200 \kms, relative to the systemic redshift.
The broad components are allowed to have $1200<\fwhm<15000\,\kms$, and to be shifted by as much as 1000 \kms.
The (narrow) absorption feature is set to have a doublet Gaussian profile with and is allowed to have $150<\fwhm<2000\,\kms$, and a center which is blue-shifted by as much as 5000 \kms, with respect to the expected (laboratory) wavelengths of the \mgii\ doublet, assuming the systemic redshift.
As with the other line-fitting parameters, we verified that these choices capture the properties of the vast majority of sources that show such absorption features, while not affecting the quality of the {\it broad} \mgii\ profile fit.

\subsection{The \CIV\ complex}
\label{app_civ}

Our procedure closely follows the one presented in \cite{Fine2010_CIV} and we refer the reader to the very thorough discussion in their Appendix A.
In particular, we do \textit{not} fit the weak \feii\ features around the \civ\ line. 

The components of our \civ\ line model are very similar to those of the \mgii\ line model (see ~\ref{app_mgii} above).
Each of the two doublet profiles consists of two broad and one narrow emission Gaussians, and a single (narrow) absorption Gaussian, with the intensity ratios following the same scaling as for \mgii.
The broad components are allowed to have line widths in the range $1200<\fwhm<15000\,\kms$, and to have shifts in the range of -3000 to +1500 \kms, thus taking into account significantly blueshited \civ\ profiles, which are often observed in type-I AGN \cite[][and references therein]{Richards2011_CIV}.
The narrow components are allowed to have $300<\fwhm<1200\,\kms$, and to be shifted by as much as 200 \kms.
We note that our choice to include a narrow emission component does not mean that such a component contributed significantly to the line profile. 
Indeed, several studies claimed that the narrow emission component in \civ\ is very weak, or non-existent altogether (see, e.g., \citealt{Wills1993}, BL05 and the discussion in \citealp{Vester_Peterson2006}).
Although our code allowed the peak intensity of the narrow component to reach 0.5 that of the broad components, 
its actual contribution to the overall fitted \civ\ line profile is indeed found to be insignificant. 
For the SDSS \mgXciv sub-sample, the median of the EW(\civ,NLR)/EW(\civ,BLR) ratio is only$\sim0.007$, while for 99\% of the sources this ratio is smaller than about 0.05 and for 25\% of the sources it is completely negligible (i.e., $<0.001$).
Thus, the inclusion of a narrow \civ\ component is not expected to significantly affect measurements of the {\it broad} component of the line, which we use throughout the present work.
This also justifies our usage of broad-\civ\ measurements that include (e.g., BL05) or don't include (e.g., Sh07) a narrow \civ\ component.
The narrow absorption feature is set in the same manner as in the case of the \mgii\ line (see \ref{app_mgii} above).

We also assigned a single broad Gaussian component to each of the \HeIIuv, \OIIIuv\ and \NIVuv\ lines, which are forced to have a common line width, in the rage of  $1200<\fwhm<8200\,\kms$, and a common shift, in the range of -1500 to +750 \kms.
As mentioned in \cite{Fine2010_CIV}, these features may in practice account for emission which originates either from \feii\ or other, unresolved, lines (e.g., Al\,{\sc ii}\,$\lambda1671$).
The derived properties of these lines are \textit{not} used to infer the physical parameters of the BLR gas.


\label{lastpage}
\end{document}